\documentclass[prd, onecolumn, nofootinbib]{revtex4}

\usepackage{epsfig} 
\usepackage{amsmath} 

\newcommand{\beq}{\begin{equation}}
\newcommand{\eeq}{\end{equation}}
\newcommand{\beqa}{\begin{eqnarray}}
\newcommand{\eeqa}{\end{eqnarray}}
\newcommand{\om}{\Omega_m}

\newcommand{\ode}{\Omega_{DE}}

\newcommand{\dls}{d_{\rm lss}}
\newcommand{\zd}{z_{\rm div}}
\newcommand{\whi}{w_{N+1}}

\def\pa{{\partial}}

\def \F{{\bf F}}

\begin{document} 

\title{To Bin or Not To Bin: Decorrelating the Cosmic Equation of State} 
\author{Roland de Putter \& Eric V.\ Linder} 
\affiliation{Berkeley Lab \& University of California, Berkeley, CA 94720} 

\date{\today} 

\begin{abstract} 
The physics behind the acceleration of the cosmic expansion can be 
elucidated through comparison of the predictions of dark energy equations 
of state to observational data.  In seeking to optimize this, we 
investigate the advantages and disadvantages of using principal component 
analysis, uncorrelated bandpowers, and the equation of state within 
redshift bins.  We demonstrate that no one technique is a panacea, with 
tension between clear physical interpretation from localization and 
from decorrelated errors, as well as model dependence and form dependence.  
Specific lessons include the critical role of proper 
treatment of the high redshift expansion history and the lack of a unique, 
well defined signal-to-noise or figure of merit.  
\end{abstract} 

\maketitle

\section{Introduction \label{sec:intro}}

The acceleration of the universe poses a fundamental mystery to 
cosmology, gravitation, and quantum physics.  Understanding the nature 
of the dark energy responsible for the acceleration relies on careful, 
robust measurements of the dark energy properties, in particular its 
equation of state (EOS), or pressure to energy density, ratio that directly 
enters the Friedmann equation for cosmic acceleration.  As scientists 
design the next generation of dark energy experiments they seek to 
optimize the measurements for the clearest insight into this unknown 
physics. 

Two critical pieces of information will be the value of the EOS at some 
epoch, such as the present, and a measure of its time variation, in much 
the way that early universe inflation theories are classified by the 
value of the spectral index and its running.  The best parametrized EOS 
are physics 
based and model independent, i.e.\ able to describe dark energy dynamics 
globally, or at least over a wide range of behaviors.  Such EOS are very 
successful at fitting to data and projecting the results of future 
experiments, and can be robust to bias against inexact parametrization. 

Other approaches seek to remove one drawback of parametrized EOS by 
not assuming a functional form for the time variation, lest the true 
dark energy model lie outside the apparently wide range of validity of 
the form, i.e.\ they aim for form independence.  Two major avenues for 
achieving this are decomposition into basis functions or principal 
components (e.g.\ \cite{HutStark03}, also see 
\cite{CritPog05,ShapTurn06,SimpBrid06,DickKnoxChu06,Ste06,HutPeir07}) 
and individual values of the EOS $w(z)$ over finite 
redshift bins, which become more general as the number of elements 
increases.  However uncertainties in estimation of the EOS properties 
also grow as the number of principal components or bins increases.  

This article begins by examining general properties of the cosmological 
data and its dependence on the EOS in \S\ref{sec:cosdep}.  Many of the 
later, detailed results will already be foreshadowed by this straightforward 
and general analysis.  In \S\ref{sec:pca} we examine principal component 
analysis of the EOS and in \S\ref{sec:band} uncorrelated bandpowers. 
Bins of EOS in redshift is investigated in \S\ref{sec:bin}, including 
figures of merit for quantifying the uncertainties.  Further concentration 
on the crucial role of the high redshift EOS, and the risk of biasing 
parameter estimation, occurs in \S\ref{sec:whi}.  We consider physical 
constraints on EOS properties in \S\ref{sec:wlimit} and summarize our 
results and conclude in \S\ref{sec:concl}.

\section{Cosmological Information and the Equation of 
State \label{sec:cosdep}} 

Cosmological observations probe the EOS through its influence on the 
cosmic expansion history and the growth history of massive structures. 
The relation involves in general an integral (or double integral) over 
the EOS.  This implies that the kernel, or response of the observables 
to the EOS, is broad in redshift, not tightly localized.  For distances, 
the EOS at one redshift formally influences distances at all higher 
redshifts, while for growth variables that EOS value influences all lower 
redshifts; this implies a certain skewness.  After setting up the 
simulated observations, we demonstrate that cosmological information is 
difficult to simultaneously localize and decorrelate, as well as 
highlighting some necessary cautions regarding treatment of data and priors. 

\subsection{Cosmological Variables} 

Information inherent in measurements of cosmological quantities regarding 
the EOS and other parameters can be estimated through the Fisher 
information matrix, 
\beq
\label{eq: fisher def}
F_{i j}=\sum_{k,k'} \frac{\pa O_k}{\pa p_i}\, COV^{-1}[O_k,O_{k'}] \, 
\frac{\pa O_{k'}}{\pa p_j}, 
\eeq 
where $\pa O_k/\pa p_i$ gives the sensitivity of observable $O_k$ to 
parameter $p_i$, and $COV$ gives the measurement covariance matrix.  
One often takes the measurement errors to be diagonal, 
$COV\to\sigma_k^2\,\delta_{kk'}$.  Alternately one could use another 
likelihood estimator 
such as a Monte Carlo Markov Chain; the general results will not change.  
Each observable depends on the EOS and other parameters such as the 
present matter density relative to the critical density, $\om$. 

For the EOS, we begin by dividing the redshift interval $(0, z_{\rm max})$ 
into $N$ bins of not necessarily equal widths $\Delta z_i$ 
($i = 1,\dots N$), where $\sum_i \Delta z_i = z_{\rm max}$.
The index $i$ is taken to increase with $z$.  The equation 
of state is written as 
\beq
\label{eq: expansion w}
w(z) - w_b(z) = \alpha_i \,e_i(z)
\eeq
(repeated indices are to be summed over), where $e_i(z) = 1$ inside the 
$i$th bin and zero outside.  
Such a binning is general, and serves as the first step for investigation 
of principal components (\S\ref{sec:pca}), decorrelated bandpowers 
(\S\ref{sec:band}), or straight binning (\S\ref{sec:bin}).  

The $N$ coefficients $\alpha_i$ are the parameters describing the EOS 
in this model.  Note that these coefficients measure the equation of state 
relative to some ``baseline'' equation of state $w_b(z)$.  We can choose 
$w_b$ to be some model, like the cosmological constant $\Lambda$ 
($w_b = - 1$), to which we want to compare the data.  We address issues 
of the baseline EOS and 
binning variable in \S\ref{sec:pca}.  For convenience we 
sometimes write $e_i(z)$ as ${\bf e_i}$, and $\alpha_i$, in the case 
where ${\bf e_i}$ is a unit box function, as $w_i$. 

For cosmological observables, we focus here on various distances, including 
as measured by Type Ia supernovae (SN), by the cosmic microwave background 
(CMB) acoustic peaks, and by baryon acoustic oscillation (BAO) patterns 
in large scale structure.  For all these the EOS enters through the 
Hubble parameter 
\beq 
H(z)/H_0=\left[\om\,(1+z)^3+\ode\,f(z)\right]^{1/2}, 
\eeq 
where the present dark energy density $\ode=1-\om$ for a spatially flat 
universe as assumed here.  The function $f(z)$ is the ratio of the dark 
energy density at redshift $z$ to its current energy density. When $z$ 
lies in the $j$th EOS bin, 
\beq
f(z) = \left( \frac{1 + z}{1 + z_j}\right)^{3(1+w_j)}\, \prod_{i = 1}^{j-1} 
\left(\frac{1 + z_{i + 1}}{1 + z_i} \right)^{3(1 +w_i)}, 
\eeq
where $z_i$ is the lower redshift bound of the $i$th bin (note $z_1 = 0$) 
and $w_i$ the fiducial value of the EOS in that bin. 

The SN luminosity distance data set extends from redshift zero to 
$z_{\rm max}=1.7$, with a distribution and systematic errors as given 
for the future SNAP mission in \cite{Kimetal04}.  CMB data is treated as a 
0.7\% constraint on the reduced distance to last scattering, $\dls= 
(\om h^2)^{1/2}\int_0^{1089} dz/H(z)$, as should be available from the 
Planck mission.  In addition to the $N$ EOS bins between 
$z = 0-z_{\rm max}$, we define a single bin for redshifts $z>z_{\rm max}$ 
having averaged, hence constant, EOS $w_{N + 1}$.  Note that freely 
marginalizing over $w_{N+1}$ when only one data point depends on this 
parameter is equivalent to not including the parameter and the data point.  
We consider BAO in \S\ref{sec:whi}. 
Thus the Fisher matrix has dimensions $(N+3)\times(N+3)$, with $\om$ 
(or equivalently $\ode$) and the parameter ${\mathcal M}$ giving the 
combination of SN absolute magnitude and Hubble constant in addition to 
the $N+1$ EOS values $w_i$.  Unless otherwise stated, results shown 
marginalize over $\om$ and ${\mathcal M}$. 

\subsection{Information Localization} 

Ideally, binned EOS would reflect an invariant measure of the information 
(or conversely, uncertainty) at its particular redshift.  Such a mapping 
between information and local variables, or bandpowers, works well for 
large scale structure (LSS), even into nonlinear scales, and we follow the 
approach of \cite{HamTeg00} but apply it to the EOS.  
To refine the localization of information one can attempt to use a large 
number of bins.  We initially consider $N=100$ EOS bins equally spaced in 
redshift. 

\begin{figure}
  \begin{center}{
  \includegraphics*[width=8.8cm]{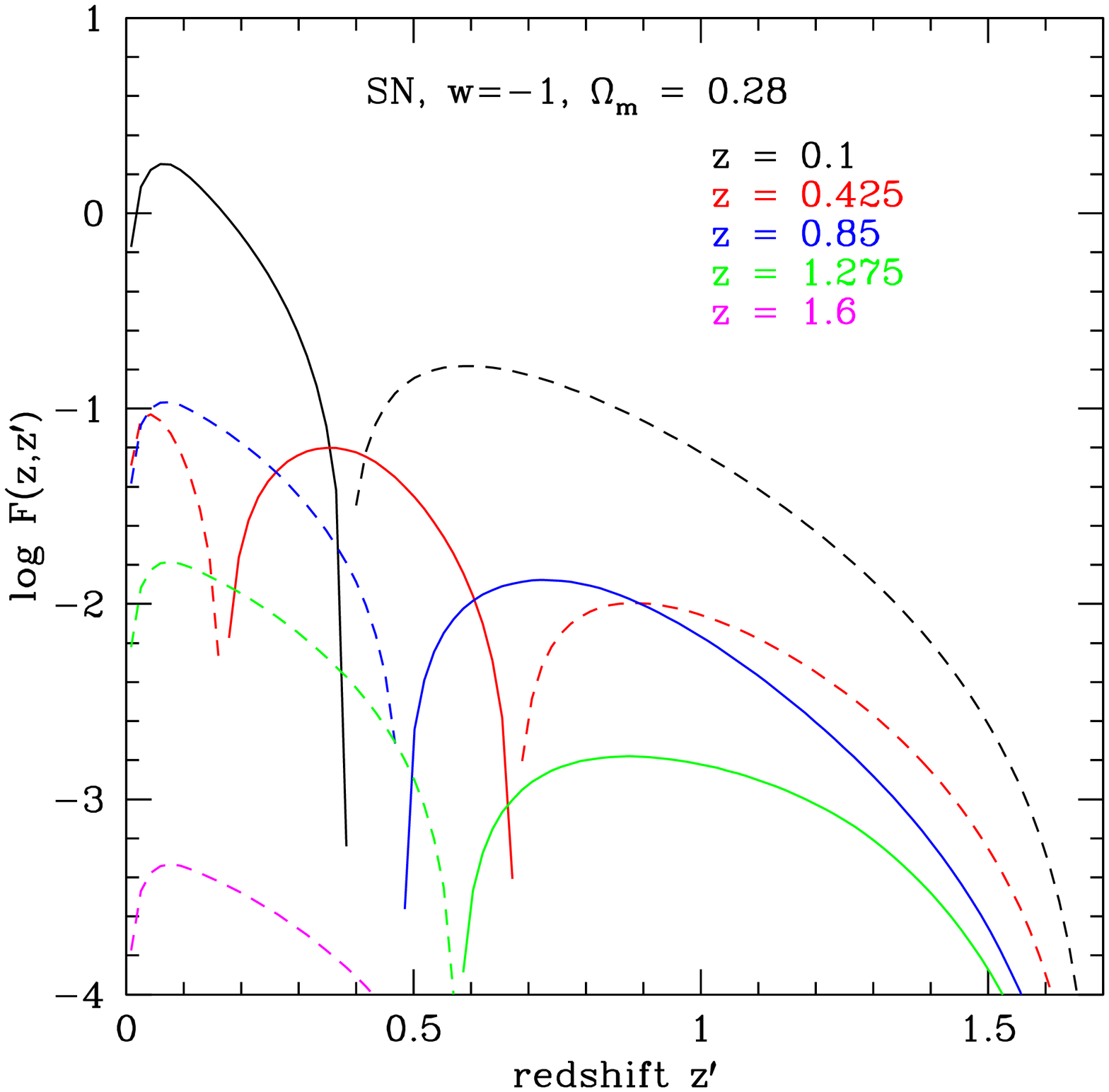}
  \includegraphics*[width=8.8cm]{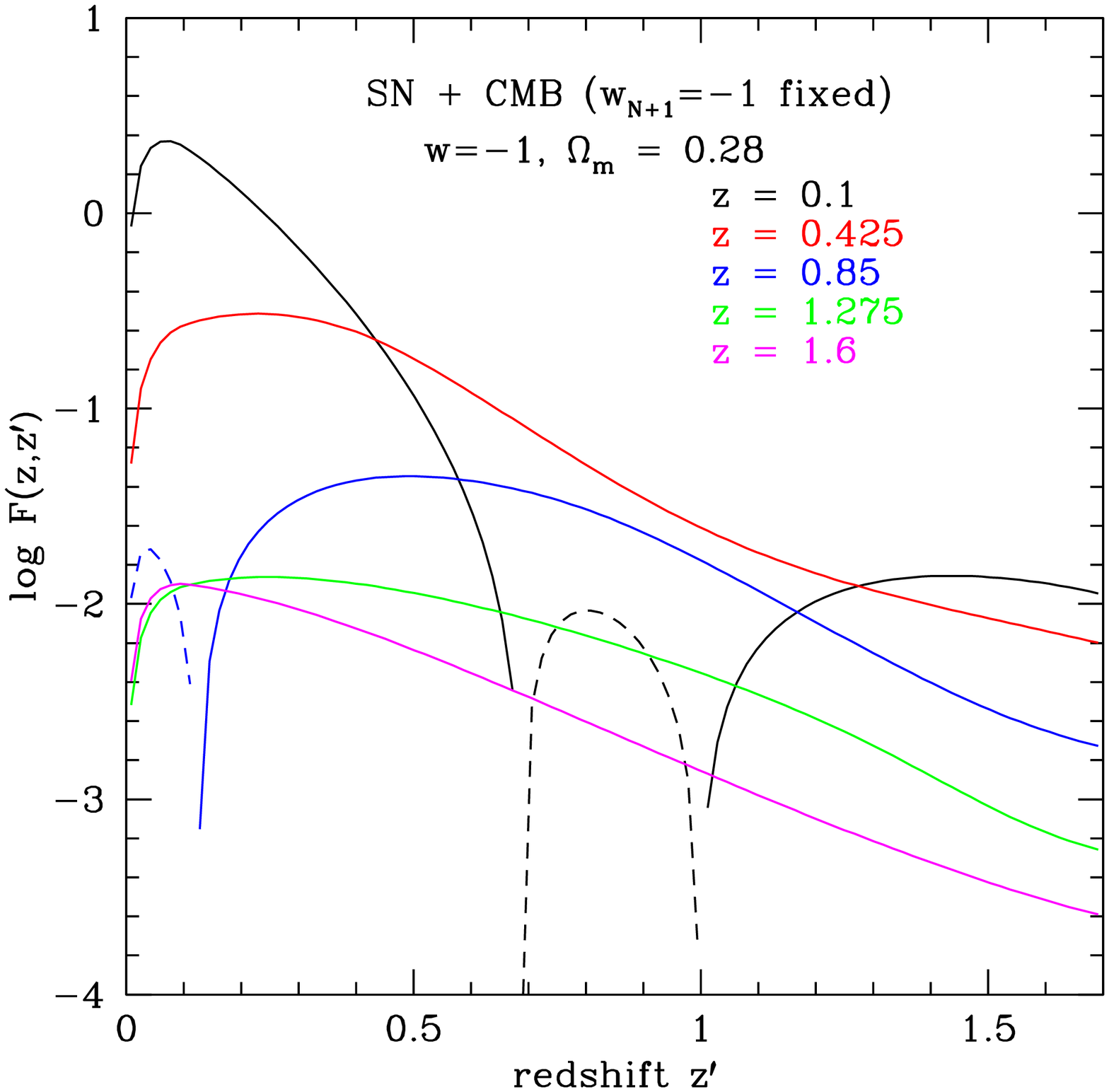}
  }
  \end{center}
  \caption{Five rows (or columns, $\F$ is symmetric) of the Fisher matrix 
calculated using a uniform binning in redshift $z$ ($N = 100$ bins), 
showing the cosmological information as a function of redshift.  Dashed 
lines show where $F_{i j}$ is negative.  The first panel uses only 
supernova data, the second panel includes the distance to CMB last 
scattering $\dls$, with the equation of state for $z = 1.7-1089$ 
fixed to the fiducial value $w_{N + 1} = -1$.  The curves of 
information are far from sharp spikes at $z=z'$, indicating the cosmological 
information is difficult to localize and decorrelate. 
  }
  \label{fig: Fisher}
\end{figure}

Figure \ref{fig: Fisher} plots five rows of the Fisher information matrix 
as a representation of the information as a function of redshift.  An 
element $F(z,z')$ denotes the Fisher matrix entry $F_{ij}$ with respect to 
parameters $p_i=w(z_i=z)$ and $p_j=w(z_j=z')$.  Note that in contrast 
to the LSS case (see, e.g., Fig.~1 of \cite{HamTeg00}), 
the information is far from localized (the peaks are broad), is not 
``faithful'' (the peaks do not generally peak at $z=z'$, especially for 
large $z$), and is skew (the matrix rows are not symmetric about the 
peaks).  In the LSS case, the peaks were sharp and on the matrix diagonal, 
with amplitudes some two order of magnitude above the broader ``continuum''. 
For the EOS case the kernels are broad without well defined peaks, and the 
above properties indicate the matrix is far from diagonal.  

\begin{figure}
  \begin{center}{
  \includegraphics*[width=8.8cm]{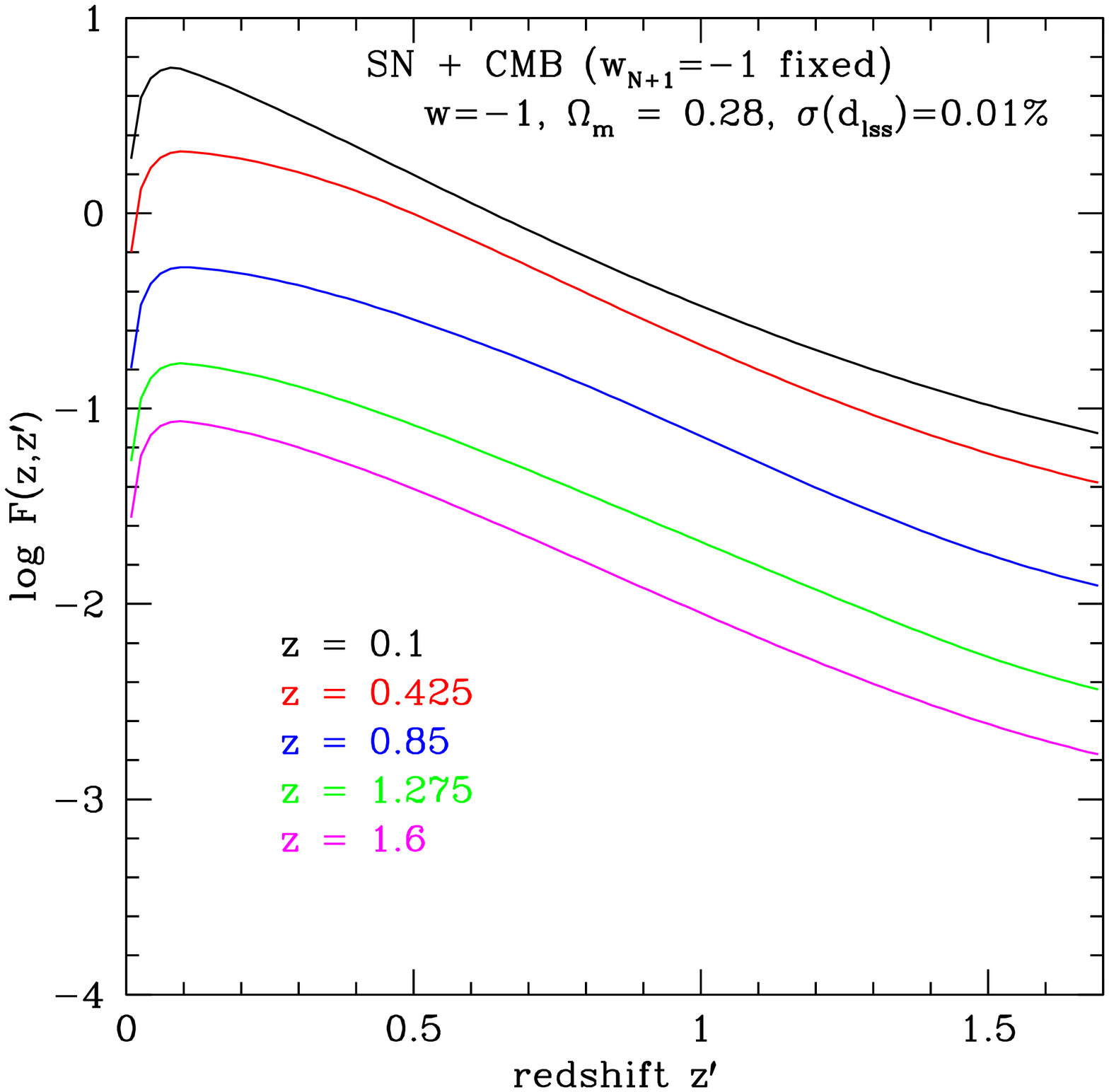}
  \includegraphics*[width=8.8cm]{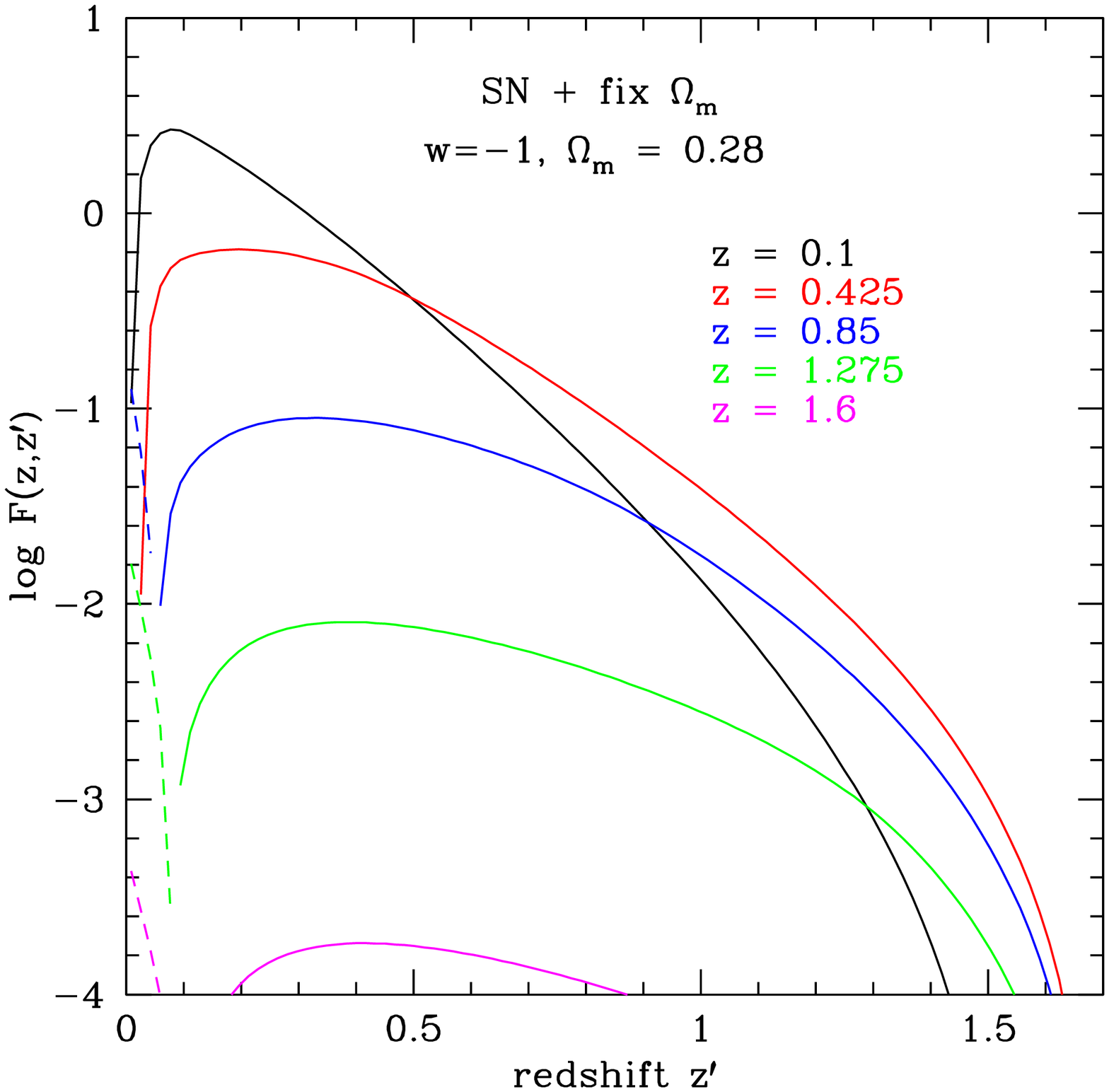}
  }
  \end{center}
  \caption{As Fig.~\ref{fig: Fisher}, but with a very tight prior on the 
CMB information $\dls$ (first panel) or fixing the matter density $\om$ 
(second panel). 
  }
  \label{fig: Fisher 2}
\end{figure}

Further difficulties arise with respect to localization or characterization 
of information for the EOS case when considering priors or additional data, 
and changes in binning or variables.  Suppose we add CMB data\footnote{We 
here simultaneously fix the value of the EOS in the one bin beyond the SN 
data, $w_{N+1}$.  As mentioned, adding one data point and marginalizing over 
the one new parameter is equivalent to not including the data and new 
parameter, i.e.\ it gives the same Fisher matrix as in the SN only case.}. 
As shown in the second panel of Fig.~\ref{fig: Fisher}, this 
has three effects:  it increases the overall amplitude of the Fisher 
matrix $\F$, broadens the peaks of the rows, and shifts the peaks to 
lower $z$, 
decreasing their ``faithfulness'' (moving the peaks further away from 
where they would be in the diagonal case).  The first effect is easy to 
understand. We add information so $\F$ becomes larger and uncertainties 
decrease.  The second and third effects can be summarized by saying 
that $\F$ is made less diagonal. This is understandable too. The CMB 
information in $\dls$ has about the same dependence on all low $z$ EOS 
parameters and thus adds to their correlation.  To check this, 
Fig.~\ref{fig: Fisher 2} shows the resulting Fisher information when 
an extremely tight prior is put on CMB data, or the matter density 
$\om$ is fixed.  Localization and faithfulness are almost completely 
lost (the EOS part of the Fisher matrix is far from diagonal).

\begin{figure}
  \begin{center}{
  \includegraphics*[width=8.8cm]{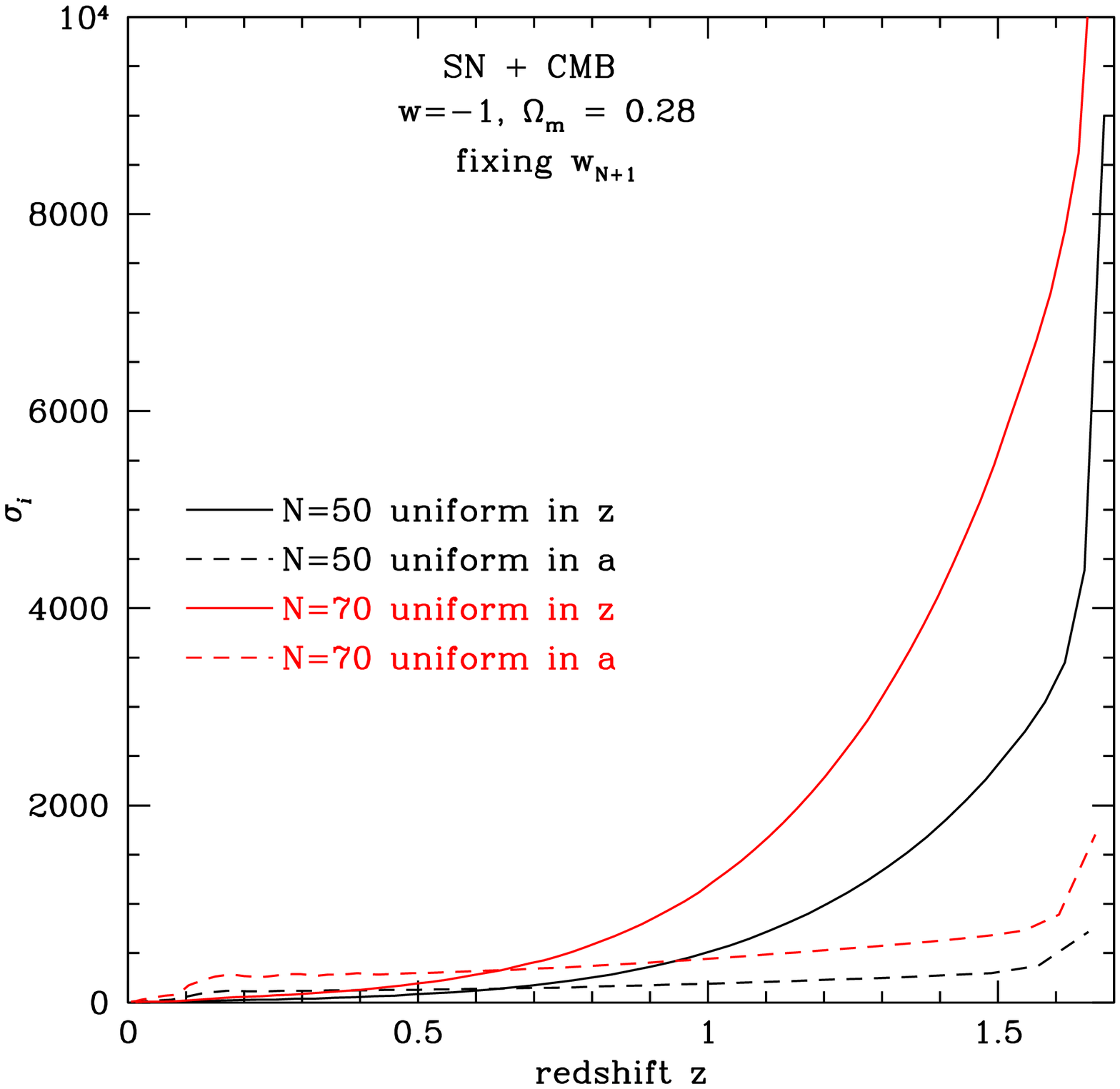}
  \includegraphics*[width=8.8cm]{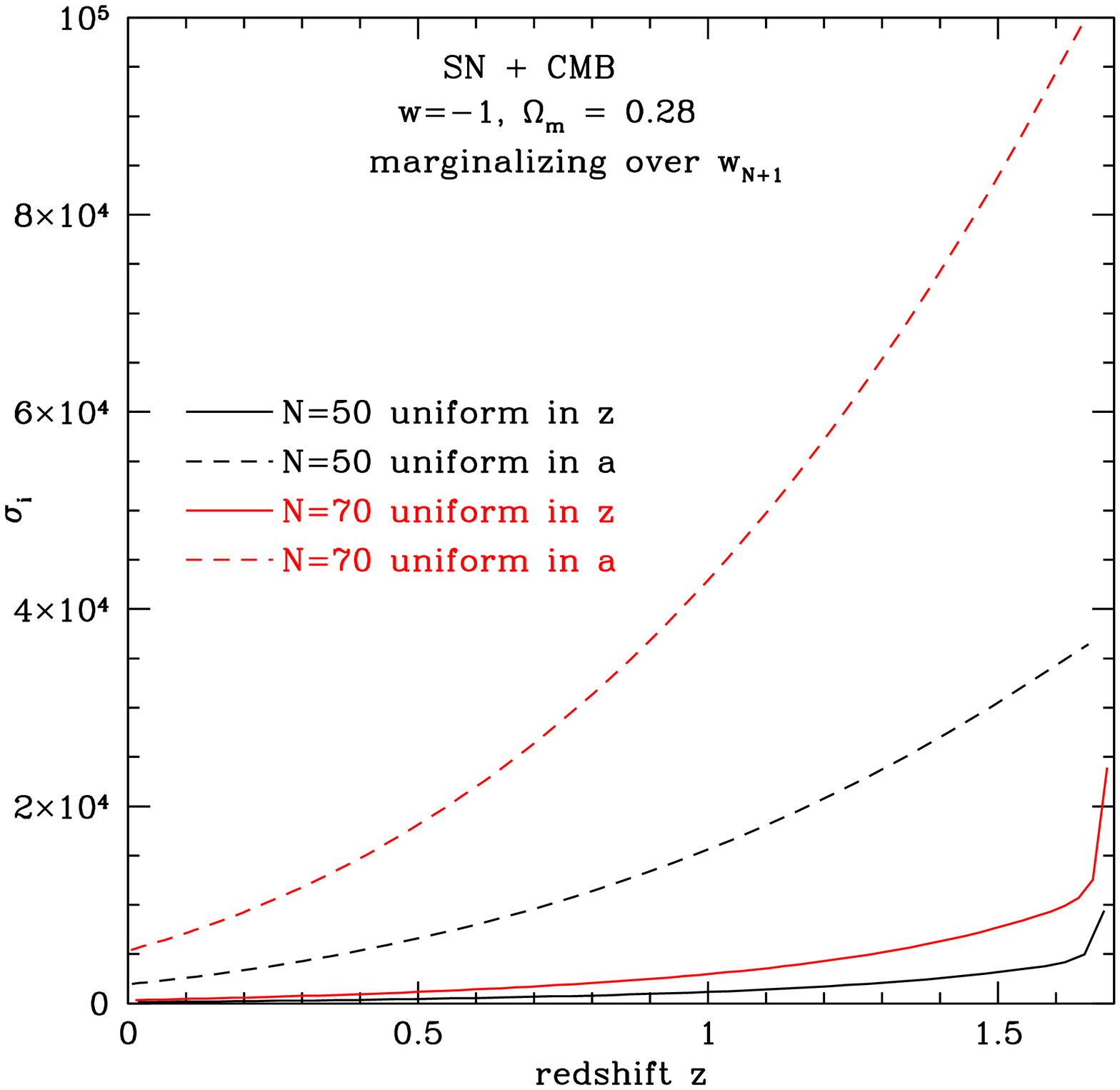}
  }
  \end{center}
  \caption{The standard deviation of the EOS in each bin for 50 and 70 bins 
uniformly spaced in redshift $z$ or scale factor $a$. The first panel shows 
the case with fixed $w_{N + 1}=-1$ and the second panel has $w_{N + 1}$ 
marginalized over. Note that the standard deviation depends on binning 
variable nontrivially and does not scale with number of bins $N$ (i.e.\ the 
inverse of the bin width) as $N^{1/2}$. 
  }
  \label{fig: sigma large N}
\end{figure}

Information within a localized region is also not invariant when 
considering changes in the number of bins or binning variable.  Note 
that changing the binning variable from redshift $z$ to scale factor 
$a=(1+z)^{-1}$ or e-fold factor $\ln a$ is equivalent to changing the 
bins to non-uniform widths in $z$.   Figure \ref{fig: sigma large N} 
demonstrates 
the variations that occur in the standard deviation of the EOS parameters 
when considering a binning uniform in $z$ vs.\ one uniform in $a$, as 
well as when changing the number of bins $N$.  A key point is that while 
the Fisher matrix behaves in a simple fashion when bin spacing is changed 
(as shown in \S\ref{sec:pca}), the uncertainties $\sigma_i$ -- which are 
square roots of the diagonal elements of the {\it inverse} of the Fisher 
matrix -- behave in a complicated manner.

First of all, when $N$ is increased from 50 to 70, and so the bin width 
is correspondingly reduced for a given 
binning variable, the $\sigma$'s do not simply scale by a factor 
$\sqrt{70/50}$ as one might have been tempted to think.  Recall that 
$N$ is the number of parameter bins not data bins.  
Thus a localized information quantity like $d\sigma^{-2}/dz$ 
does not have any real meaning, being dependent on the number of bins 
and the binning variable. 
Second, when considering a change in the binning variable, in the case 
where we marginalize over $w_{N + 1}$ (here equivalent to using only SN data), 
a binning uniform in $a$ gives larger $\sigma$'s across all redshifts 
when compared to the binning uniform in $z$, even if we use the same 
total number of bins in both cases.  This is counterintuitive since at 
low redshift the bins uniform in $a$ are smaller, and at high redshift 
they are larger than the uniform $z$ bins so we would expect the EOS 
uncertainties to be relatively larger, then smaller, respectively.  
This indeed occurs when we fix $w_{N + 1}$.  

Exploring this behavior, we find that changes in the present dark energy 
density overwhelm the EOS parameters.  For the higher redshift bins of 
EOS, the Fisher information is only contributed by the relatively few 
high redshift data points, and there the Fisher sensitivity to $\om$ can 
be more than an order of magnitude greater than to $w_i$.  Computations 
show that only when $\om$ is fixed or restricted to a degeneracy surface 
by the CMB $\dls$ constraint does the natural behavior of the EOS bin 
parameters with changes in binning become manifest.  We conclude that 
changes of binning variables, or equivalently non-uniform bin widths, 
affect EOS uncertainties in a nontrivial manner, and the treatment of the 
high redshift EOS needs care as well. 

\subsection{Extracting the Equation of State} 

The key lesson of this section has been that there is no well-defined 
measure for localized information on the EOS.  Unlike for the LSS power 
spectrum, the cosmological EOS information has a very broad kernel and 
the Fisher matrix is far from diagonal.  While one can always adopt a 
basis to transform the Fisher matrix to diagonal form, we will see that 
this does not help with localization and so the results cannot be 
interpreted as actual EOS values at a certain redshift.  Another issue 
is the problem of defining a measure of uncertainty in the EOS estimation 
that does not depend on the specific binning chosen. 

This general analysis foreshadows the problem of actually deciding how 
to quantify measurement of the EOS and any figure of merit to go along 
with that.  In the following sections we investigate three concrete 
proposals for the meaning behind EOS measurement.  One approach is 
principal component analysis (PCA; see, 
e.g.~\cite{HutStark03,HuOka04,Leach06,Kadotaetal05,MortHu07}), 
effectively making the number of bins very large, diagonalizing the 
Fisher matrix and using its eigenvectors as a basis $e_i(z)$ in 
Eq.~(\ref{eq: expansion w}).  A second approach is uncorrelated 
bandpowers, using a small number of bins, diagonalizing and scaling the 
Fisher matrix in an attempt to localize the decorrelated EOS parameters 
(see, e.g., \cite{HutCoor05, Riessetal06, SulCooHolz07}).  Finally, one 
can exactly localize the EOS parameters using a few bins, at the price of 
retaining correlations in their uncertainties.  
Advantages for a method will come from giving robust insight into the 
physical nature of dark energy.

\section{Principal Components} \label{sec:pca}

It is important to recognize that PCA the way it is normally applied in 
astrophysics, e.g.\ to spectra, is very different from the qualities 
desired in measuring the EOS.  In conventional PCA one wants to maximize 
the variance, essentially the signal, while for the application of PCA 
to cosmological parameter estimation 
(\cite{HutStark03,HuOka04,Leach06,Kadotaetal05,MortHu07}) 
one wants to minimize the variance because it represents the observational 
uncertainty. In the former case, using a basis of eigenvectors (or 
eigenmodes) is very useful because it extracts the specific linear 
combinations of parameters that have the most signal. In the latter 
case, at least when applied to the dark energy EOS where we 
want the small variations of data to be revelatory, i.e.\ arise from 
very different EOS and so point to the physics, 
we will see that it 
is less obvious what the quantitative advantages of PCA are beyond 
decorrelating the parameter uncertainties.  
(PCA is still useful in obtaining impressions of sensitivity, i.e.\ 
what qualities of the data are best constrained.)  
For example, for CMB analysis one still prefers 
to work with quantities having clear physical interpretations rather than 
principal components, despite the decorrelation \cite{EfBond99}. 

To decorrelate the EOS characteristics, one diagonalizes the Fisher 
(or inverse covariance) matrix by applying a basis transformation to
a basis of eigenmodes.
In this new basis ${\bf e'_i}$, 
\beq
w(z) - w_b(z) = \alpha'_i\,e'_i(z),
\eeq
such that the uncertainties in the new parameters $\alpha'_i$ are 
uncorrelated. It is important to note that in general the basis vectors, or modes,
tell us how to interpret the uncertainties in the parameters $\alpha'_i$ 
in terms of their effect on the equation of state function $w(z)$ through 
\beq
e'_i(z) = \frac{\pa w(z)}{\pa \alpha'_i}. \label{eq:modedrv}
\eeq

We discuss various important mathematical properties regarding modes 
in Appendix~\ref{sec:apxpca}; here we summarize the most relevant 
characteristics and results. 

\begin{itemize}

\item There are an infinite number of bases that decorrelate the 
coefficients $\alpha'_i$ 

\item Because the Fisher matrix transforms nontrivially under change 
of basis, the eigenvectors are not invariant.  They are not equivalent 
between different binning variables or bin widths. 

\item Each eigenvector has arbitrary normalization and so the meaning 
of uncertainty in measuring a mode is not well defined. 

\end{itemize}

\subsection{Eigenmodes}
\label{sec: eigen results}

Despite the first point in the list above, we can of course choose a 
particular basis and 
work from there.  We proceed to do this and illustrate the second and 
third points.  Starting with the unit box basis ${\bf e_i}$ introduced in 
\S\ref{sec:cosdep} we calculate the eigenmodes (but remember that this 
set depends on this particular starting point).  The fiducial model is 
$\Lambda$CDM: $w=-1$ with $\om=0.28$ and we consider initial binnings 
uniform in $z$, $a$ and $\ln(1 + z)$. 

\begin{figure}
  \begin{center}{
  \includegraphics*[width=8.8cm]{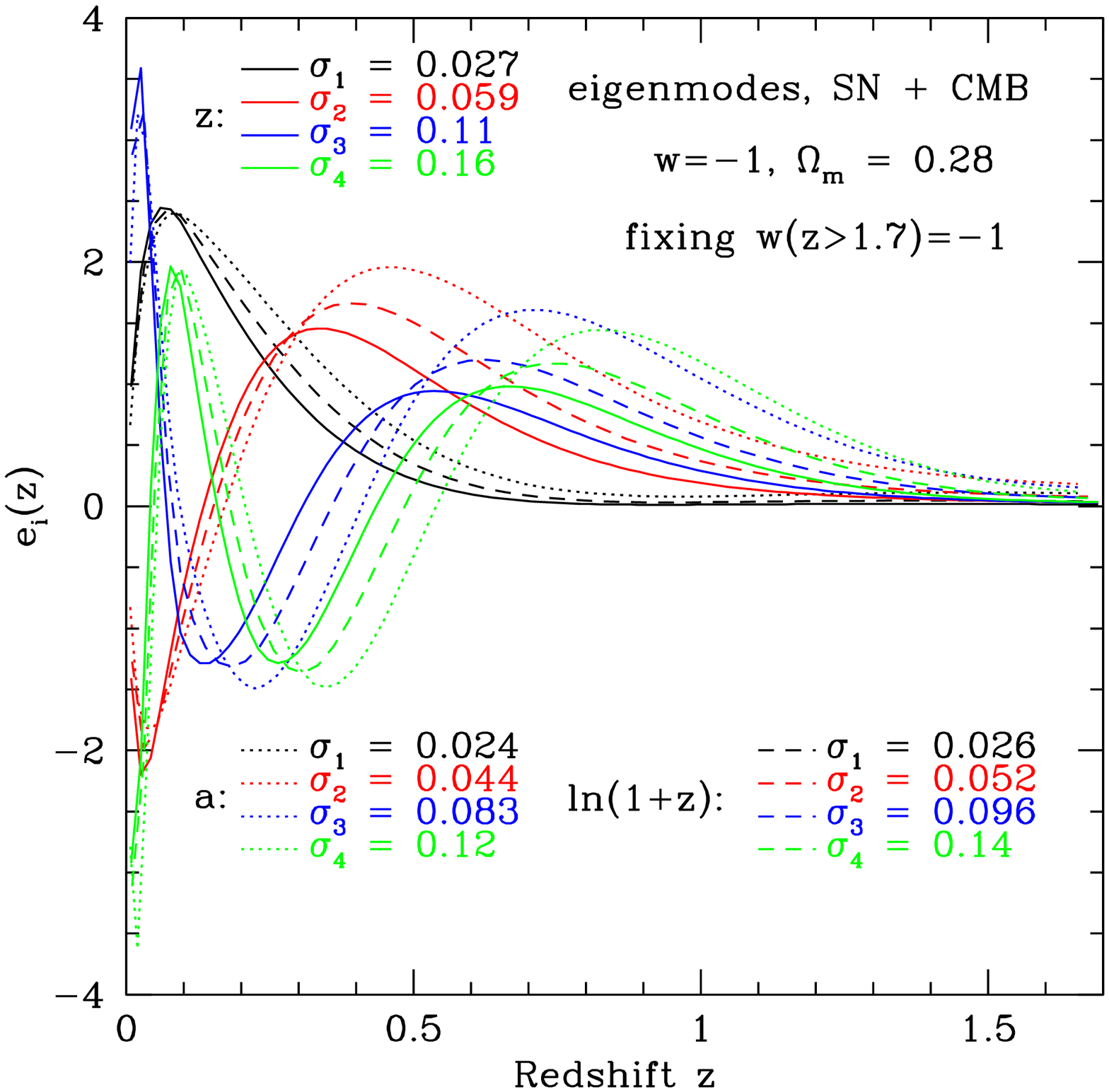}
  \includegraphics*[width=8.8cm]{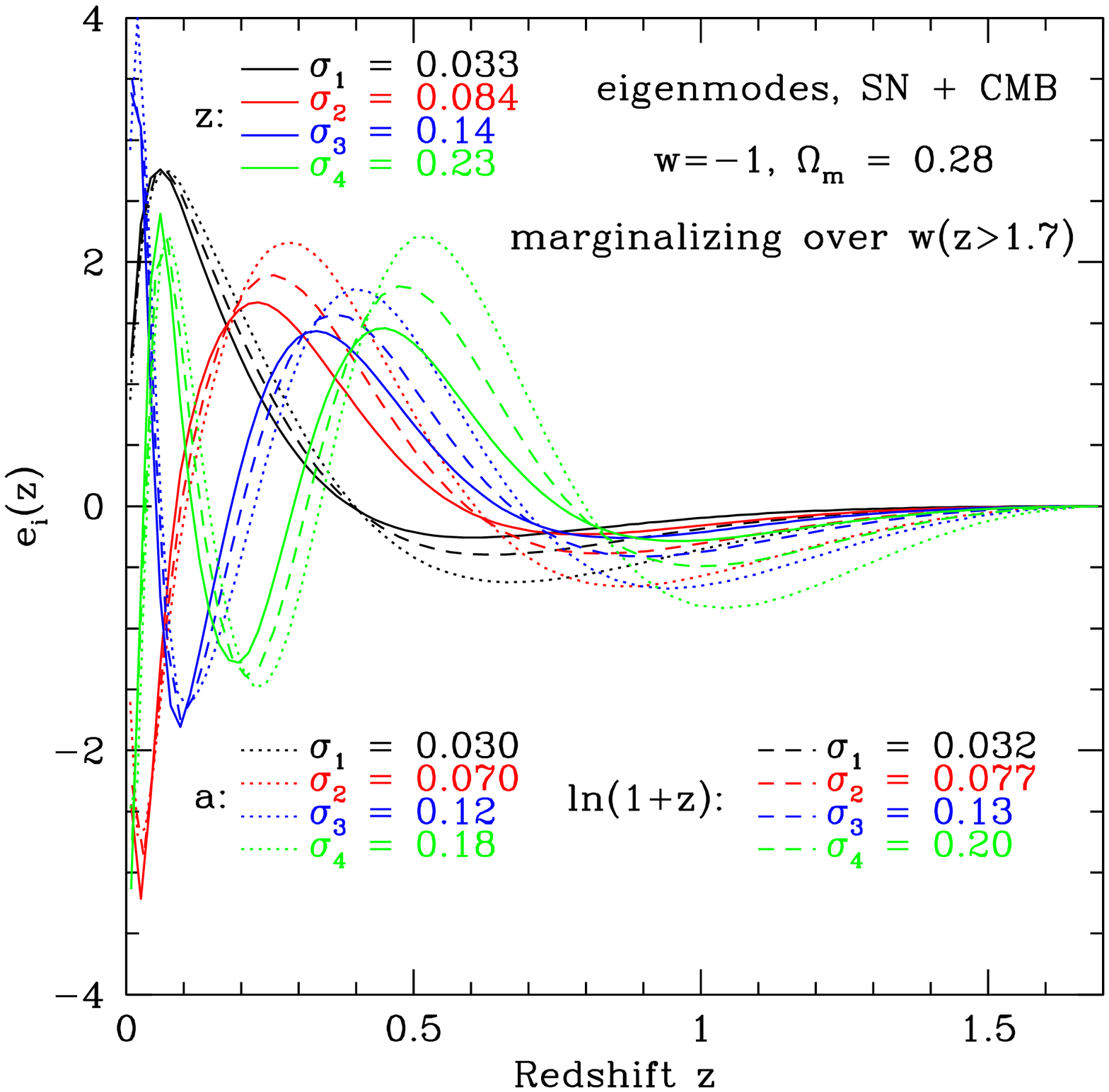}
  }
  \end{center}
  \caption{The first four eigenmodes and their uncertainties calculated 
using uniform binning in $z$, $a$ and $\ln(1 + z)$. In the first panel 
$w_{N+1}$ is fixed, in the second panel it is marginalized over. 
For $a$ and $\ln(1 + z)$, the number of EOS bins $N = 50$, for $z$ we 
use $N = 100$, enough for the modes to converge.   Note the modes, and 
their uncertainties, depend on binning variable (even modulo normalization).} 
  \label{fig: eigenmodes}
\end{figure}

Figure~\ref{fig: eigenmodes} illustrates the first four modes, after 
marginalization over $\Omega_{m}$ and $\mathcal{M}$.  For convenience 
we suppress the primes indicating the new basis.  The first panel 
has the EOS at $z > 1.7$ fixed to its fiducial value, $w_{N + 1} = -1$; 
in the second panel, $w_{N + 1}$ is treated as a free parameter and 
marginalized over.  For each binning variable or coordinate $x = z$, $a$, 
or $\ln (1+z)$, we normalize the modes according to $\int dx\, e_i^2(x)=1$.  
Although completely arbitrary, this choice is common.

Note that, as discussed above, the modes (and their respective standard 
deviations) are different for the different coordinates, even modulo 
normalization.  As the coordinate changes from $z$ to $\ln(1 + z)$ to 
$a$ the modes spread out more, gaining more power at large $z$ as 
expected from the relative bin widths.  The difference between a binning 
uniform in a coordinate $x$ and one uniform in $z$ depends on the 
coordinate transformation $dx/dz$ over the relevant redshift range. 
Since $da/dz = -1/(1 + z)^2$, $d\ln(1 + z)/dz = 1/(1 + z)$, and 
$dz/dz=1$, this explains the progression. 

Also, comparing the modes with $w_{N + 1}$ fixed to those where it is 
fit from the data shows two things. The uncertainties 
$\sigma_i\equiv\sigma(\alpha_i)$ are smaller of course.  Second, and 
perhaps less obvious a priori, the modes are more confined to low 
redshifts when $w_{N + 1}$ is made a free parameter. This becomes easier 
to understand when we remember that marginalizing over 
$w_{N + 1}$ is equivalent to not using $\dls$ at all. Thus only when 
$w_{N+1}$ is fixed does the inclusion of CMB $\dls$ data count, and 
this spreads the eigenmodes out towards higher $z$.

\subsection{Number of Eigenmodes and their Uncertainties} 

The eigenmode approach becomes completely general in the limit of an 
infinite number of bins, $N \to \infty$, as all continuous functions 
can be constructed from the complete basis.  The downside of this is 
that the uncertainties approach infinity.  One compromises by selecting 
a small set of the best determined modes, i.e.\ the principal components, 
and throwing away the others.  We face two problems when we try to adopt 
this approach.  The first is the question which set of eigenmodes to begin 
with (i.e.\ calculated using which coordinate).  The second problem is that 
``best determined'' is not well defined. We elaborate on this below. 

As we have demonstrated,
calculating the eigenmodes in a different coordinate yields a different 
set of modes so it is not clear which basis of eigenmodes to choose.
Although each {\it full} basis spans the same space of functions,
restricting oneself to the first few eigenmodes with respect to $z$ gives an 
essentially different parametrization than using the first few modes with 
respect to $a$ or any other coordinate that is not a linear function of $z$.  
The uncertainties will be different and there is the risk that how good one 
experiment is compared to another will be judged differently.

Even if we have chosen a certain basis, say the eigenmodes arising from 
uniform binning in the coordinate $z$, there is still the issue of 
quantifying how well determined a mode is.  That is, we would like to 
calculate a measure for how constrained a mode ${\bf e}_i$ is.  (Again, 
we suppress the primes as we will always be interested in the new basis.) 
An obvious choice seems to be the standard deviation of its coefficient, 
$\sigma_i \equiv \sigma(\alpha_i)$.  However, if we rescale ${\bf e}_i$ 
by a factor $A$, $\sigma_i$ is rescaled by $A^{-1}$. Thus, $\sigma_i$ 
only has meaning if we also specify the normalization of the mode, and 
the normalization is arbitrary, we have no physics guidance in choosing 
one normalization over another. In fact, it is perfectly legitimate to 
rescale all modes such that their (coefficients') uncertainties $\sigma_i$ 
are equal to one. Yes, this way it appears many modes have very large 
fluctuations, but without putting in any physical constraints on $w(z)$, 
i.e.\ a priori restrictions on the EOS, the word large is meaningless.  

Another approach to measuring how well determined a mode is involves using 
not a pure uncertainty criterion but a signal to noise criterion.  This 
was the approach advocated by \cite{LinHut05} but is also problematic.  
Consider the ratio of the standard deviation $\sigma_i$ over the coefficient 
$\alpha_i$.  At first sight, this seems to solve the problem of 
normalization as $\sigma_i/\alpha_i$ is invariant under changes of 
normalization.  However, this approach has its own problems.  From the 
mode expansion 
\beq
w(z) - w_b(z) = \alpha_i \, e_i(z)
\eeq
we see that the expectation values of the $\alpha_i$'s depend on which 
baseline function $w_b(z)$ we expand our measured EOS with respect to.  
For example, if we use $w_b = -1$ and the true EOS (or simulated EOS 
if projecting the leverage of a future survey) is also $w = -1$, then the 
expectation values of the $\alpha_i$'s are all zero.  Thus the 
noise-to-signal $\sigma_i/\alpha_i$ blows up. 

The reason why the quantities $\sigma_i$ and $\sigma_i/\alpha_i$ 
suggested above do not work as measures for how well (or how poorly) 
determined a mode is, is simple.  We have an estimate 
of the {\it noise} in the uncorrelated parameters $\alpha_i$, but not of 
the typical {\it signal} and thus cannot define a proper signal to noise 
ratio to tell us which modes are well-constrained and which ones are not. 
It may be tempting to simply throw out modes with large 
uncertainties, say $\sigma_i > 1$, but then we are implicitly making 
the assumption that the coefficients $\alpha_i$ are typically of order 
$1$ in the particular normalization -- and baseline model -- 
one has chosen for the modes. We have little knowledge on which to base 
such an assumption. 

The method {\it would\/} be useful if in addition to knowing the 
observational uncertainties $\sigma_i$, we knew the typical ranges of 
the $\alpha_i$'s. For example, if we knew the expectation values 
$\left\langle\alpha_i\right\rangle$ and the typical deviations from 
their expectation values $\sqrt{\left\langle(\alpha_i - \left\langle 
\alpha_i \right\rangle)^2\right\rangle}$ (brackets here denote averages 
over realizations of the parameters, they have nothing to do with 
observational uncertainties, given by $\sigma_i$), we can call 
$\alpha_i$ (and the 
corresponding mode) well-constrained if the signal to noise ratio 
\beq
{\rm SNR} \equiv \frac{\sqrt{\left\langle(\alpha_i - \left\langle 
\alpha_i\right\rangle)^2\right\rangle}}{\sigma_i}
\eeq
is large.

There are two scenarios in which one has knowledge about quantities 
like $\left\langle\alpha_i\right\rangle$ and $\sqrt{\left\langle 
(\alpha_i - \left\langle \alpha_i \right\rangle)^2\right\rangle}$, 
both quite common in physics. One is when one can observe a 
(large) sample of realizations of the parameters. If for example the
function of interest is a source spectrum (e.g.\ of quasars or supernovae 
\cite{Suz06,Davetal07}), the sample size is 
equal to the number of observed sources. Unfortunately, we can only 
observe one universe and thus only one equation of state. The other 
scenario is where one knows what the 
underlying physics is and what natural values are for the parameters 
of the theory (e.g.\ for the ionization fraction see \cite{MortHu07}).  
For example, if we knew dark energy was described by 
a scalar field model described by a set of $n$ parameters and in 
addition we had a prior probability distribution on those parameters, 
we could propagate this distribution to the parameters $\alpha_i$.  
Again unfortunately, we have a large number of possible theories for 
dark energy and little guidance as to the parameter values within 
those theories.  We return to the question of placing physical constraints 
on the EOS and its modes in \S\ref{sec:wlimit}. 

In conclusion, it is always possible to select a subset of modes and work 
with those, but it should be realized that what one is doing at that point 
is putting in assumptions of what the equation of state should look like -- 
precisely what we were trying to avoid by switching to PCA from a functional 
form -- and one cannot call the approach truly form independent anymore.

\section{Uncorrelated Bandpowers \label{sec:band}}

While using a large number of bins for the EOS increases the generality 
of functional forms $w(z)$, one ends up with a large number of poorly 
determined parameters.  Instead one could use a small number of bins but 
perform a basis transformation to decorrelate the parameters.  In large 
scale structure and CMB applications in cosmology this is often called 
uncorrelated bandpowers, e.g.\ where the functions are the matter power 
spectrum binned in wavenumber, $P(k)$, or the photon power spectrum binned 
in multipole, $C(\ell)$. 

To increase the localization of the modes within the bins, or bands, 
\cite{HamTeg00} proposed letting the ``square root'' of the Fisher matrix 
define the transformation. See \cite{HutCoor05} for application specifically 
to the EOS.  Such a transform has the advantage that, in the ideal case, 
the weights defining the new parameters in terms of the old ones 
are localized and mostly positive.  This would 
make the new parameters easier to interpret, as true bandpowers, i.e.\ 
giving the values of the EOS in a given redshift interval, with uncertainties 
uncorrelated between bins.  Unfortunately we will find that, as presaged 
in \S\ref{sec:cosdep}, the cosmological EOS analysis is far from the ideal 
case due to the broadness and skewness of the kernel, in contrast to the 
LSS case. 

\subsection{Modes and Weights} 

We briefly present the procedure for finding the square root of the Fisher 
matrix and the corresponding transformation.  This is placed in the main 
text because it highlights the important distinction between the properties 
of the eigenvectors and the weights, which has not always been clear in the 
literature. 

The transformation of interest is given by the symmetric matrix 
${\bf W}$ (see Appendix~\ref{sec:apxpca} for our conventions)
that transforms the Fisher matrix into the identity matrix: 
\beq
{\bf W} \, \F \, {\bf W^{\rm T}} = {\bf 1}.
\eeq
This matrix is constructed using the matrix ${\bf O}$ of which the rows 
are the (normalized) eigenvectors of ${\bf F}$, i.e.\ the orthogonal 
(${\bf O^{\rm T}} = {\bf O^{-1}}$) matrix that diagonalizes the Fisher matrix 
\beq
{\bf O} \, \F \, {\bf O^{\rm T}} = {\bf D}.
\eeq
${\bf W}$ is now given by
\beq
{\bf W} = {\bf O^{\rm T}} \, {\bf D^{-1/2}} \, {\bf O},
\eeq
note that the square root of the Fisher matrix ${\bf F^{1/2}} \equiv 
{\bf W^{-1}}$ is also symmetric and it squares to the Fisher matrix 
(hence the name).

The new basis vectors ${\bf e'_i}$ are now given by the rows of ${\bf W}$ 
(see Appendix~\ref{sec:apxpca}) and their coefficients $\alpha'_i$ are 
\beq
\label{eq: transform coeff 2}
\alpha'_i = W^{-1}_{j i} \alpha_j. 
\eeq 
We follow \cite{HutCoor05} and rescale the basis vectors and thus 
${\bf W}$ such that the $\alpha'_i$ are weighted averages of the 
$\alpha_i$, i.e.\ we rescale the rows of ${\bf W}$ such that 
\beq
\sum_{j=1}^{N} \, W^{-1}_{j i} = 1.
\eeq 
For notational convenience we use the same name for the rescaled 
transformation matrix ${\bf W}$  as for the original one, but note that 
after the rescaling ${\bf W}$ is no longer symmetric. 
The $\alpha'_i$ are now uncorrelated and their uncertainties are given by 
\beq
\sigma'_i = \left( \sum_{j = 1}^N \, F^{1/2}_{j i}\right)^{-1}.
\eeq 
In summary, the rows of ${\bf W}$ contain the new basis vectors ${e'_i}$ 
and the rows of ${\bf (W^{-1})^{\rm T}}$ contain the weights. 

An important point is that even though the weights tell us how to construct 
the new parameters out of the old ones, as discussed in \S\ref{sec:pca},
to interpret the meaning of the uncertainties $\sigma'_i$ for the EOS 
one needs to look at the basis vectors $e'_i(z) = \pa w(z)/\pa \alpha'_i$ 
and not at the weights.  That is, 
\beq 
\sigma^2[w(z)]=\sum_i \sigma'_i{}^2 e'_i{}^2(z). \label{eq:sigwe} 
\eeq 
We emphasize that plots of the weights alone 
cannot be directly interpreted as values of the EOS.  To some extent this 
confusion has been exacerbated by sometimes writing the weights as 
${\mathcal W}_i$ -- these are {\it not\/} the EOS $w_i$. 
The distinction between vectors and weights exists because the uncorrelated 
bandpowers correspond to a non-orthogonal transformation (the only 
orthogonal transformation decorrelating the parameters is the one to a 
basis of eigenvectors, as already considered).  This distinction 
will be important to the 
question of localization and physical interpretation of the parameters.  
As illustrated in the next section, one can have weights that 
are all positive while the corresponding basis vectors have significant 
negative contributions, clouding the interpretation.

\subsection{Decorrelated Estimates of the Equation of 
State \label{sec:decorr}} 

Since the matrix of weights is defined as the square root of the Fisher 
matrix (up to a rescaling to make the weights sum to one),
the positivity and localization of the weights depends on how 
positive and localized the Fisher matrix itself is.  The idea of the 
square root scaling is that the square root is typically narrower, so 
the weights gain some localization relative to the Fisher matrix.  
However, we saw in \S\ref{sec:cosdep} that even next generation data 
probing the EOS involves a very nondiagonal Fisher matrix.  
This is inherent to the cosmological properties and degeneracies and 
does not arise from any particular binning or parametrization. 

We now calculate the modes described in the previous section.  
To facilitate comparison to the literature, specifically \cite{HutCoor05}, 
we choose four low redshift bins with the following ranges: $z = 0 - 0.2$, 
$z = 0.2 - 0.4$, $z = 0.4 - 0.6$ and $z = 0.6 - 1.7$. The bins define four 
EOS parameters $w_1$ to $w_4$.  The other cosmological parameters, fiducial 
values, and data sets are as before.  The Fisher matrix for the low $z$ 
EOS parameters in the case where $w_5\equiv w(z>1.7)$ is fixed is given by
\beq 
{\mathbf F}_{\rm fix5}=
\begin{pmatrix}
\ 205 & 84 & 17 & 5.8 \ \\
\ 84 & 65 & 27 & 19 \ \\
\ 17 & 27 & 21 & 21 \ \\
\ 5.8 & 19 & 21 & 35 \ 
\end{pmatrix}
\label{eq:fix5} 
\eeq 
and when $w_5$ is marginalized over, 
\beq 
{\mathbf F}_{\rm marg5}=
\begin{pmatrix} 
\ 146 & 33 & -14 & -33 \ \\
\ 33 & 22 & 0.66 & -13 \ \\
\ -14 & 0.66 & 4.4 & 0.93 \ \\
\ -33 & -13 & 0.93 & 10 \  
\end{pmatrix} 
\label{eq:marg5}
\eeq 
It is evident that the Fisher matrix is far from diagonal and furthermore 
that proper treatment of the high redshift EOS behavior, rather than 
assuming a fixed value for $w_{N+1}$ (here $w_5$), has a significant effect.  
For one thing, marginalizing over $w_5$ introduces negative entries in the 
Fisher matrix and we will see this causes some of the weights in the 
decorrelated basis to be negative.  

\begin{figure}
  \begin{center}{
  \includegraphics*[width=8.8cm]{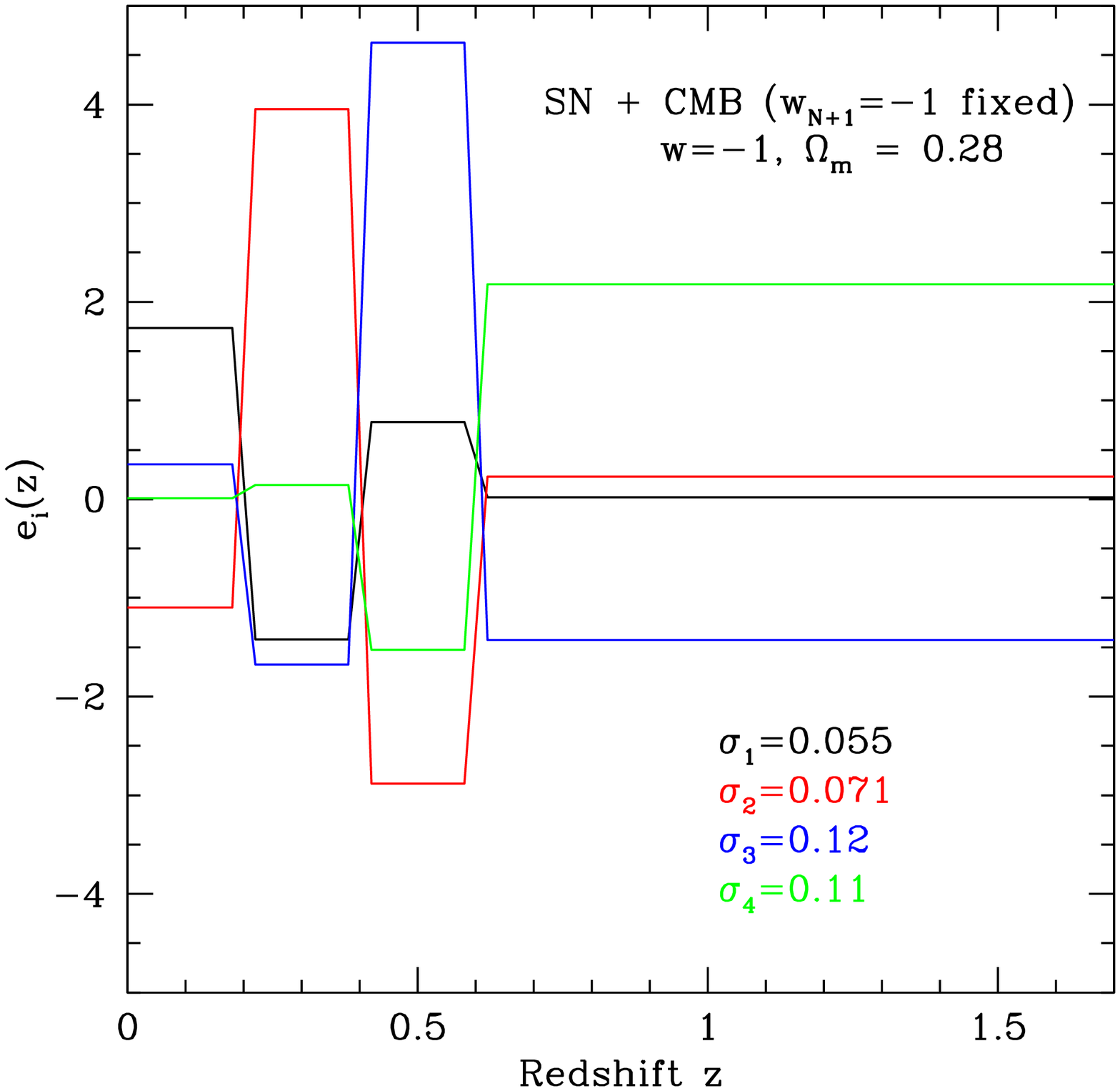}
  \includegraphics*[width=8.8cm]{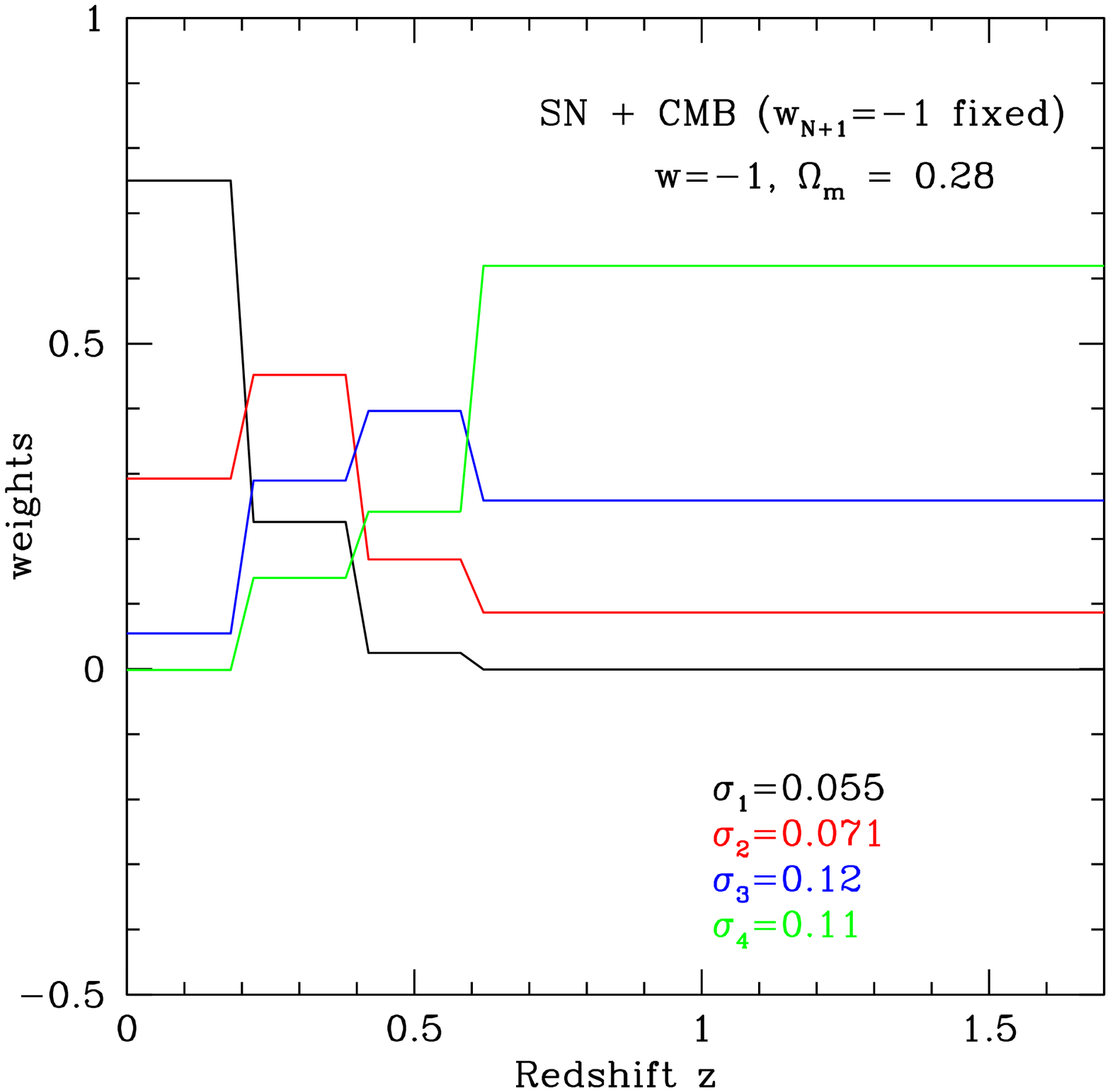}
  }
  \end{center}
  \caption{Uncorrelated basis functions, or modes, (first panel) and weights 
(second panel) obtained from the square root of the Fisher matrix.  Here 
$w_5\equiv w(z>1.7)$ is fixed to its fiducial value ($w_5 = -1$).  
Note that the modes have quite different shapes than the plots of the 
weights; the modes are what gives the impact on EOS $w(z)$ of an uncertainty 
$\sigma_i$.  The weights are only moderately localized (a consequence of the 
cosmological properties of the original Fisher matrix). 
 } 
  \label{fig: bands 4 nomarg}
\end{figure}

\begin{figure}
  \begin{center}{
  \includegraphics*[width=8.8cm]{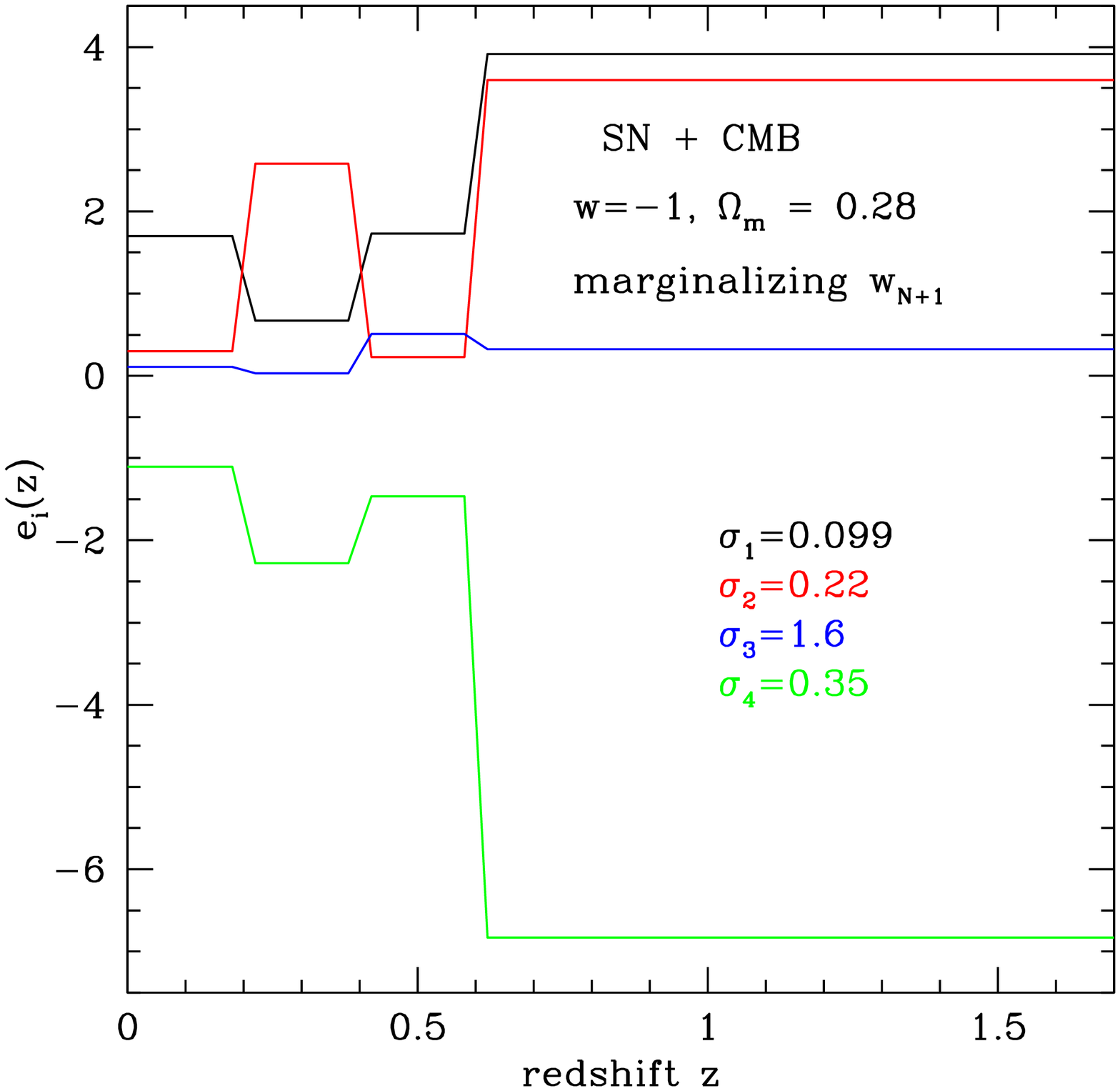}
  \includegraphics*[width=8.8cm]{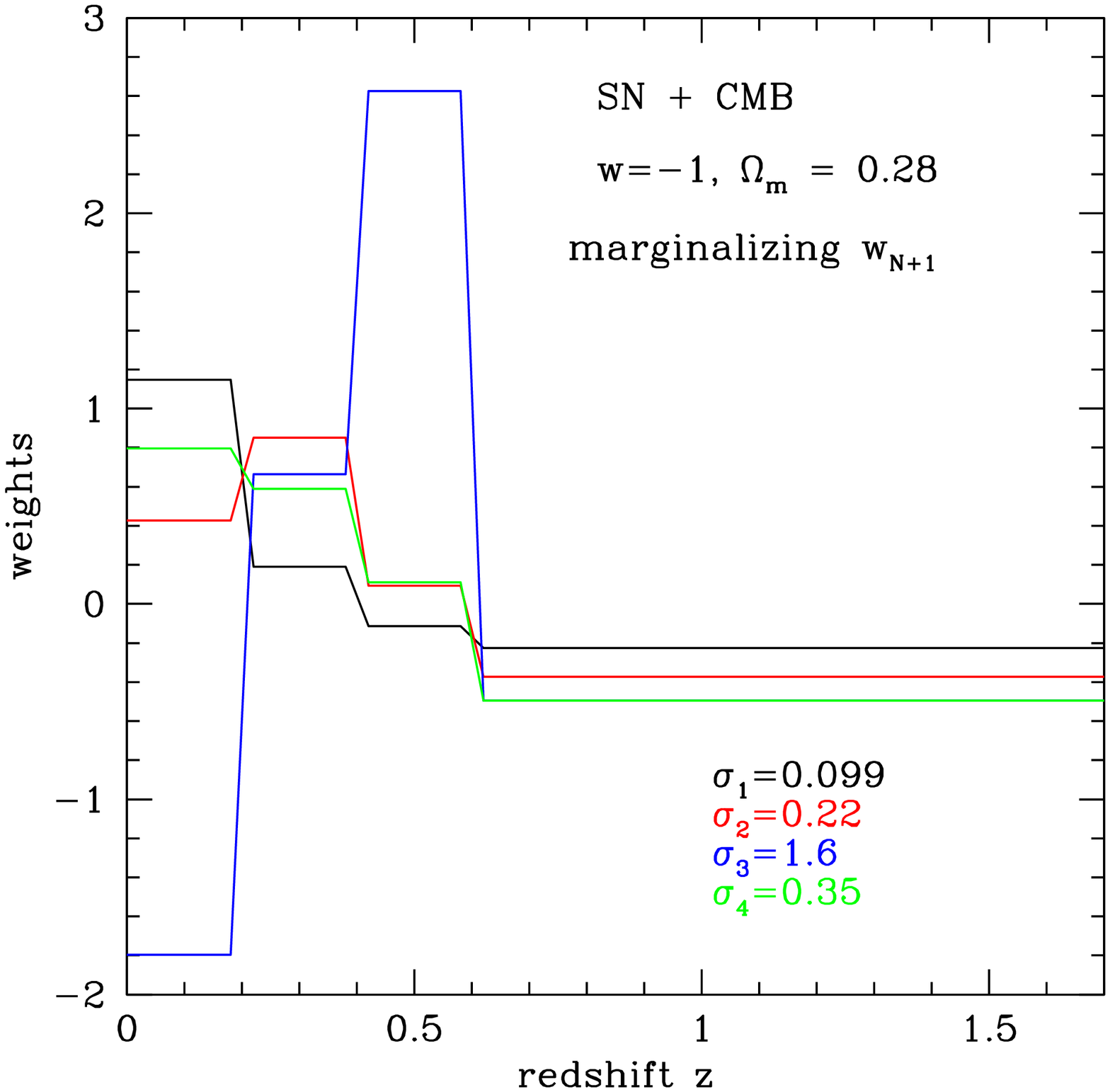}
  }
  \end{center}
  \caption{As Fig.~\ref{fig: bands 4 nomarg}, but with $w_5$ marginalized 
over. With $w_5$ as a free parameter, the weights and modes substantially 
lose their desired properties (being positive and localized).} 
  \label{fig: bands 4 marg}
\end{figure}

The main results of this section are illustrated in 
Figs.~\ref{fig: bands 4 nomarg} and \ref{fig: bands 4 marg}, giving the 
uncorrelated modes and the corresponding weights.  First consider 
Fig.~\ref{fig: bands 4 nomarg} where $w_{N+1}$ is fixed.  Previous results 
(e.g.\ \cite{HutCoor05, Riessetal06}) showed weights that were almost always 
positive and strongly localized, i.e.\ the weights defining the $i$th 
parameter were predominantly peaked in the $i$th bin.  This implies that 
the Fisher matrix of the original parameters, including priors, must have 
been close to 
diagonal to begin with in those cases.  In Fig.~\ref{fig: bands 4 nomarg} 
the weights are indeed essentially all positive and substantially localized 
(slightly less than in the works referred to above but differences in the 
fiducial model and data could account for this.)

While the characteristics of the weights in the case where $w_{N+1}$ is 
fixed look promising, recall that it is the eigenvectors that 
tell us how to interpret the results in terms of the EOS (see 
Eqs.~\ref{eq:modedrv} and \ref{eq:sigwe}).  Each (uncorrelated) uncertainty 
$\sigma_i=\sigma(\alpha_i)$ derived from the data corresponds to a 
variation in the EOS behavior $w(z)$ of the form of the eigenfunction 
$e_i(z)$.  We see that the basis functions have quite different shapes 
than the weights; in particular they have large negative contributions 
and large oscillations, far from being localized. For example,
if $\alpha_1$ is $1\sigma$ larger than its fiducial (and the other 
coefficients are exactly equal to their fiducial values), the EOS in the 
first bin, $w_1$, deviates by $+ 1.75 \times 0.055$ from its fiducial 
value $-1$, while the EOS in the second bin, $w_2$, deviates almost as 
strongly but {\it negatively} by $- 1.45 \times 0.055$ from $-1$.

For a deviation in the third coefficient, $\alpha_3$, by $1\sigma$, the 
consequences are even more dramatic: a bump in $w_3$ by $+4.7 \times 
0.12$ and a dip in $w_2$ by $-1.7 \times 0.12$.  
Note that while the $\alpha_i$ are decorrelated, the impact on the EOS 
is not localized, so the values of $w_i$ remain correlated.  
Such information is hard 
to get from just looking at the apparently well-behaved weights (which are 
often the only quantities plotted).  

Much of the good behavior of the weights is an artefact of fixing 
the high redshift behavior of the EOS, i.e.\ imposing a form (in a 
supposedly form independent approach).  When we instead allow freedom 
in $w_{N+1}$ and marginalize over it, the effects are dramatic as seen 
in Fig.~\ref{fig: bands 4 marg}.  This is not surprising given the 
differences in the respective Fisher matrices, Eqs.~(\ref{eq:fix5}) and 
(\ref{eq:marg5}).  Some of the weights now have considerably negative 
values and the modes are certainly not localized in the expected bin. 
Instead, all of them have substantial power in the highest redshift bin 
shown. 

\begin{figure}
  \begin{center}{
  \includegraphics*[width=8.8cm]{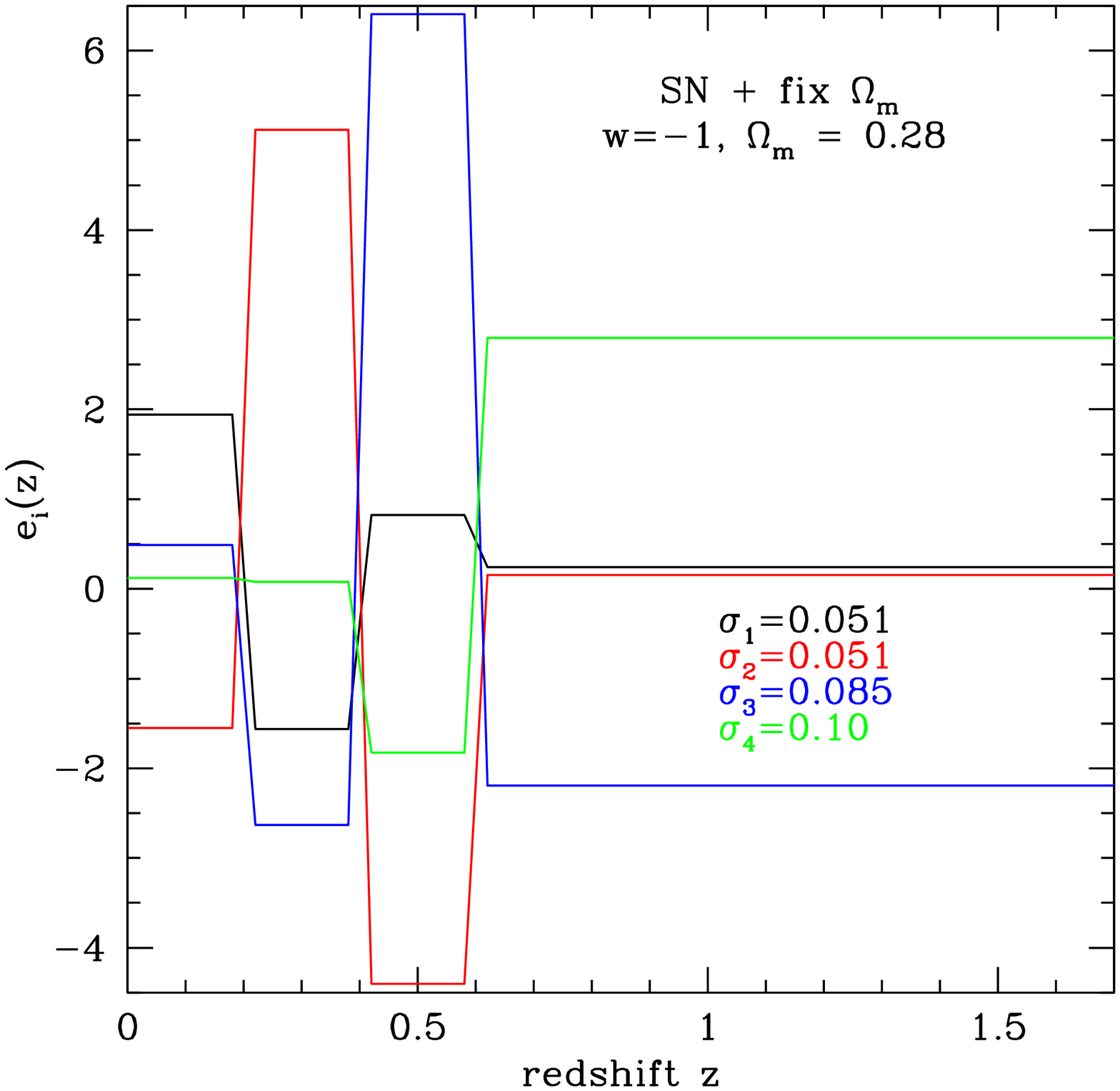}
  \includegraphics*[width=8.8cm]{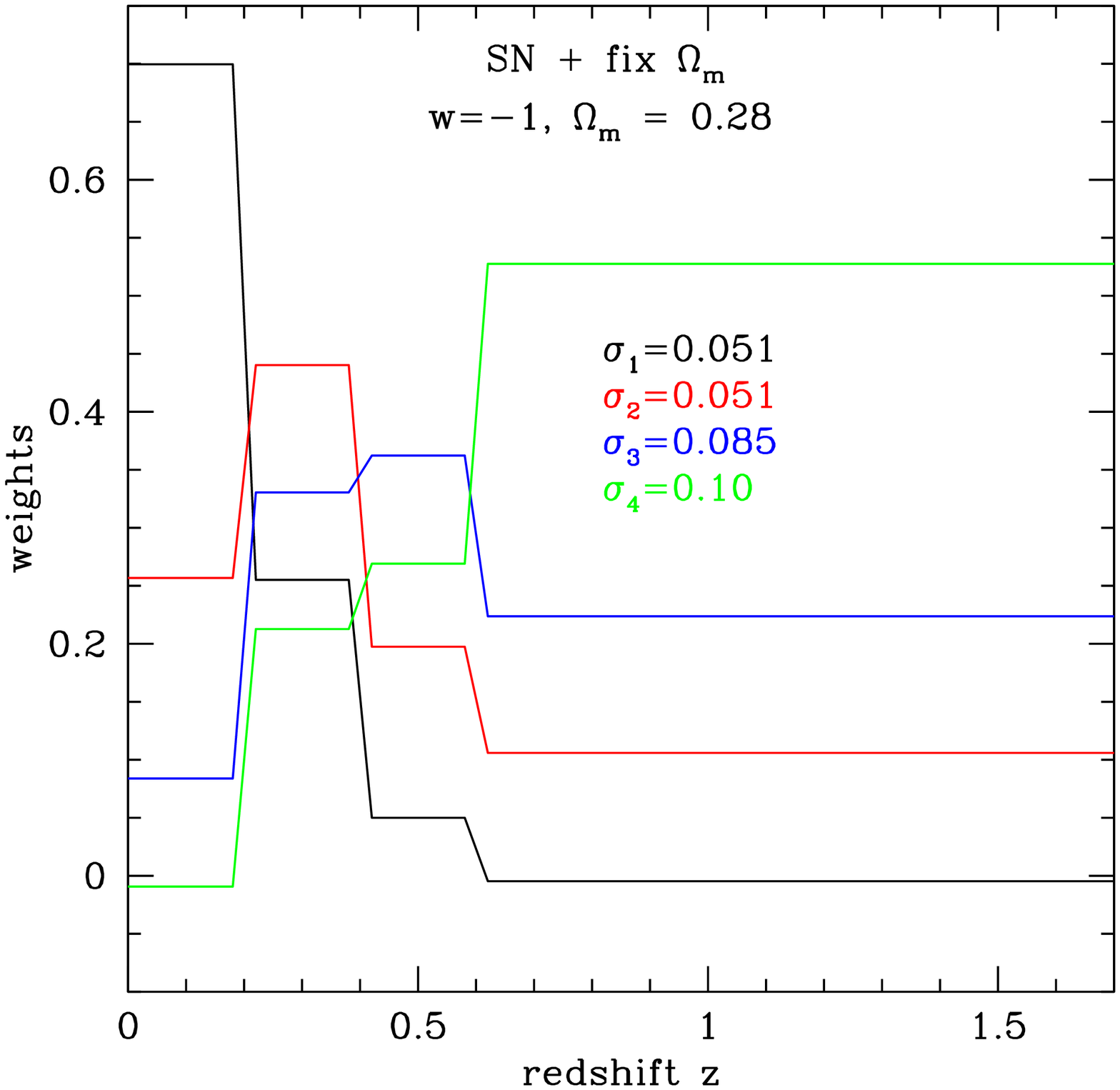}
  }
  \end{center}
  \caption{As Figs.~\ref{fig: bands 4 nomarg} and \ref{fig: bands 4 marg}, 
but instead of using the CMB data point (thus $w_{N+1}$ does not enter) 
we fix the matter density $\Omega_m$.  This illustrates the effect 
of a tight prior. } 
  \label{fig: bands 4 priorOm}
\end{figure}

To verify that it is the strength of the prior information, and not 
the square root of the Fisher matrix scaling per se, that causes 
the weights in Fig.~\ref{fig: bands 4 nomarg} (and the literature 
examples) to look so well behaved, we imposed ever tighter priors on 
$\om$.  When the prior is weak, the weights are both positive and 
negative.  As the prior tightens, the weights become progressively 
more positive and localized.  Figure~\ref{fig: bands 4 priorOm} shows 
the limit as we fix $\om$.

\subsection{Continuum Limit}

\begin{figure}
  \begin{center}{
  \includegraphics*[width=8.8cm]{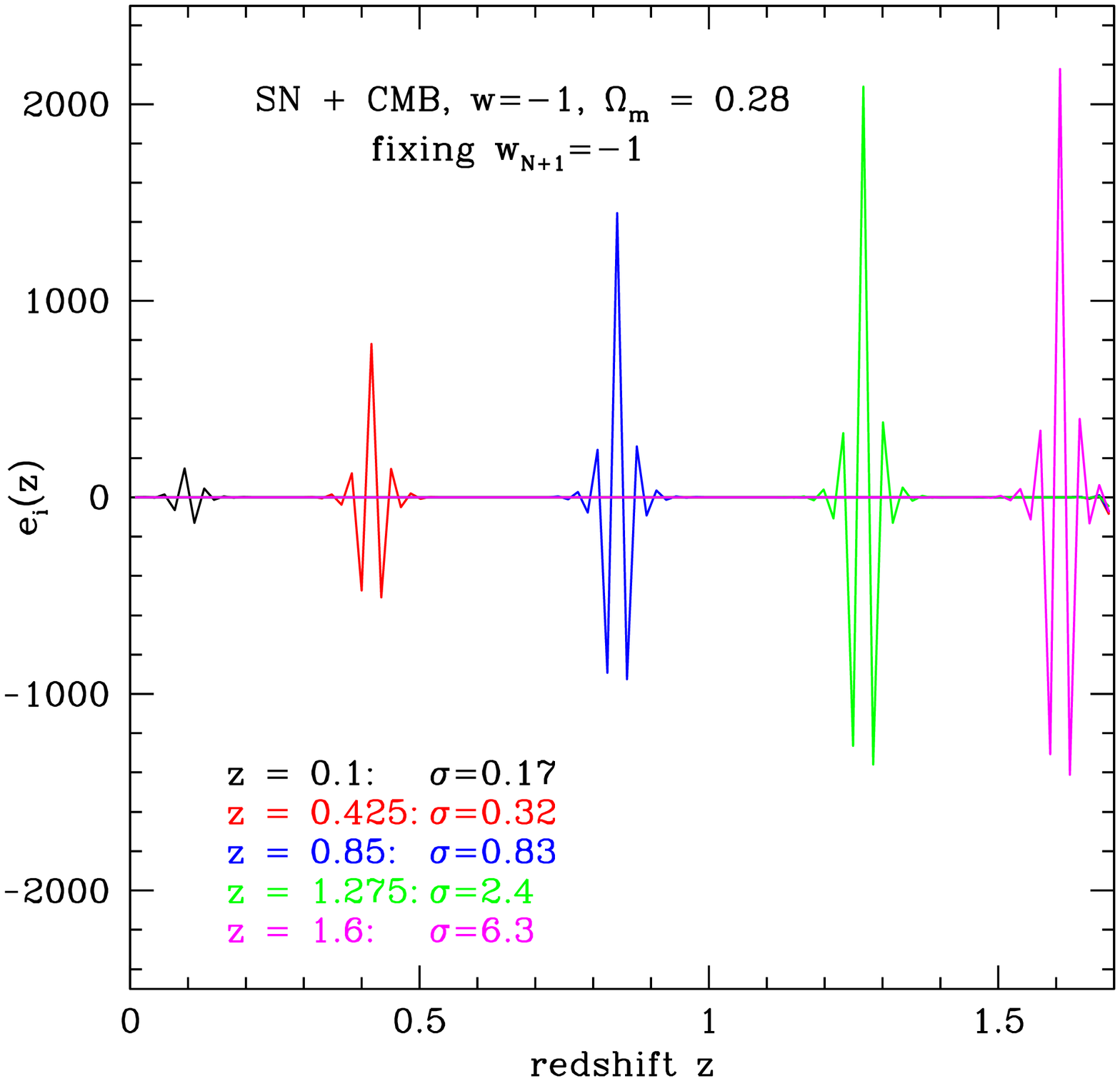}
  \includegraphics*[width=8.8cm]{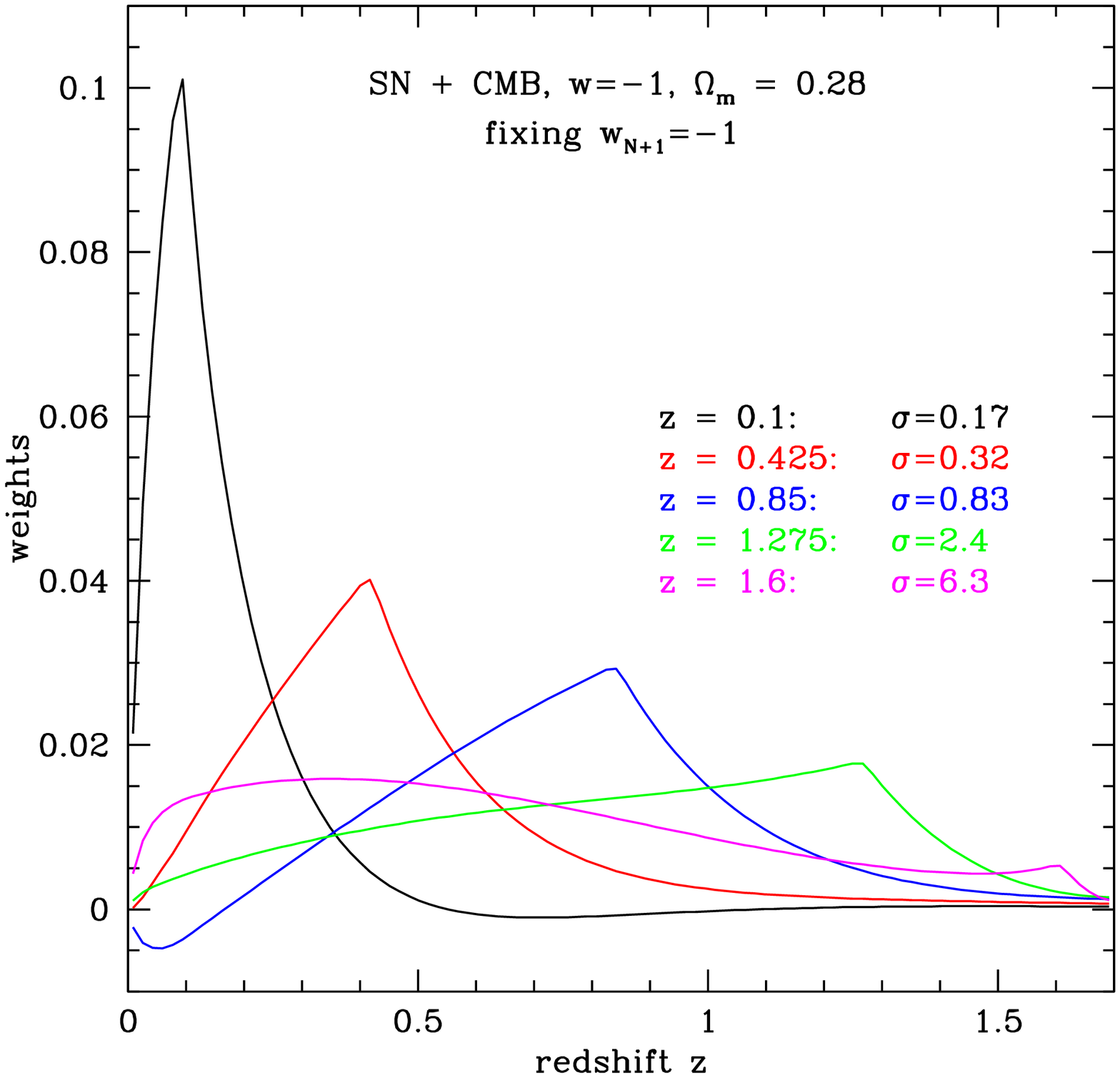}
  }
  \end{center}
  \caption{Illustration of what the modes (first panel) and weights (second 
panel) based on ${\bf F^{1/2}}$ look like in the large $N$ case, here 
$N = 100$.  Here we fix $w_{N+1}=-1$. } 
  \label{fig: root large N nomarg}
\end{figure}

\begin{figure}
  \begin{center}{
  \includegraphics*[width=8.8cm]{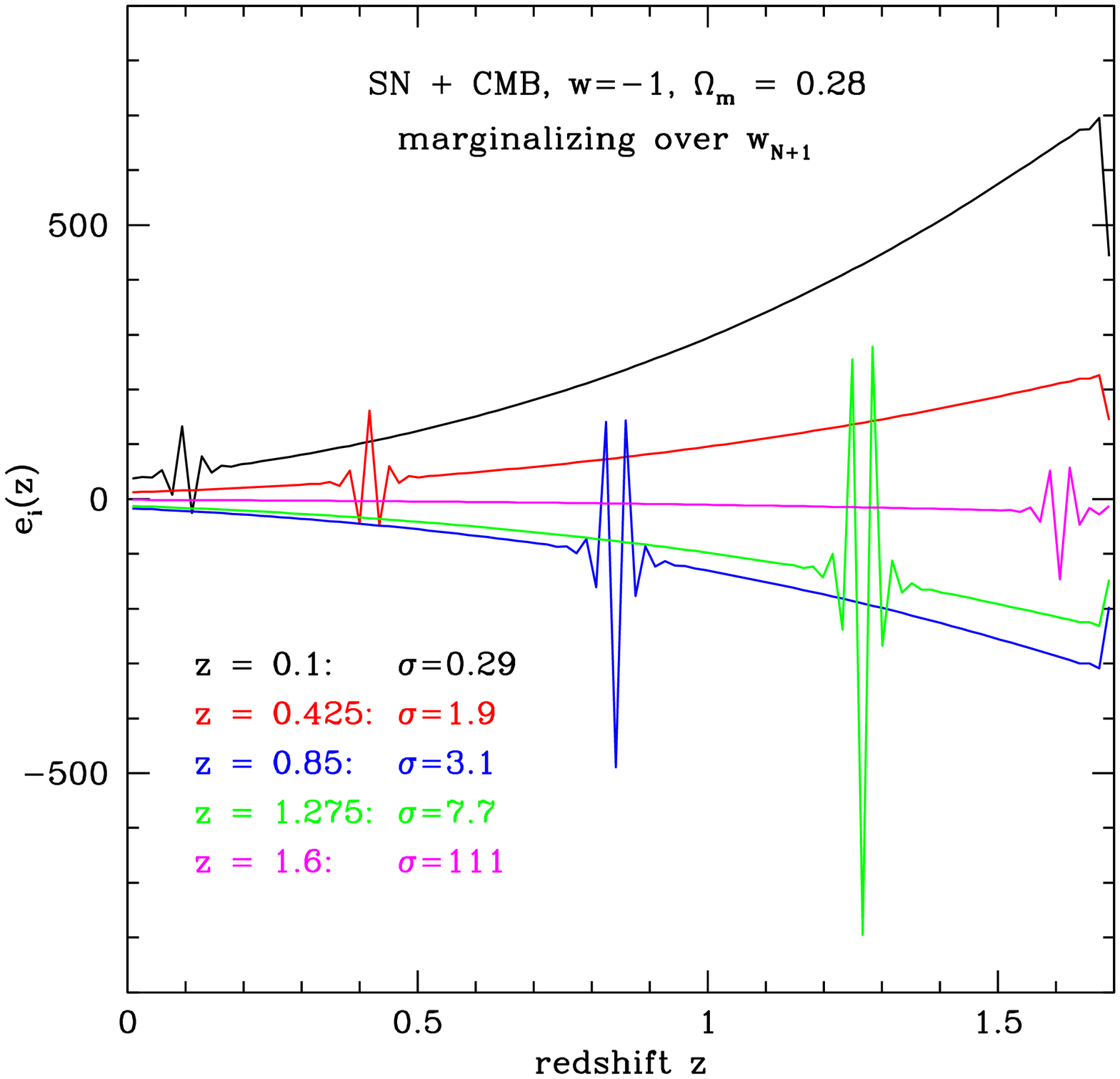}
  \includegraphics*[width=8.8cm]{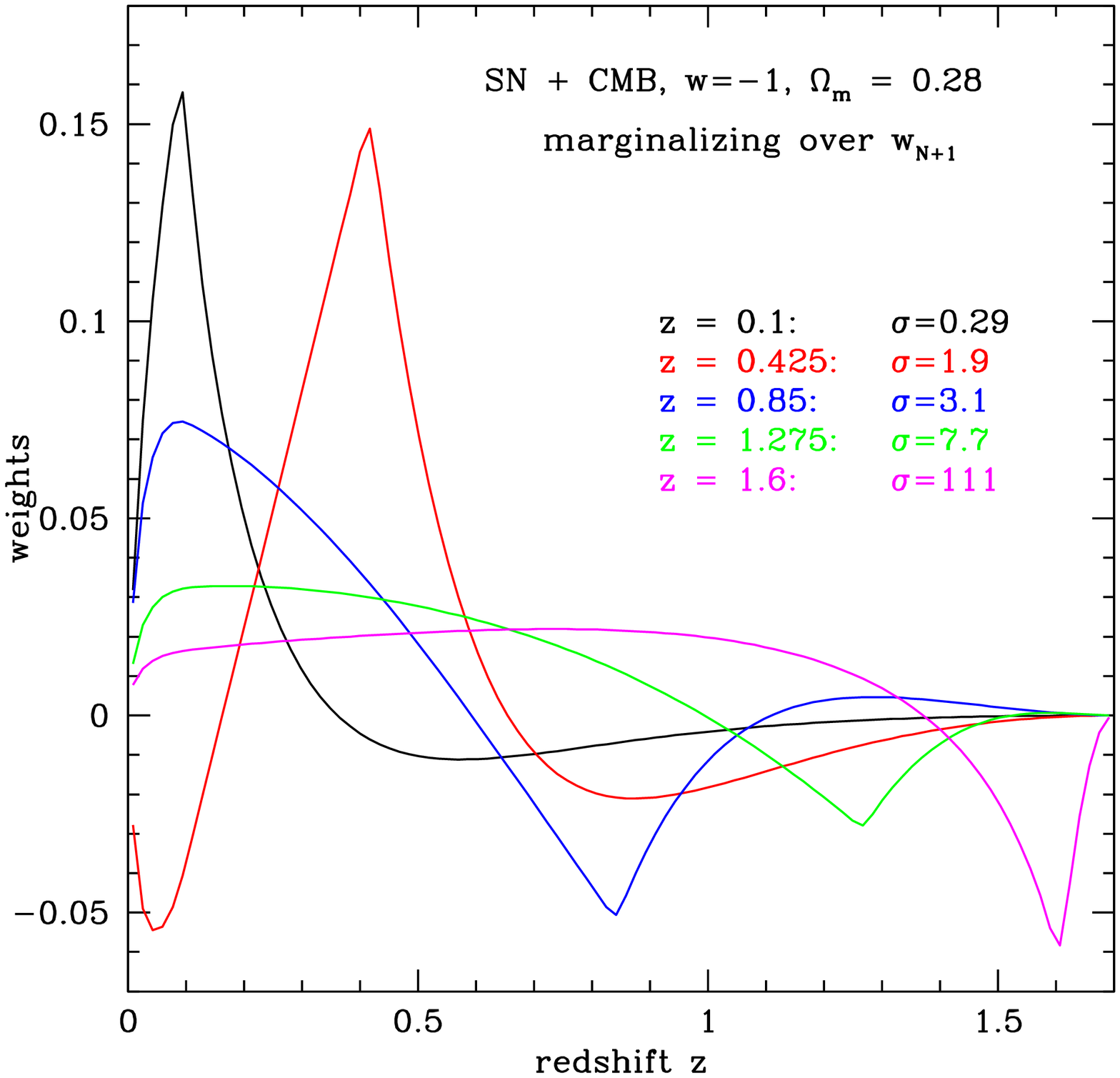}
  }
  \end{center}
  \caption{As Fig.~\ref{fig: root large N nomarg}, but marginalizing 
over $w_{N+1}$.}
  \label{fig: root large N marg}
\end{figure}

To ensure that the breakdown in positivity and locality of the weights 
is not an artefact of the binning, but rather is inherent to the 
cosmological data probing the EOS, we take the continuum limit, $N\gg1$. 
Figures~\ref{fig: root large N nomarg} and \ref{fig: root large N marg} 
plot the uncorrelated modes and weights corresponding to the square root 
of the Fisher matrix for $N=100$.  We see that even in this limit the modes 
fluctuate heavily and the weights are not very localized (which makes 
sense because they are given by the square root of the Fisher matrix 
depicted in Fig.~\ref{fig: Fisher}) though they are more faithful, i.e.\ 
peak at the given redshift.  Again, the physically appropriate 
act of marginalizing over $w_{N + 1}$ removes most vestiges of the 
desired positivity and locality.

The conclusion is that to obtain truly localized weights implies that one 
already started with a substantially localized (peaked, with a narrow 
kernel) inverse covariance matrix $\F$.  In such a case the EOS parameters 
are already easy to interpret without decorrelating them. Conversely, 
having weights that do not become tightly localized (and we have shown they 
may not without a strong external prior) implies that the new basis 
parameters are hard to interpret -- one might as well stick to the original 
correlated parameters.  Thus, like PCA, using the square root of the 
Fisher matrix in an attempt to obtain uncorrelated bandpowers is not a 
panacea in the quest for understanding dark energy.

\section{Binned Equation of State \label{sec:bin}} 

The third approach to understanding the EOS is simply considering the 
values in a small number of redshift bins.  That is, one defines 
piecewise constant EOS in some redshift range, e.g.\ $w(z)=w_i$ when 
$z_i<z<z_{i+1}$ (like in \S\ref{sec:band}, but without decorrelating).  
This guarantees localization and straightforward physical interpretation, 
at the price of some correlation in the uncertainties.  As we have seen, 
however, one cannot in practice generally have both localization and no 
correlation.  

\subsection{Uncertainties and Correlations} 

Calculation of the EOS estimation is straightforward.  Here we concentrate 
on questions of sensitivity to changes in binning and to treatment of 
the high redshift bin, rather than specific numbers for the uncertainties. 
To see the trends most clearly, we consider only two bins below $z=1.7$ 
along with the one at higher redshift.  

The quantities of interest are the uncertainties $\sigma_i$ on the EOS values 
(marginalizing over the other cosmological parameters), the correlation 
coefficients between EOS values, 
\beq
r_{i j} = \frac{C_{i j}}{\sigma_i \sigma_j}\,, 
\eeq
and the global correlation coefficients \cite{glcorr} 
\beq
r_i = \sqrt{1 - \frac{1}{C_{ii} F_{ii}}}\,,
\eeq 
which give the maximum correlation of $w_i$ with a linear combination of 
all the other EOS bins.  The covariance matrix ${\mathbf C}$ is the inverse 
of the Fisher matrix.   The high redshift value $w_3$ can either be 
fixed to the fiducial value (see \S\ref{sec:whi} for consequences of the 
true value being different than the fiducial assumed) or marginalized over. 

\begin{figure}
  \begin{center}{
  \includegraphics*[width=8.8cm]{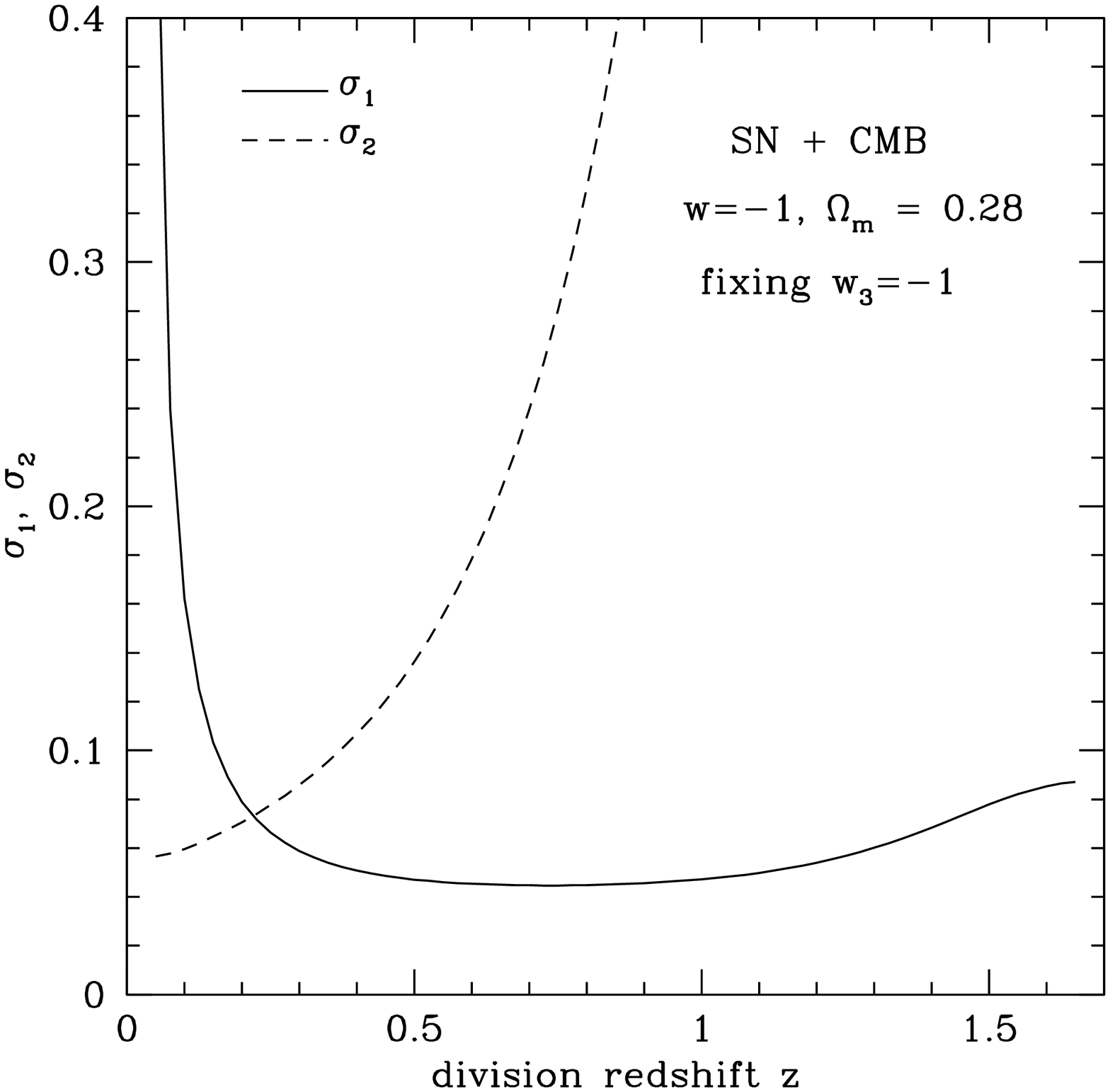}
  \includegraphics*[width=8.8cm]{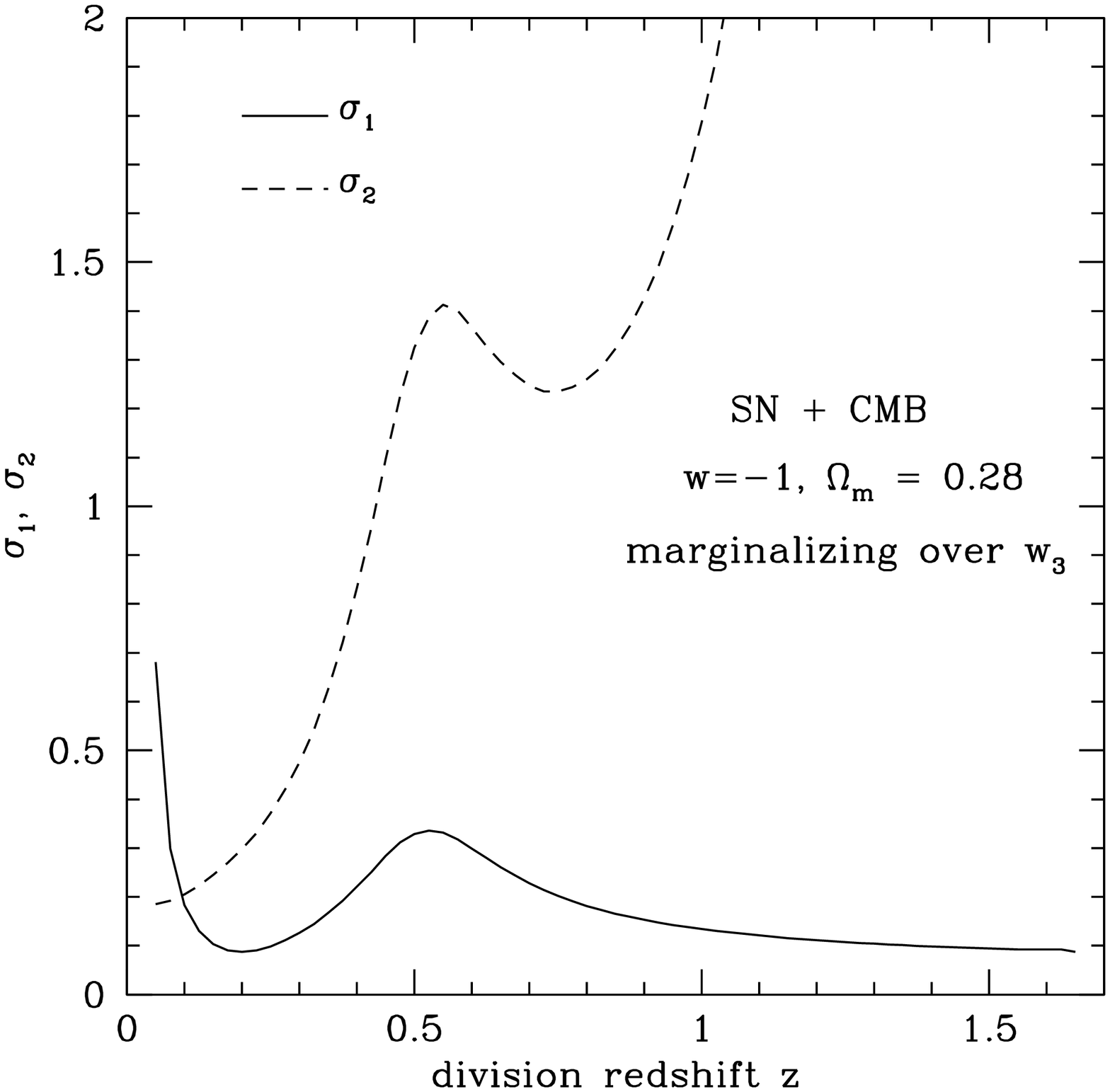}
  }
  \end{center}
  \caption{Uncertainties in the EOS values for two bins between $z = 0-1.7$ 
as a function of the redshift dividing the two bins.  The first panel has 
fixed $w_{N+1}=-1$, the second panel has $w_{N+1}$ marginalized over.  
Note the different scales. } 
  \label{fig: sigmas 2 bins}
\end{figure}

\begin{figure}
  \begin{center}{
  \includegraphics*[width=8.8cm]{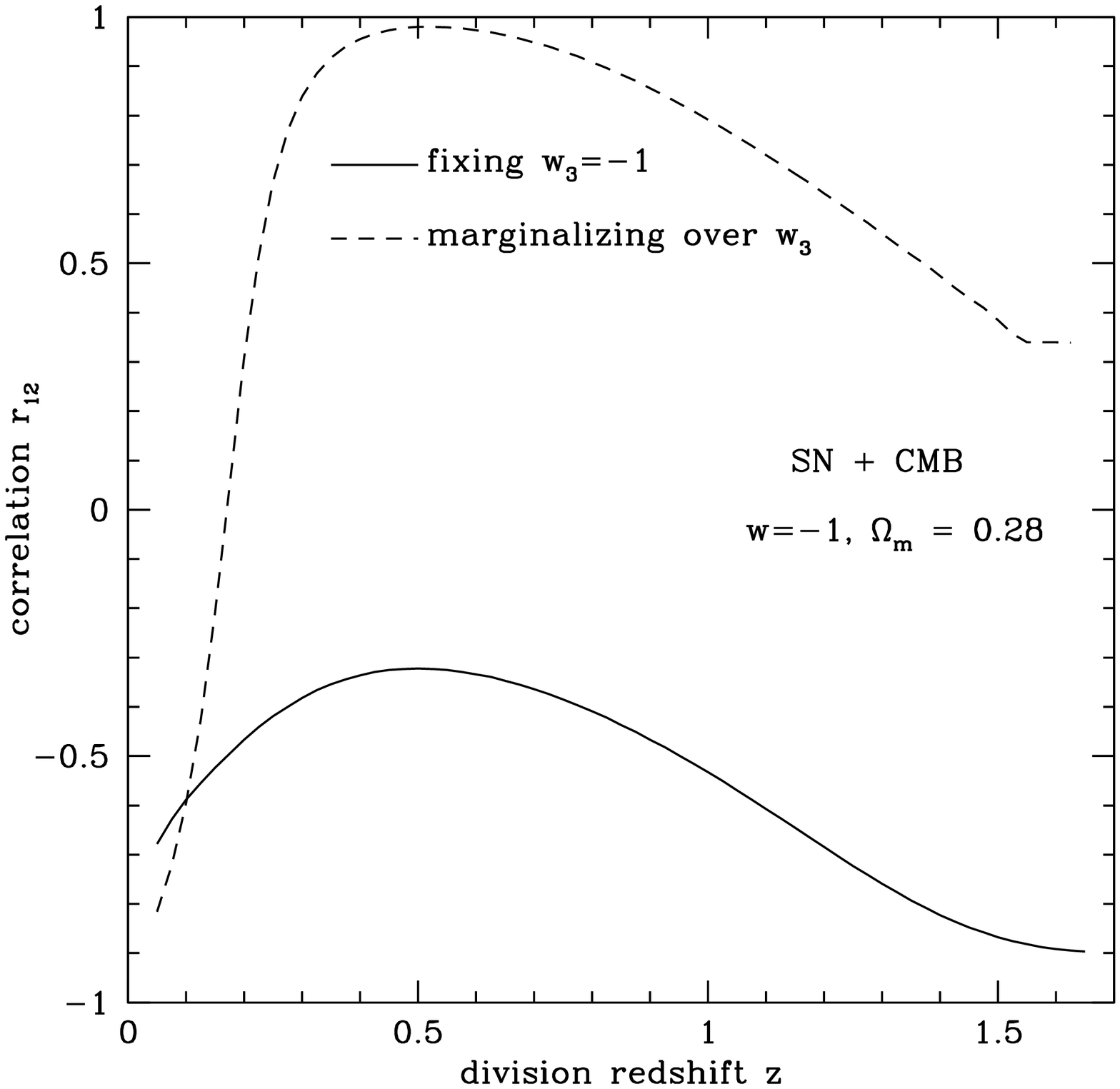}
  \includegraphics*[width=8.8cm]{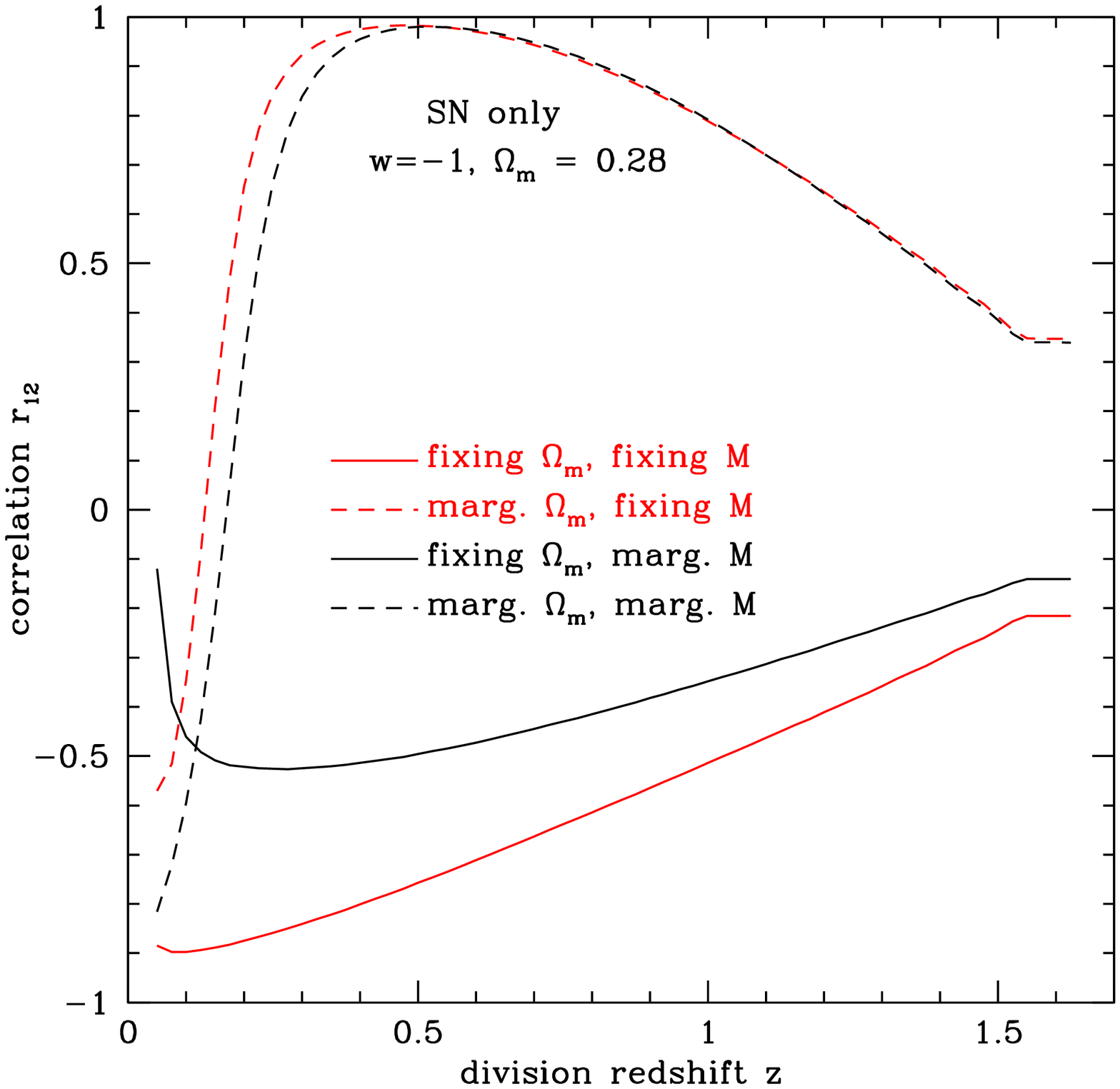}
  }
  \end{center}
  \caption{As Fig.~\ref{fig: sigmas 2 bins} but showing the correlation 
coefficient $r_{12} = C_{12}/(\sigma_1 \sigma_2)$ of the two $z < 1.7$ 
bins as a function of the division redshift.  The first panel uses 
SN and CMB data and compares fixing and marginalizing over $w_{N+1}$.  
The second panel shows that a tight prior on $\om$ (without adding CMB 
data) has a similar effect on the correlations as adding CMB data and 
fixing $w_{N+1}$, i.e.\ one must be wary of priors dominating the behavior. } 
\label{fig: corr bins 2}
\end{figure}

Figures~\ref{fig: sigmas 2 bins} and \ref{fig: corr bins 2} illustrate 
several interesting points.  Both the bin positions and the treatment of 
$w_3$ have a big impact on the uncertainties and correlations.  Regarding 
the uncertainties, when $w_3$ is kept fixed, the effect of making the first 
bin larger is to decrease $\sigma_1$ (and increase $\sigma_2$). 
(The slight rise in $\sigma_1$ when the first bin gets very wide is due 
to covariance with the matter density and goes away with a tight $\om$ 
prior.)  They are of 
comparable size when the boundary between the two bins lies around 
$z = 0.2$.  Note that there is only a very narrow region where the two 
parameters are determined to better than 0.1, so there is virtually no 
possibility of determining three EOS parameters to better than 0.1 with 
realistic next generation SN+CMB data -- and this is in the most optimistic 
case of fixing $w_3$.  

The correlation between estimates of $w_1$ and $w_2$ (still fixing $w_3$) 
is not very strong, with minimum correlation at $z_{\rm div}\approx 0.5$.

When the high redshift behavior of the dark energy EOS, represented by 
$w_3$, is not fixed a priori (after all, we want to probe dark energy 
properties, not assume them), significant changes occur.  Examination of 
the global correlation coefficient for $w_3$ shows this must happen: 
$r_3$ ranges between 0.97 and 1, i.e.\ the high redshift behavior is 
extremely highly correlated with the low redshift behavior.  This 
immediately tells us it that it is dangerous to fix $w_3$ because if it 
is fixed to the wrong value, it can strongly affect the values derived 
for the other parameters (see \S\ref{sec:whi}). 

Another consequence of the strong correlation $r_3$ is that including $w_3$ 
as a fit parameter makes the uncertainties in $w_1$ and $w_2$ increase, 
by factors up to 10.  When the first bin is small ($\zd = 0.1$), it is 
hardly correlated with $w_3$ and the change in its uncertainty is negligible, 
whereas $\sigma_2$ is increased by a factor of almost four. However, as the 
boundary redshift is moved up, the first bin grows more correlated with the 
third bin until at $\zd = 0.5$ both $w_1$ and $w_2$ have quite strong 
correlations with each other and with $w_3$, e.g.\ $r_{13}=r_{23}=-0.99$, 
and both $\sigma_1$ and $\sigma_2$ 
degrade considerably due to $w_3$. The effect is so strong that the trend of 
$\sigma_1$ decreasing as the bin widens is broken: $\sigma_1 = 0.33$ for 
$\zd = 0.5$ compared to $\sigma_1 = 0.09$ for $\zd = 0.2$.

Interestingly, when marginalizing over $w_3$ there is a division redshift 
for which the estimations of the low redshift EOS values are uncorrelated, 
$\zd\approx0.18$.  This decorrelation, or pivot, redshift arises without 
any need for using the square root of the Fisher matrix. But for a broad 
choice of $\zd$ the correlation is near unity.  
The strong correlation goes away when fixing $w_3$, but this is an example 
of prior information rather than data determining our view of the dark 
energy properties, as we saw in \S\ref{sec:decorr}.  For example, in the 
second panel of Fig.~\ref{fig: corr bins 2}, we recreate the same behavior 
of breaking the strong correlation $r_{12}$ by fixing $\om$.  One must be 
cautious that priors do not overwhelm the data, to see a true picture 
of dark energy. 

\subsection{Figures of Merit \label{sec: FOM}} 

In attempting to comprehend the nature of dark energy, some researchers 
advocate condensing the information down to a single figure of merit (FOM) 
related to the uncertainties in the parameter estimation.  In \S\ref{sec:pca} 
we saw some difficulties of defining this in a robust manner. 
Indeed, FOM's for binned EOS typically depend sensitively on both the 
binning adopted (which has nothing to do with the cosmology within the 
data) and, again, the treatment of the high redshift EOS. We now analyze 
some possible FOM's for binned equations of state. 

\begin{figure}
  \begin{center}{
  \includegraphics*[width=8.8cm]{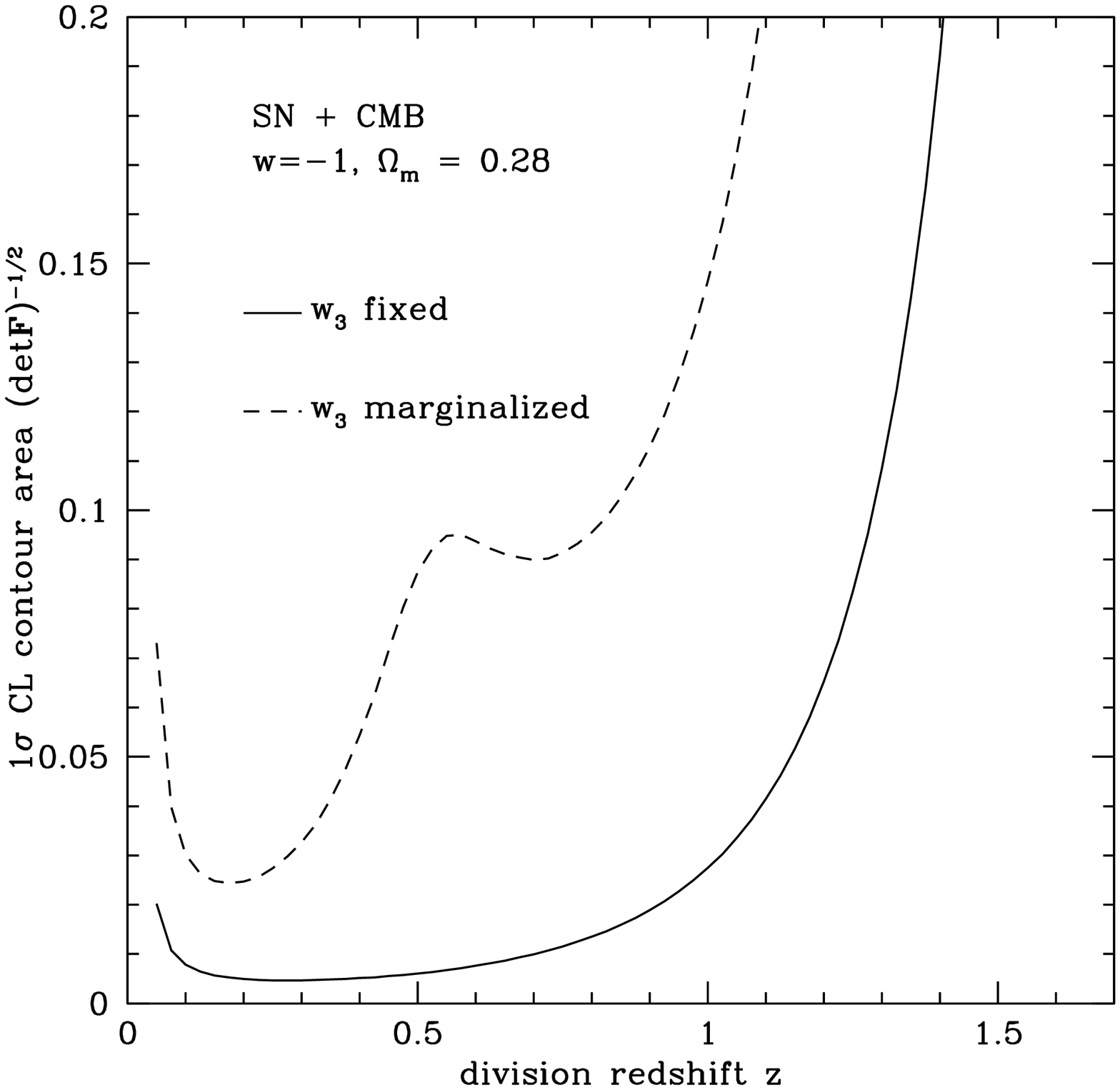}
  \includegraphics*[width=8.8cm]{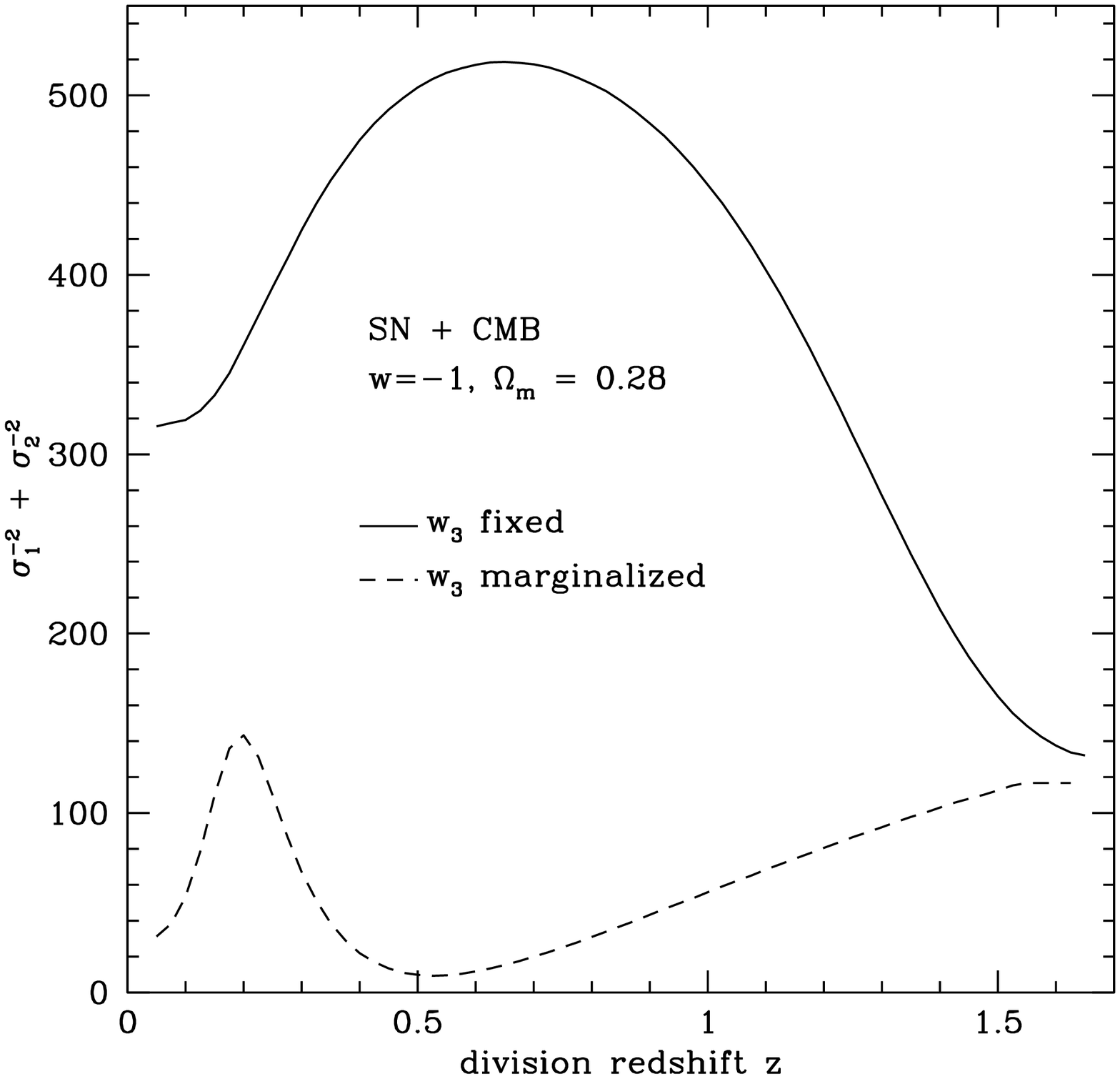}
  }
  \end{center}
  \caption{As Fig.~\ref{fig: sigmas 2 bins} but showing two suggested 
figures of merit.  The first panel shows $(\det{\F})^{-1/2} = A/\pi$ as 
a function of the bin division redshift $z$, where $A$ is the area enclosed 
by the 68\% confidence level contour in the $w_1 - w_2$ plane for the 
model with two bins between $z = 0$ and $z = 1.7$.   The second panel shows 
$\sum_i \sigma_i^{-2} = \sigma_1^{-2} + \sigma_2^{-2}$. 
  }
  \label{fig: area 2 bins}
\end{figure}

Figure~\ref{fig: area 2 bins}, first panel, plots the area (taking out 
a factor $\pi$) 
enclosed by the $1\sigma$ confidence level contour in the $w_1$-$w_2$ plane, 
as a function of the bin division redshift.  This area is proportional to 
$(\det{\F})^{-1/2}$, which is invariant under any transformation ${\bf W}$ 
with $(\det{{\bf W}})^2 = 1$ (see Eq.~\ref{eq: transform F} or 
\cite{HutTur01}), and in 
particular under any orthogonal transformation.  When $w_3$ is fixed 
(first panel), the area is minimized at a division redshift of $\zd 
\approx 0.25$. One might interpret 
this as saying that we obtain the most information (in the $N = 2$ case) 
with one bin from $z = 0 - 0.25$ and one from $z = 0.25 - 1.7$. When $w_3$
is marginalized over, the behavior changes somewhat but 
there is still a clear minimum, this time at slightly lower redshift 
$\zd \approx 0.18$. 

As more bins are added, individual bin parameters can become extremely 
uncertain and the volume $(\det{\F})^{-1/2}$ in the $N$-dimensional 
space of $w_1$-\dots -$w_N$ (see, e.g., \cite{AlbBern07}) 
will be dominated by these poorly determined 
parameters.  In an attempt to ``cut off'' the highly uncertain parameters, 
a figure of merit like 
\beq
\label{eq: FOM 1}
{\rm FOM}_{\rm corr} \equiv \sum_i \sigma_i^{-2}
\eeq
has been proposed (see e.g.\ \cite{SulCooHolz07}).

We first consider the $\sigma_i$ in Eq.~(\ref{eq: FOM 1}) as the 
uncertainties in the (correlated) bin parameters $w_i$. The behavior of 
this FOM as a function of division redshift in the two bin case is shown 
in Fig.~\ref{fig: area 2 bins}, second panel (note that now a large value 
is good).  
Such a measure would advocate -- for the same data -- using $\zd\approx0.65$ 
when $w_3$ is fixed.  In contrast, when $w_3$ is marginalized, this peak 
in the FOM becomes a strong dip, saying the experiment is weak.  Comparing 
to Fig.~\ref{fig: sigmas 2 bins}, this FOM can give high marks to choices 
that lose almost all the information on the second parameter. 

The FOM discussed above does not take into account correlations between 
parameters.  As an alternative, we could use the uncertainties in 
the {\it decorrelated} weighted averages $\alpha'_i$ described in 
\S\ref{sec:band}.  It is actually this choice, or rather its inverse, 
that is advocated in \cite{SulCooHolz07}.  
To be consistent with our previous notation, we should now write
\beq
\label{eq: FOM 2}
{\rm FOM}_{\rm decorr} \equiv \sum_i \sigma'_i{}^{-2}
\eeq
(note that this is the trace of the decorrelated Fisher matrix $\F'$).
This FOM has a very simple, but slightly disappointing interpretation: 
\beq
{\rm FOM}_{\rm decorr} = \sigma(w)^{-2}, 
\eeq 
i.e.\ the FOM is the inverse square uncertainty on a constant $w$, or 
equivalently when there is only one bin.\footnote{
To see this, first note that in terms of the $N$ decorrelated parameters 
$\alpha'_i$, the constant mode
\beq
e_{\rm const}(z) = 1, \quad\ 0 < z < z_{\rm max},
\eeq
which is the only mode present in the mode
expansion when $N = 1$,
is given by the $N$-dimensional vector ${\bf e'_{\rm const}} = (1,1,\dots,1)$ because the  $\alpha'_i$ are weighted averages
of the original parameters. Hence,
using the transformation law Eq.~(\ref{eq: transform F}) for the Fisher matrix,
the diagonal element of the Fisher matrix corresponding to the coefficient of the constant mode (i.e.\ the Fisher information of the constant mode) is 
\beq
{\bf e'^{T}_{\rm const} \,F'\, e'_{\rm const}} = \sum_{i j} F'_{i j} = \sum_i \sigma'_i{}^{-2}.
\eeq
But by definition this quantity is the inverse variance of the coefficient of
the constant mode in the case of $N = 1$ bins.} 

We have confirmed numerically that, as it must be from the single bin 
interpretation, this trace FOM is independent of the 
division redshift(s) and number of bins.  
In conclusion, this FOM only captures the information that was already 
contained in the standard deviation of $w$ when using the simplest 
parametrization, namely $w = $ constant.

Note that neither the area (determinant) nor trace FOM's takes particular  
advantage of physical foundations.  We have seen that the trace FOM neglects 
all dark energy dynamics, reducing to a constraint on a static EOS.  
For the area FOM, as discussed 
in \cite{Linder0604}, the area of the error contour is the Snarkian, or blank 
map, approach where all dynamics is equal.  Instead, \cite{Linder0604} 
advocates that the FOM must be 
adapted to the physics objective, e.g.\ whether one wants to distinguish 
the EOS from the cosmological constant or thawing behavior from freezing 
behavior, and depends on dark energy properties.  We revisit physical 
bases for discerning the nature of dark energy in \S\ref{sec:wlimit}.

\section{High Redshift Equation of State and Bias \label{sec:whi}} 

For each method of analysis considered the high redshift value of the 
EOS has been shown to be a crucial ingredient; fixing the value of 
$w_{N+1}=w(z>1.7)$ has significant effects on the derived properties 
of the dark energy.  A similar point has been made for functional forms 
by \cite{Linder0708}.  In addition to misestimating the uncertainties by 
fixing $w_{N+1}$, if it is fixed to the wrong value\footnote{Treating 
the EOS between $z = 1.7$ and $z = 1089$ as constant may introduce a bias 
in itself, but here we focus on the bias introduced by using the wrong 
constant value.}  
(and a priori we don't 
know what the correct value is) then the {\it values\/} themselves of 
all the cosmological parameters are biased -- we will derive a picture of 
dark energy skewed from reality. 

Bias in derived parameters can be calculated from offsets in observables 
within the Fisher matrix formalism by (see, e.g., \cite{Linder0604}) 
\beq
\label{eq: bias}
\delta p_i = (F^{-1})_{i j}\, \sum_k \frac{\pa O_k}{\pa p_j} 
\,\frac{1}{\sigma_k^2} \,\Delta O_k,
\eeq
where $\delta p_i$ is the difference of the estimated parameter value 
from its true value, 
$\delta p_i \equiv p_{{\rm e}, i} - p_{{\rm t}, i}$, and 
$\Delta O_k$ is the offset in the $k$th observable.  For bias arising 
from choosing the wrong value for $w_{N + 1}$ (which then propagates into 
the expected, i.e.\ simulated, observation), the expression becomes
(see Appendix~\ref{sec:apxbias}) 
\beq
\label{eq: bias 3}
\frac{d p_i}{d p_{N + 1}} = - \sum_{j = 1}^N (F^{(N)})^{-1}_{i j}\, 
(F^{(N + 1)})_{j, N + 1},
\eeq
where $d p_{N + 1}$ is the difference of the value $w_{N + 1}$ is fixed to 
from its true value, $\delta p_{N + 1} = p_{{\rm fix}, N + 1} - p_{{\rm t}, 
N + 1}$.

To give concrete examples of the induced parameter bias, we choose two 
EOS models that we will fit with binned piecewise constant EOS.  
We use three low redshift bins $z = 0 - 0.2$, $0.2-0.5$, $0.5-1.7$ and a 
high redshift bin from $z_{\rm max}= 1.7$ to $z_{\rm lss}= 1089$, 
and define a weighted average 
\beq 
w_{N+1}=\frac{1}{\Delta\ln(1+z)}\int_{z_{\rm max}}^{z_{\rm lss}} 
\frac{dz}{1+z} w(z). 
\eeq 
(For consistency, we use an appropriately defined weighted average for 
each bin.) 

The first model is based on a pseudoNambu-Goldstone 
boson (PNGB) model \cite{Friemanetal95}, which has $\whi\approx-1$, 
\beq
\label{eq: EoS PNGB}
w(z) = -1 + (1 + w_0) (1 + z)^{- F}
\eeq
with $w_0 = -0.8$ and $F = 1.5$.  The second model is based on a so called 
bending model \cite{Wett04} motivated by dilaton fields, giving a 
nonnegligible contribution of early dark energy density (here $\sim2\%$ 
relative to the matter density) and $\whi$ far from $-1$:  
\beq
\label{eq: EoS bending}
w(z) = \frac{w_0}{\left[1 + b \ln(1 + z)\right]^2}\,,
\eeq
with $w_0 = -0.9$ and $b = 0.415$ (this is very similar to the model 
$w(z)=w_0 + w_a (1-a)$, with $w_0 = -0.9$ and $w_a = 0.7$). In both cases, 
$\Omega_{m} = 0.28$. 

Table \ref{tab: bias} shows the EOS values in each bin for both models and 
also the amount of bias of the estimated parameter values per offset of 
$w_{N + 1}$ relative to the a priori assumption, as calculated from 
Eq.~(\ref{eq: bias 3}).  For example, if one assumed that $w_{N+1}=-1$, 
then the bias in $w_2$ for the bending model would be $\delta w_2= 
0.22=-0.26\times (-1+0.16)$; 
that is, instead of measuring the true value $w_2=-0.72$ one 
would think $w_2=-0.50$.  Assuming cosmological constant behavior at high 
redshift has very little effect on the PNGB model, since at high redshift 
it indeed is close to $w=-1$.  But we don't know a priori what the true 
dark energy behavior will be.

\begin{table*}
\begin{center}
\begin{tabular}{c|c|c|c|c|c|c}
parameter $p_i$ & $w_1$ ($z = 0 - 0.2$) & $w_2$ ($z = 0.2 - 0.5$) & $w_3$ ($z = 0.5 - 1.7$) & $w_4$ ($z = 1.7 - 1089$) & $\mathcal{M}$ & $\Omega_{DE}$\\
\hline
PNGB true value & -0.83 & -0.87 & -0.93 & -0.995 & anything & 0.72\\
\hline
$d p_i/d w_4$ & -0.015 & -0.019 & -0.063 & {\bf x} & -0.00022 & -0.0097\\
\hline
$\sigma(p_i)$ fixing $w_4$ & 0.10 & 0.16 & 0.15 & {\bf x} & 0.016 & 0.012\\
\hline
\hline
parameter $p_i$ & $w_1$ ($z = 0 - 0.2$) & $w_2$ ($z = 0.2 - 0.5$) & $w_3$ ($z = 0.5 - 1.7$) & $w_4$ ($z = 1.7 - 1089$) & $\mathcal{M}$ & $\Omega_{DE}$\\
\hline
Bending true value & -0.84 & -0.72 & -0.55 & -0.16 & anything & 0.72\\
\hline
$d p_i/d w_4$ & -0.21 & -0.26 & -0.39 & {\bf x} & -0.0022 & -0.14\\
\hline
$\sigma(p_i)$ fixing $w_4$ & 0.096 & 0.16 & 0.11 & {\bf x} & 0.016 & 0.012
\end{tabular}
\end{center}
\caption{Biases in cosmological parameter estimation due to fixing $w(z>1.7)$ 
to an incorrect value.  The top half of the table considers a PNGB dark 
energy model, which has $w(z>1.7)\approx-1$, and the bottom half considers 
a bending dark energy model, where $w(z>1.7)$ differs substantially from $-1$. 
The amount of bias $d p_i$ per how much $w_4$ is misestimated is shown in 
the middle row of each set.} 
\label{tab: bias}
\end{table*}

To avoid bias, we must leave $\whi$ as a fit parameter.  However, this 
greatly increases the uncertainties, since adding a single parameter 
and a single data point, with only that data point constraining that 
parameter, is equivalent to adding neither the parameter nor the data as 
far as the uncertainties 
in the original parameters are concerned -- essentially 
throwing away the high redshift bin.  The solution that allows for 
control of both bias and uncertainty is to obtain more, useful data 
that depends on $\whi$.  Such data could be higher redshift distances, 
such as from baryon acoustic oscillation (BAO) measurements using 
quasars or the Lyman alpha forest, or from matter density growth factors 
such as enter into weak gravitational lensing measurements.  While we 
note that SNAP, which we took to provide the supernova sample, includes 
highly precise weak lensing measurements, here we continue to concentrate 
on distances and illustrate the effect of a 1.2\% measurement of the 
reduced angular distance $\tilde d$ (transverse BAO scale) at 
$z=3$ such as the BOSS experiment \cite{BOSS} could provide.  

Table~\ref{tab: corr bending} shows the effects on the EOS 
uncertainties from fixing $\whi$ (and so incurring bias), fitting for it 
with only a CMB $\dls$ measurement (and so effectively using SN 
alone), and fitting for it with both $\dls$ and $\tilde d(z=3)$ measurements. 
We see that not only do the uncertainties greatly decrease when data give 
constraints on the high redshift expansion history, but the correlations 
between EOS parameters greatly diminish.  Again we emphasize that weak 
lensing measurements have the same or better effect.  The key point is 
that assuming high redshift behavior for dark energy leads to bias -- to 
overcome this requires accurate measurements (beyond CMB data alone) of 
the high redshift universe, e.g.\ through direct $z>1.7$ observations or 
through weak lensing observations involving the growth factor.  Given 
such measurements, one recovers almost the full leverage on the EOS as 
when $\whi$ was assumed, but without bias.  For the two very different 
models we considered, the EOS parameter estimation by doing a global fit 
including $w_{N+1}$ 
is degraded by less than 15\% and the risk (the uncertainty and the bias 
summed in quadrature) is improved by factors up to 3. 
Of course if with the additional data one attempts to fit additional high 
redshift EOS parameters, then the constraints do not improve as much.

\begin{table*}
\begin{center}
\begin{tabular}{c|c|c|c|c|c|c|c|c|c|c}
{\bf PNGB}& $\sigma_1$ & $\sigma_2$ & $\sigma_3$ & $\sigma_4$ & $r_{12}$ & $r_{13}$ & $r_{14}$ & $r_{23}$ & $r_{24}$ & $r_{34}$\\
\hline
fixing $w_4$ & 0.10 & 0.16 & 0.15 & {\bf x} & -0.76 & 0.45 & {\bf x} & -0.72 & {\bf x} & {\bf x}\\
fitting $w_4$ (CMB) & 0.32 & 0.42 & 1.3 & 20 & 0.79 & 0.96 & -0.95 & 0.89 & -0.92 & -0.99\\ 
fitting $w_4$ (CMB+$d_3$) & 0.10& 0.16& 0.17& 2.1& -0.73& 0.48& -0.21& -0.59& -0.13& -0.53\\ 
\hline
\hline
{\bf Bending}& $\sigma_1$ & $\sigma_2$ & $\sigma_3$ & $\sigma_4$ & $r_{12}$ & $r_{13}$ & $r_{14}$ & $r_{23}$ & $r_{24}$ & $r_{34}$\\
\hline
fixing $w_4$ & 0.096 & 0.16 & 0.11 & {\bf x} & -0.73 & 0.39 & {\bf x} & -0.84 & {\bf x} & {\bf x}\\
fitting $w_4$ (CMB)& 0.98 & 1.2 & 1.8 & 4.7 & 0.98 & 1.00 & -1.00 & 0.98 & -0.99 & -1.00\\ 
fitting $w_4$ (CMB+$d_3$)& 0.098& 0.15& 0.12 & 0.07 & -0.77& 0.43& -0.20& -0.81& 0.20& -0.48 
\end{tabular}
\end{center}
\caption{As Table~\ref{tab: bias}, showing the EOS uncertainties and 
correlation coefficients.  Fitting for $w_4\equiv w(z>1.7)$, 
which removes the bias calculated in Table~\ref{tab: bias}, increases the 
uncertainties 
and correlations, but the addition of further high redshift data (here 
illustrated with $d_3\equiv \tilde d(z=3)$) can substantially restore them. 
}
\label{tab: corr bending}
\end{table*}

\section{Physical Constraints on Equation of State \label{sec:wlimit}} 

\subsection{Eigenmode Expansion \label{sec: princ comp sub}} 

We pointed out at the end of \S\ref{sec:pca} that to reduce the parameter space by throwing out poorly
determined modes in the eigenmode expansion, we need to make assumptions 
about the appropriate range of values for
the parameters/coefficients $\alpha_i$. One way to do this is to take constraints on $w(z)$ based on theory
(if we have any such constraints) and convert these into constraints on the 
parameters $\alpha_i$ (see for example \cite{DickKnoxChu06}).  If for example we then find that $-\alpha_i^{\rm max} < \alpha_i < \alpha_i^{\rm max}$, we may want to
throw out the $i$th mode if $\alpha_i^{\rm max} < \sigma_i$ (or perhaps $\alpha_i^{\rm max} < 2 \sigma_i$)
because $\alpha_i^{\rm max} = \sigma_i$ means that the maximum physical signal
in $\alpha_i$ is equal to its observational uncertainty and thus we cannot get a convincing signal in this parameter.

As an example, imagine we expect the equation of state to be $w = -1$ 
and have some reason to believe that $-2 < w(z) < 0$ is required, for all $z$.
In other words,
if we choose the baseline equation of state (see Eq.~\ref{eq: expansion w}) 
to be $w_b = -1$, we want the magnitude of
the deviation from the baseline to be smaller than one:
\beq
\label{eq: constraint}
|w(z) - w_b(z)| = \left|\sum_i \alpha_i \,e_i(z)\right| < 1.
\eeq
This constraint of course defines some complicated volume in the
$\alpha_1$-\dots-$\alpha_N$ space (correlating the constraints on the different $\alpha_i$),
but we can get simple maxima $\alpha_i^{\rm max}$ for the individual $\alpha_i$
by treating the constraint~(\ref{eq: constraint}) less rigorously.

One way of doing this is to demand that the contributions of the individual modes do not exceed one, i.e.\ $|\alpha_i \, e_i(z)| < 1$
for all $z$
for each $i$ individually. This gives
\beq
\alpha_i^{\rm max} = 1/|e_i(z)|_{\rm max}
\eeq
and we have checked that (using the criterion $\alpha_i^{\rm max} < \sigma_i$) this allows us
to throw out all but the first five modes for the case depicted in 
Fig.~\ref{fig: eigenmodes} (left), independent of
whether the binning is uniform in $z$, $a$ or $\ln(1 + z)$. One of the problems with this approach is that if one mode
locally causes an unacceptably large deviation from $w = -1$,
this deviation may be canceled by another mode with large amplitude so in those cases the constraint is
stricter than Eq.~(\ref{eq: constraint}). The inverse is also 
true, that a mode that has an acceptably small deviation might be augmented 
by another mode so as to exceed our desired constraint.

An alternative approach that does not suffer from the first of the two problems mentioned above is
discussed in \S3 of \cite{MortHu07}, where it is applied to the reionization
history of the universe instead of the dark energy EOS (note the 
ionization fraction is bounded in [0,1]). In this approach, maxima are 
calculated such that if {\it any} coefficient violates
$|\alpha_i| < \alpha_i^{\rm max}$, Eq.~(\ref{eq: constraint}) is violated as well. The converse is not true. All modes
satisfying $|\alpha_i| < \alpha_i^{\rm max}$ does not guarantee that the original constraint is satisfied so this approach does suffer from the
second problem mentioned in the previous paragraph. The $\alpha_i^{\rm max}$'s calculated in this approach are greater than (or equal to, in the
limiting case of a constant mode) the ones in the approach discussed 
above and thus give a larger range of allowed values. 
When applied to the case at hand, the maxima for the approach discussed 
in \cite{MortHu07} are given by
\beq
\alpha_i^{\rm max} = \int d z \, |e_i(z)|.
\eeq
We have checked, again for the case depicted in Fig.~\ref{fig: eigenmodes} 
(left), that if we require this $\alpha_i^{\rm max} > \sigma_i$ then 
this very conservative criterion means we can eliminate modes beyond 
the first 9 or 10 (depending on if we calculate the eigenmodes with 
respect to $z, a$ or $\ln(1 + z)$). 

Note that even if we throw out a large number of modes using the methods described above, the remaining parameters
still carry a lot of uncertainty.
Also, to illustrate our ideas we have assumed an expected $w = -1$ with $-2 < w(z) < 0$, but in reality
we have very little knowledge to base such assumptions on (but see the 
next subsection).  
Finally, please recall that in \S\ref{sec:pca} we identified two main problems with the eigenmode
approach. Above, we considered the problem of how to quantify which modes are well-determined and which ones are not.
However, there was another problem, namely that different binnings give a different set of modes.
This implies that, after throwing out poorly determined modes, essentially different models remain. For example,
the first five modes with respect to $a$ span a different
set of equations of state than the first five modes with respect to $z$.

\subsection{Time Variation}
\label{sec: time var}

The EOS $w(z)$ has physical constraints not just on its value but also 
its time variation.  The effective mass of scalar field dark energy is 
related to the curvature of the potential and can be written in terms of 
$w$, $w'$, and $w''$, as in \cite{Cald00,Linder06}, where a prime 
denotes a derivative with respect to $\ln a$.  (Note there is a typo 
in the first term of Eq.~46 in \cite{Linder06} where $2q$ should be $q/2$.)  
If the mass exceeds the 
Hubble parameter, $m\gg H$, then the Compton wavelength for fluctuations 
in the scalar field will be less than the Hubble length and dark energy 
will exhibit clustering \cite{Maetal99}.  If we wish to disallow such models 
(ideally through observational constraints, although high energy physics 
such as supergravity can lead to limits on mass scales \cite{Kalloshetal02}) 
then this imposes the condition 
\beq 
\frac{m}{H}\lesssim 1 \quad\ \Longrightarrow \quad \ 
\left|\frac{w'}{1+w}\right|\lesssim 1, \label{eq:mH} 
\eeq 
unless the relation between $w$, $w'$, and $w''$ is fine tuned.  For 
example, this imposes constraints on oscillatory behavior, saying the 
variation cannot be too extreme.  For EOS expanded in a Fourier basis 
in $\ln a$, say, all terms $\cos(B\ln a)$ with $B\gg1$ would give 
inhomogeneities so the physical condition of smoothness would limit which 
modes should be included. 

In terms of binned EOS, the condition~(\ref{eq:mH}) reads 
\beq 
\left|\frac{w_{i+1}-w_i}{1+(w_{i+1}+w_i)/2}\right|\, 
\frac{1}{\ln[(1+z_{i+1})/(1+z_i)]} \lesssim 1. 
\eeq 
To help satisfy this we want a large distance between bin centers.  Taking 
the extreme case of $z_1\approx0$, $z_2\approx1.7$, then $|\Delta w|\lesssim 
1+\bar w\lesssim 1$.  That is, bin values should not jump by of order unity. 
For bins closer together the jump constraint is tighter.   Dark energy 
lying within the thawing and freezing regions defined by \cite{CaldLind05} 
automatically satisfies the mass constraint.  For effective dark energy 
without a physical 
fluid, as in extended gravity origins, constraints on $w'$ from inhomogeneity 
considerations may not apply.  Other possibilities for constrained EOS 
behavior can arise within a particular class of models; \cite{albrechtgroup} 
explores this for some potentials using PCA and \cite{CritPog05} chooses a 
correlation function over redshift for $w(z)$.

\subsection{Testing the Equation of State} 

Finally, one might want to apply several tests for physical properties to 
the EOS, which can be phrased simply in terms of the EOS bin values.  To 
check consistency with the cosmological constant, $w=-1$, to a confidence 
level of $S\sigma$, one looks for $(1+w_i)/\sigma(w_i)>S$.  To look for 
departures from a constant EOS, one probes whether 
\beq 
\frac{w_i-w_j}{\sigma(w_i-w_j)}=\frac{w_i-w_j}{\sqrt{\sigma_i^2+\sigma_j^2 
-2\,C_{ij}}}>S, 
\eeq 
for any $i$, $j$.  This also gives a necessary but not sufficient condition 
for distinguishing thawing vs.\ freezing behavior: whether $w$ decreases 
or increases with larger redshift.  

Another interesting property would be 
nonmonotonicity in the EOS.  This could be indicated by having $w_{i+p}-w_i$ 
of opposite sign from $w_{i+r}-w_{i+q}$, where $p<q<r$.  (Note we do not 
only consider consecutive bins since low $\sigma$ differences between 
neighboring bins could add up to statistically significant deviations over 
a wider range.)  That is, one tests whether 
\beq 
\frac{w_{i+p}-w_i}{\sigma(w_{i+p}-w_i)}<-S \quad \ {\rm and} \quad \ 
\frac{w_{i+r}-w_{i+q}}{\sigma(w_{i+r}-w_{i+q})}>S, 
\eeq 
or the opposite.  

While from the above points it would appear that for testing $\Lambda$, 
say, the FOM should be minimizing $\sigma(w_i)$ in any one bin, this 
in fact does not hold.  Such a criterion would drive us to create a 
single bin over 
the entire data redshift range, indeed giving a minimal $\sigma(w_i)$, 
but erasing any dynamics, taking a constant $w$.  This averaged $w$ can 
in fact under certain circumstances be driven to appear as $w=-1$ 
despite real time variation 
\cite{Linder0708}, so such a FOM is not useful.  For checking constancy, 
monotonicity, and related properties, one might advocate a FOM involving 
$\sigma(w_{i+p}-w_i)$.  This effectively takes a further derivative of 
the cosmological expansion and tends to yield large errors (while of course 
being a highly unstable procedure if applied directly to the data). 

Table~\ref{tab: corr 4 bins extra 2} demonstrates the lack of precision 
in determining $w_{i+p}-w_i$ or the variation $w'=dw/d\ln a$, even when 
fixing the high redshift behavior $\whi$ ({\it not\/} recommended), 
within the binned EOS approach.  Even for this optimistic case with next 
generation data, fitting four EOS parameters is too much: the dynamics 
represented by $w'$ cannot be seen.  This agrees with \cite{LinHut05} that 
next generation data will only allow 
physical insight into two EOS parameters.  For the two bin case we 
considered in \S\ref{sec:bin}, one can obtain $\sigma(w'_{12})=0.23$.

\begin{table*}
\begin{center}
\begin{tabular}{c|c|c|c|c|c||c|c|c}
$\sigma(w_2 - w_1)$ & $\sigma(w_3 - w_2)$ & $\sigma(w_3 - w_1)$ & 
$\sigma(w_4-w_3)$ & $\sigma(w_4-w_2)$ & $\sigma(w_4-w_1)$ & 
$\sigma(w'_{12})$ & $\sigma(w'_{23})$ & $\sigma(w'_{34})$ \\
\hline
0.47 & 0.94 & 0.57 & 0.88 & 0.36 & 0.35 & 2.8 & 6.6 & 2.4 
\end{tabular}
\end{center}
\caption{Uncertainties in the EOS jumps between bins and the derivatives 
$w'\equiv dw/d\ln a$ for the four redshift bins covering $z<1.7$ of 
Eq.~(\ref{eq:fix5}).  Note $w_{N+1}$ is fixed to $-1$. } 
\label{tab: corr 4 bins extra 2}
\end{table*}

\section{Conclusions \label{sec:concl}}

The dark energy equation of state properties contain clues crucial to 
understanding the nature of the acceleration of the cosmic expansion. 
Deciphering those properties from observational data involves a combination 
of robust analysis and clear interpretation.  We considered three 
approaches -- principal components, uncorrelated bandpowers, and binning; 
none of the approaches provides a panacea. 

In particular, we identify issues of dependence on basis functions, binning 
variables, and baseline models.  The three approaches are not truly 
nonparametric and physical interpretation (not merely the values) of the 
results in the two decorrelated basis techniques depends on model, priors, 
and data, indeed even on an implicitly assumed functional form. 
Nevertheless, principal components can give a useful guide to the 
qualitative sensitivity, the best constrained aspects, of the data.  

The uncorrelated bin approach unfortunately does not truly deliver 
uncorrelated bandpowers for the equation of state.  This approach using 
the square root of the Fisher matrix does not tightly localize the 
information (without a strong prior), making the interpretation nontrivial.  
This property of nonlocality is inherent in the cosmological 
characteristics.  One might prefer to stay with the original binned 
equations of state used as the initial step 
for this technique, which are readily interpreted.  Conversely, if the 
modes can be localized, the interpretation is easy, but in that case the 
original Fisher matrix is close to diagonal and thus the original bins 
almost uncorrelated.  Hence, again, one might as well stay with the bin 
parameters which have a clear meaning. 

Indeed the goal is understanding the physics, not obtaining particular 
statistical properties.  Decorrelated parameters that are not readily 
interpretable physically are of limited use; for example one still 
prefers to analyze the cosmic microwave background in terms of physical 
quantities such as physical matter density and spectral tilt rather than 
the principal axes of the eigenvectors.  
Note that the uncertainty on the EOS behavior $\sigma(w(z))$ is the same 
whether calculated by PCA (if all modes are kept), uncorrelated bands, 
or binned EOS, since the same information is in the data. 
We also emphasize that the 
modes most clearly determine the effect on the equation of state, not 
the weights, which are often the only quantity displayed.  Moderately 
localized, even all positive, weights do not guarantee a localized 
physical effect.  A further caution is that locality and positivity of 
weights can owe more to prior restrictions, especially the treatment 
of the high redshift equation of state, than to the data itself. 

Assuming a fixed value for the high redshift equation of state has 
major, widespread impacts on the results, ranging from strongly misestimated 
uncertainties to spurious localization to bias in the derived cosmology. 
We emphasize that it is essential to fit for the high redshift behavior 
in order not to be misled.  Adding CMB data and marginalizing over a new, 
high redshift bin removes the ill effects of bias but ``cancels out'', 
providing no new constraints; multiple data points for $z>2$ are required, 
such as from high 
redshift distances or weak lensing measurements of the mass growth behavior. 
Assuming that dark energy is negligible at $z>2$ is also effectively 
assuming a functional form -- precisely what the use of eigenmodes was 
supposed to avoid. 

Indeed, functional forms do not have many of the basis, model, binning, 
etc.\ dependences of eigenmodes, while principal components are in turn 
not fully form independent.  
If one assumes a functional form to obtain informative constraints on 
the equation of state,  one must indeed choose the form to represent 
robustly the physical behavior (as has been shown to be widely the 
case for $w(a)=w_0+w_a(1-a)$ by \cite{Linder03,Linder0708}), and 
carefully check the range of validity of the conclusions by examining 
other forms.  A good complementary analysis tool would be the binned 
equation of state approach examined here. 

Regardless of the form of analysis, only a finite amount of information can 
be extracted from even next generation data.  As has been concluded for 
functional equations of state and principal component analysis 
\cite{LinHut05}, the analysis here in terms of binned equation of state 
indicates that only two physically informative parameters can be fit 
with realistic accuracy.  However, we identify several issues in 
the PCA and uncorrelated bin approaches that cause accuracy or signal to 
noise criteria to be ill defined.  Similar difficulties arise in 
condensing the physical information on dark energy to a single figure 
of merit; the number is quite sensitive to cosmologically irrelevant 
aspects like the binning used (as well as very dependent on the treatment 
of the high redshift dark energy behavior). 

In conclusion, physically motivated fitting of the equation of state 
such as the $w_0$-$w_a$ parametrization in complement with a binned 
equation of state approach (perhaps with physical constraints such as 
outlined in \S\ref{sec:wlimit}) have the best defined, clearest to 
interpret, and robust insights of the approaches we considered.  
With any method, one must use caution regarding the influence of priors 
and fit the dark energy physics over the entire expansion history.

\acknowledgments

We thank Dragan Huterer for helpful discussions. 
This work has been supported in part by the Director, Office of Science, 
Department of Energy under grant DE-AC02-05CH11231.

\appendix 

\section{Properties of Decorrelated Modes \label{sec:apxpca}} 

In this Appendix, we first introduce some definitions and discuss some useful general properties of decorrelated modes (\S\ref{sec:basisexp}).
We then show that eigenvectors are formally ill-defined for a Fisher matrix 
(\S\ref{sec:basisdep}) and that the eigenmodes (eigenvectors in the
limit of a large number of bins) depend on the coordinate (redshift $z$, scale factor $a$, etc.)
one uses to write the EOS $w$ as a function of (\S\ref{sec:coorddep}). We consider the latter to be the main result of this Appendix.

\subsection{Basis Expansion \label{sec:basisexp}} 

The matrix ${\bf W}$ defines a basis transformation by
\beq
\label{eq: transform e}
{\bf e'_i} = W_{i j} {\bf e_j}, 
\eeq
so that the rows of ${\bf W}$ contain the new basis vectors as expressed 
with respect to the old basis\footnote{Note that in some literature 
(e.g.\ \cite{HutCoor05, Riessetal06}) the transformation matrix is 
defined as the matrix transforming the coordinates: our ${\bf W}$ is the 
inverse transpose of that matrix.}.  
The coefficients, or components, ${\bf \alpha} = (\alpha_1, ..., \alpha_N)$ 
then transform according to
\beq
\label{eq: transform coeff}
{\bf \alpha}' = {\bf (W^{-1})^{\rm T} \alpha}.
\eeq
If the transformation is orthogonal, ${\bf W^{\rm T}} = {\bf W^{-1}}$, the basis vectors and the coefficients
transform in the same way. However, this is not the case in general.

Since the Fisher matrix is a Hessian matrix, i.e.~it is defined in terms of
second order partial derivatives,
\beq
\label{Fisher def}
F_{i j} = \left\langle - \frac{\pa^2 \ln L}{\pa \alpha_i \pa \alpha_j} \right\rangle,
\eeq
it transforms according to
\beq
\label{eq: transform F}
{\bf F'} = {\bf W}\, {\bf F}\, {\bf W^{\rm T}}.
\eeq
It will become clear below that one of the main points of \S\ref{sec:pca},
namely that eigenmodes depend on the binning used to calculate them in,
is essentially a consequence of this transformation behavior.

Diagonalizing ${\bf F}$ comes down to finding a matrix ${\bf W}$ such that
\beq
{\bf W} \,{\bf F}\, {\bf W^{\rm T}} = {\bf D}
\eeq
is diagonal. In such a basis the uncertainties in the coefficients $\alpha'_i$ are uncorrelated.
It is straightforward to show that there is an infinite number of bases that achieve this.  
The remainder of this Appendix focuses on the particular choice of eigenvectors as basis (see also \S\ref{sec:pca}).

\subsection{Basis Dependence of Eigenmodes\label{sec:basisdep}} 

If a set of eigenvectors is orthonormal (which can always be arranged), the eigenvalues are equal to
the diagonal elements of the diagonal Fisher matrix, i.e.\ the inverse variances.
Eigenvectors are defined by
\beq
\label{eq: eigen}
{\bf F} {\bf v} = \lambda {\bf v}, 
\eeq 
and their components transform according to Eq.~(\ref{eq: transform coeff}).  
However, since the Fisher matrix transforms according to Eq.~(\ref{eq: transform F}), this
is not a covariant statement:
\beq
{\bf F'} {\bf v'} = {\bf W} \,{\bf F}\, {\bf W^{\rm T}} \,{\bf (W^{-1})}^{\rm T} \,{\bf v} = \lambda {\bf W} {\bf v}.
\eeq
This is only equal to
\beq
\lambda {\bf v'} = \lambda {\bf (W^{-1})}^{\rm T} {\bf v}
\eeq
if the coordinate transformation is orthogonal, i.e.~${\bf W^{\rm T}} = {\bf W^{-1}}$, but not
in general! This means that, formally, eigenvectors of a Fisher matrix are not well-defined.

Of course,
we can take a pragmatic approach and just compute the eigenvectors (for lack of a better word,
we will still call them eigenvectors) in a particular basis and work with those.
This is what we will do, but it is important to remember that
the set of eigenvectors found in this way depends on
the particular basis we chose to compute them in.

\subsection{Coordinate Dependence of Eigenmodes\label{sec:coorddep}}

We now turn our attention to the eigenmodes in the $N \to \infty$ limit, 
where $N$ is the number of EOS bins.  
We start with the basis of modes $e_i(z)$ discussed in \S\ref{sec:cosdep} 
that are equal to one inside the $i$th bin and zero everywhere else.  
In the limit $N \to \infty$ (keeping the {\it relative} bin widths the same) 
the eigenvectors approach a set of continuous functions (eigenmodes) and 
these eigenmodes and the corresponding standard deviations converge 
(see for example \cite{HutStark03}).  

In this section, we address the question of whether the eigenmodes are 
independent of which coordinate we use to write $w$ as a function of.  
For example, we may choose a binning that is uniform in terms of the 
scale factor $a = 1/(1 + z)$, i.e.\ $\Delta a$ constant instead of 
uniform in $z$, i.e.\ $\Delta z$ constant. Note that this is 
equivalent to a non-uniform binning in $z$,
\beq
\Delta z_i \approx \frac{d z}{d a}(z_i) \,\Delta a,
\eeq
where $z_i$ is a redshift inside the $i$th bin.
Since we saw before that
eigenvectors of the Fisher matrix are basis dependent, it should not be too surprising if the eigenmodes turn out to depend on the relative bin sizes.
Indeed, we find this is the case. We will explain this in the remainder of this section (specific examples are shown in
\S\ref{sec:pca}).

Assume a binning that is uniform in a variable $x = x(z)$, which is either monotonically increasing or decreasing as a function of $z$ in
the relevant redshift range. For example, $x$ could be the scale factor $a$ or perhaps its logarithm.  To see if the eigenmodes calculated using $x$
are the same as the ones calculated using $z$, we will need to make use of the following results.

Let ${\bf F}$ be the Fisher matrix for a set of $N$ bins with widths $\Delta z_i$ and
${\bf F}'$ be the one for a set of $N'$ bins with widths $\Delta z'_i$. Then
for large enough $N$ and $N'$, 
\beq
\label{eq: thm 1}
F'(z, z') \approx \frac{\Delta z'}{\Delta z}(z)\,\frac{\Delta z'}{\Delta z}(z') F(z, z'),
\eeq
where we have replaced discrete indices by the redshifts of the corresponding bins. For example,
$F(z, z') \equiv F_{i j}$ where the $i$th bin contains $z$ and the $j$th 
bin contains $z'$.  
Eq.~(\ref{eq: thm 1}) follows from the fact that derivatives with respect 
to EOS bin parameters should scale with the bin width for small enough bins. 
If we apply the above result to the cases of a binning with $\Delta z$ 
constant and one with $\Delta x$ constant, we get 
\beq
\label{eq: transform F x a}
F^{(x)}(z, z') \approx \left(\frac{\Delta x}{\Delta z}\right)^2 \frac{d z}{d x}(z) \frac{d z}{d x}(z') F^{(z)}(z, z'),
\eeq
where the superscript on $F$ denotes in which binning the Fisher matrix 
is calculated.

We can now apply the results from the previous paragraph to the eigenmodes 
discussion. 
Let us assume that $v(z)$ is an eigenmode calculated using $z$, i.e.
\beq
\label{eq: eigen z}
\sum_{z' (\Delta z)} F^{(z)}(z, z') \,v(z') = \lambda\, v(z),
\eeq
where the $(\Delta z)$ below the summation symbol indicates that the sum is supposed to be carried out over
the bins (labeled by $z'$) uniformly spaced in $z$.
Then,
\beqa
\sum_{z' (\Delta x)} F^{(x)}(z, z')\, v(z') 
&=& \left(\frac{\Delta x}{\Delta z}\right)^2 \sum_{z' (\Delta x)} 
\frac{d z}{d x}(z) \frac{d z}{d x}(z') F^{(z)}(z, z')\,v(z') \nonumber \\
&=& \left(\frac{\Delta x}{\Delta z}\right)^2 \sum_{z' (\Delta z)} \frac{\Delta z}{\Delta x}
\frac{d x}{d z}(z') \,\frac{d z}{d x}(z) \frac{d z}{d x}(z') F^{(z)}(z, z')\, v(z')  \nonumber \\
&=& \frac{\Delta x}{\Delta z} \frac{d z}{d x}(z)
\sum_{z' (\Delta z)} F^{(z)}(z, z') \,v(z') \nonumber \\
&=& \lambda \,\frac{\Delta x}{\Delta z}\, \frac{d z}{d x}(z)\, v(z), 
\eeqa
where in the first equality we have used Eq.~(\ref{eq: transform F x a}), 
in the second equality we went from
the binning uniform in $x$ to the binning uniform in $z$, which forced us to put in a factor 
$\frac{\Delta z}{\Delta x} \frac{d x}{d z}(z')$, and in the fourth equality we use the fact that
$v(z)$ is an eigenmode in the binning uniform in $z$, i.e.\ Eq.~(\ref{eq: eigen z}).
What the above shows is that $v(z)$ is an eigenmode in the $x$-binning only if $\frac{d z}{d x} =$ const
(recall that $\Delta x$ and $\Delta z$ are just constants by construction). 
Hence, using the scale factor $a$ or
any other coordinate that is not a linear function of $z$ will result in a different set of eigenmodes.  
(We illustrate this with numerical results in Fig.~\ref{fig: eigenmodes} of 
\S\ref{sec: eigen results}).
The above has strong implications when we try
to decide how many modes/parameters are well determined,
an issue that is explored further in \S\ref{sec: princ comp sub}.

\section{Model Dependence \label{sec:modeldep}} 

In addition to the modes and weights depending on basis, binning 
variable, and specific binning choice, we now consider dependence on 
the fiducial model.  
We analyze how the uncorrelated bandpowers that were discussed in 
\S\ref{sec:band} change as the fiducial EOS is changed from the $w = -1$ 
$\Lambda$CDM cosmology to the (discretized) PNGB and bending models 
discussed in \S\ref{sec:whi}.  

\begin{figure}
  \begin{center}{
  \includegraphics*[width=8.8cm]{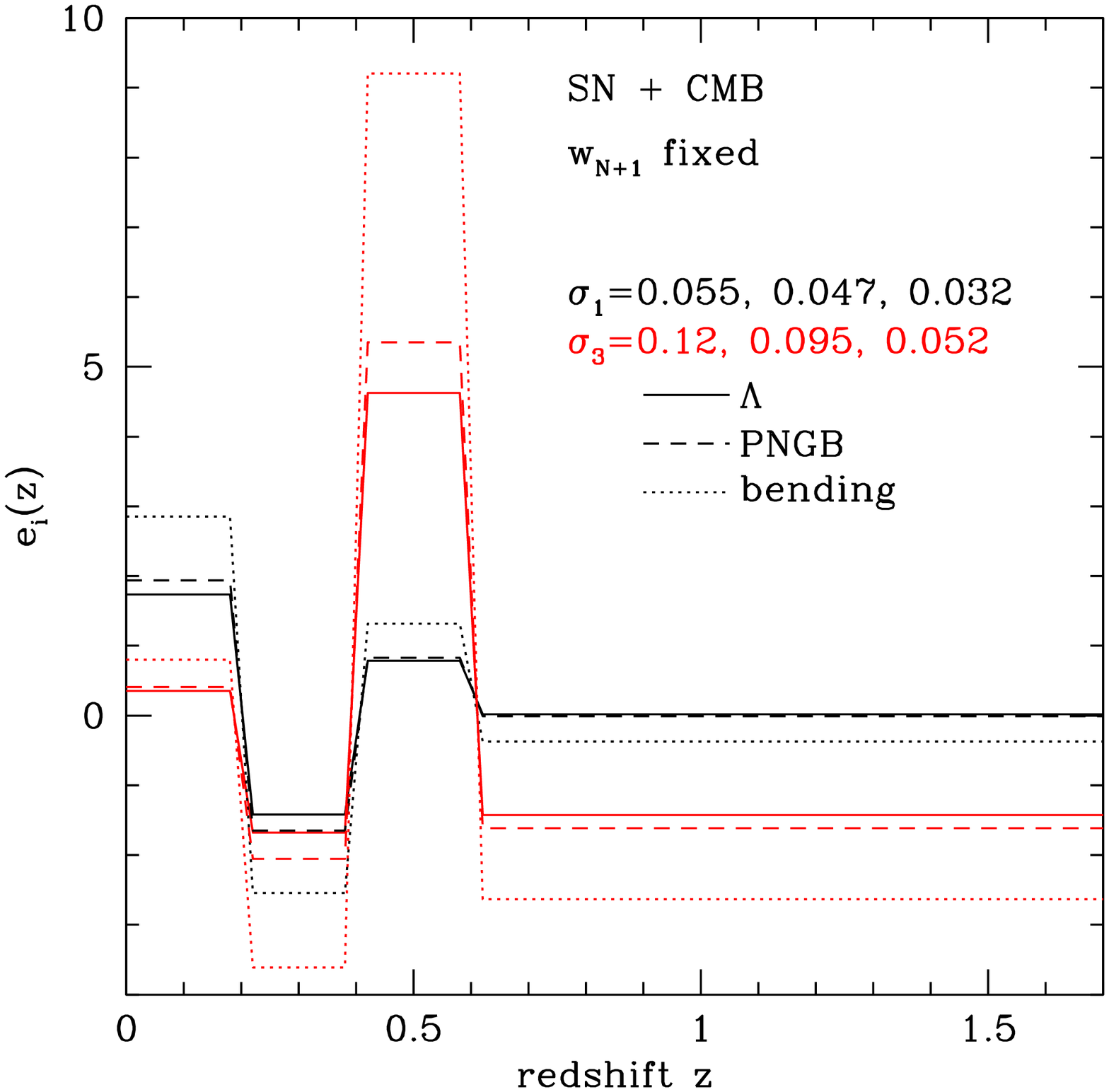}
  \includegraphics*[width=8.8cm]{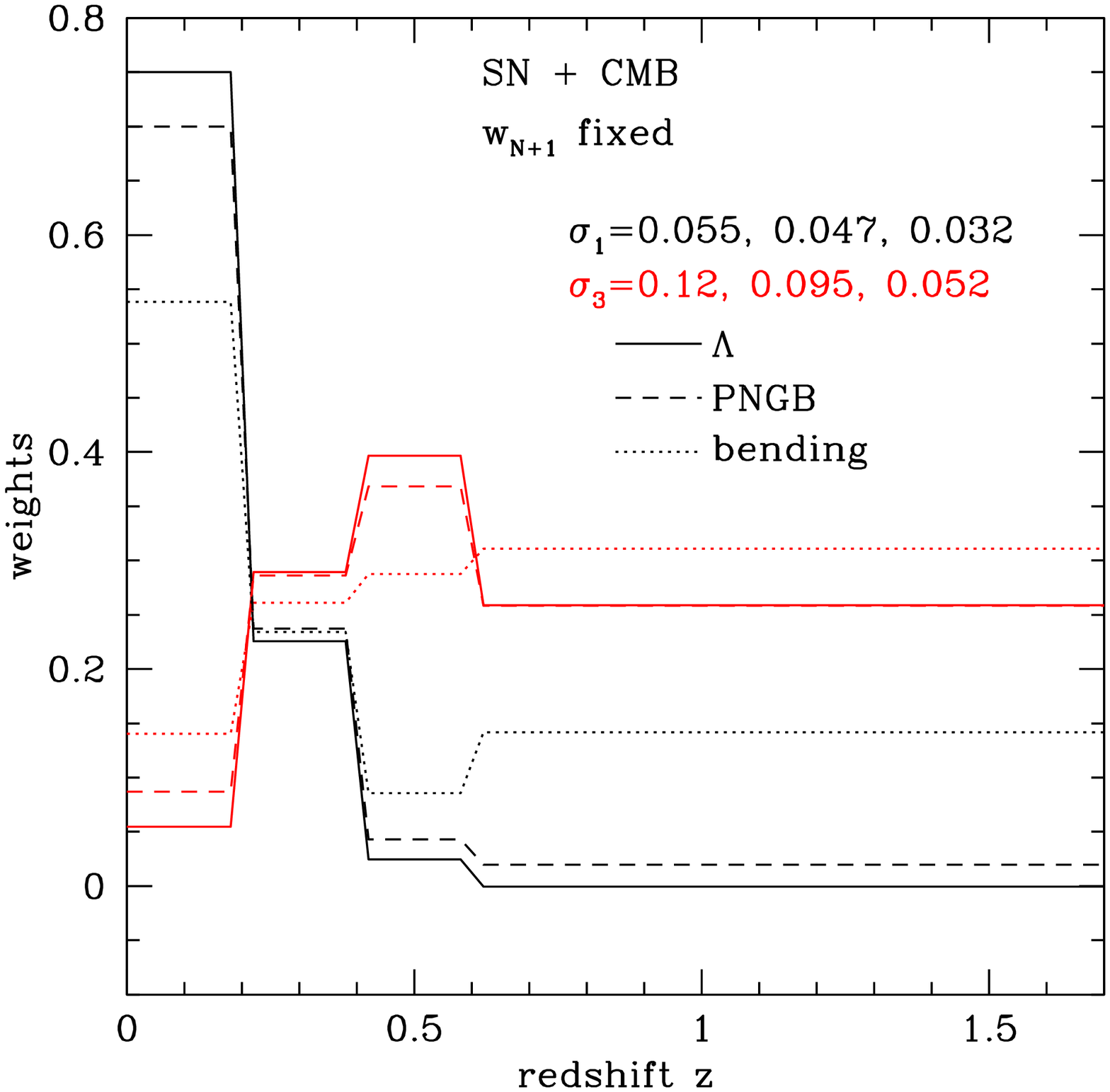}
  }
  \end{center}
  \caption{As Fig.~\ref{fig: bands 4 nomarg}, but comparing the first 
and third modes (left panel) and associated weights (right panel) for 
three dark energy fiducial models: cosmological constant $\Lambda$, 
PNGB, and bending (see \S\ref{sec:whi}).  Here we fix $w_{N+1}$ to its 
appropriate fiducial value for each model. } 
  \label{fig: bands 3 models nomarg}
\end{figure}

\begin{figure}
  \begin{center}{
  \includegraphics*[width=8.8cm]{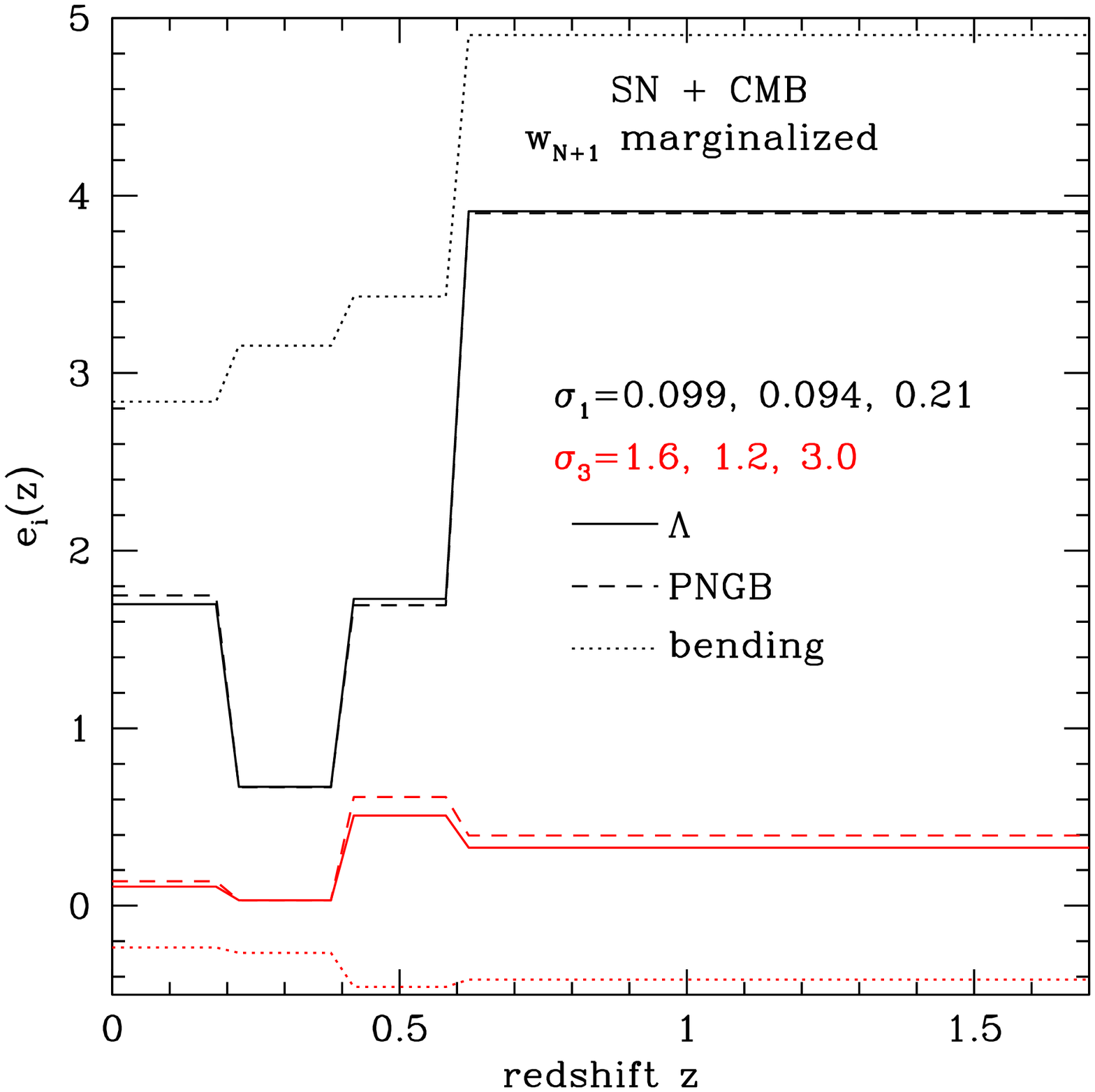}
  \includegraphics*[width=8.8cm]{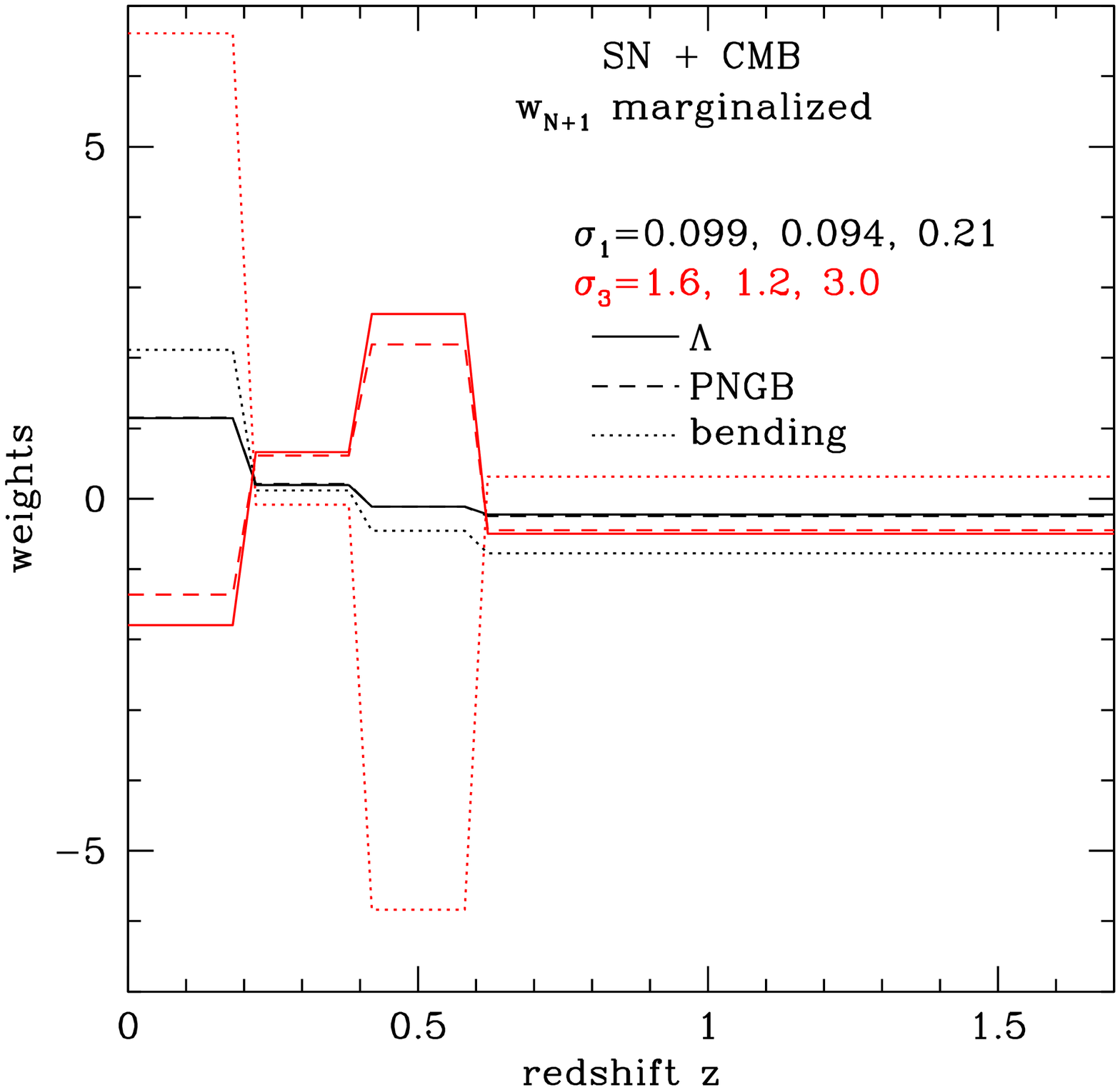}
  }
  \end{center}
  \caption{As Fig.~\ref{fig: bands 3 models nomarg}, but marginalizing 
over $w_{N+1}$. } 
  \label{fig: bands 3 models marg}
\end{figure}

Figures \ref{fig: bands 3 models nomarg} and \ref{fig: bands 3 models marg} 
show how the first and third modes, the corresponding weights, 
and the uncertainties change between these models. We again show results 
both for the case where we fix the EOS at $z > 1.7$ to the respective 
fiducials (Fig.~\ref{fig: bands 3 models nomarg}) and for the case where 
we treat it as a free parameter (Fig.~\ref{fig: bands 3 models marg}).  
While the PNGB results lie rather close to the $\Lambda$CDM ones, the 
bending model results in significantly different bandpowers and 
uncertainties.   Fiducial models deviating appreciably from each other 
will induce appreciable model dependence in the mode analysis.  
Note that changing the fiducial does not make the weights look ``better'', 
i.e.\ they do not get significantly more localized or positive.

\section{Parameter Bias Formula \label{sec:apxbias}}

In this section we derive Eq.~(\ref{eq: bias 3}), which tells us how much we misestimate the other parameters
when we fix one of the parameters to the wrong value.
Consider the general case where the observables $O_k = O_k(\{p_i\}_{i=1}^{n+1})$ depend on $n + 1$ parameters $p_i$. We call the true values
of the parameters $p_{{\rm t}, i}$. Now imagine that, instead of
fitting all $n + 1$ parameters to the data, we first fix $p_{n + 1}$ to $p_{{\rm fix}, n + 1}$
and then fit the resulting $n$ parameters to the data. To get the
correct values for these parameters, the observables would have to be given by $O_k(\{p_{{\rm t}, i}\}_{i=1}^n, p_{{\rm fix}, n + 1})$.
In reality, ignoring observational uncertainties (we do not want to write ``the expectation values of'' over and over),
the data are given by $O_k = O_k(\{p_{{\rm t},i}\}_{i=1}^{n+1})$. Hence, if $p_{{\rm fix}, n + 1} \neq p_{{\rm t}, n + 1}$, the $n$ parameter
values $p_{{\rm e}, i}$ derived from the data will be different from the actual values.

If we define
\beq
\label{eq: delta O}
\Delta O_k \equiv O_k(\{p_{{\rm t}, i}\}_{i=1}^{n+1}) - O_k(\{p_{{\rm t}, i}\}_{i=1}^n, p_{{\rm fix}, n + 1}) = - \frac{\pa O_k}{\pa p_{n + 1}}\,\delta p_{n + 1},
\eeq
where $\delta p_{n + 1} = p_{{\rm fix}, n + 1} - p_{{\rm t}, n + 1}$,
we can use
Eq.~(\ref{eq: bias}),
\beq
\label{eq: bias app}
\delta p_i \equiv p_{{\rm e}, i} - p_{{\rm t}, i} = (F^{(n)})^{-1}_{i j}\, \sum_k \frac{\pa O_k}{\pa p_j} 
\,\frac{1}{\sigma_k^2} \,\Delta O_k,
\eeq
where the superscript $(n)$ means that we need the $n \times n$ Fisher matrix calculated using the first $n$ parameters
(the ones that we have not fixed). Inserting Eq.~(\ref{eq: delta O}) into Eq.~(\ref{eq: bias app}) gives
\beq
\label{eq: bias 2}
\delta p_i = - \delta p_{n + 1} \times \sum_{j = 1}^n (F^{(n)})^{-1}_{i j}\,(F^{(n + 1)})_{j, n + 1}
\eeq
($i = 1, ..., n$), where we have used Eq.~(\ref{eq: fisher def}) to substitute the $(n + 1) \times (n + 1)$ Fisher matrix.  We then obtain 
Eq.~(\ref{eq: bias 3}), 
\beq
\label{eq: bias 3 appendix}
\frac{d p_i}{d p_{n + 1}} = - \sum_{j = 1}^n (F^{(n)})^{-1}_{i j}\,(F^{(n + 1)})_{j, n + 1}.
\eeq
Note that since Eq.~(\ref{eq: bias}) is only valid to first order, we 
can calculate the Fisher matrix using the true parameter values.

\bibliography{refs}

\begin{thebibliography}{35}
\expandafter\ifx\csname natexlab\endcsname\relax\def\natexlab#1{#1}\fi
\expandafter\ifx\csname bibnamefont\endcsname\relax
  \def\bibnamefont#1{#1}\fi
\expandafter\ifx\csname bibfnamefont\endcsname\relax
  \def\bibfnamefont#1{#1}\fi
\expandafter\ifx\csname citenamefont\endcsname\relax
  \def\citenamefont#1{#1}\fi
\expandafter\ifx\csname url\endcsname\relax
  \def\url#1{\texttt{#1}}\fi
\expandafter\ifx\csname urlprefix\endcsname\relax\def\urlprefix{URL }\fi
\providecommand{\bibinfo}[2]{#2}
\providecommand{\eprint}[2][]{\url{#2}}

\bibitem[{\citenamefont{{Huterer} and {Starkman}}(2003)}]{HutStark03}
\bibinfo{author}{\bibfnamefont{D.}~\bibnamefont{{Huterer}}} \bibnamefont{and}
  \bibinfo{author}{\bibfnamefont{G.}~\bibnamefont{{Starkman}}},
  \bibinfo{journal}{Phys.Rev.Lett.} \textbf{\bibinfo{volume}{90}},
  \bibinfo{pages}{031301} (\bibinfo{year}{2003}),
  \eprint{arXiv:astro-ph/0207517}.

\bibitem[{\citenamefont{{Crittenden} and {Pogosian}}(2005)}]{CritPog05}
\bibinfo{author}{\bibfnamefont{R.~G.} \bibnamefont{{Crittenden}}}
  \bibnamefont{and}
  \bibinfo{author}{\bibfnamefont{L.}~\bibnamefont{{Pogosian}}},
  \eprint{arXiv:astro-ph/0510293}.

\bibitem[{\citenamefont{{Shapiro} and {Turner}}(2006)}]{ShapTurn06}
\bibinfo{author}{\bibfnamefont{C.}~\bibnamefont{{Shapiro}}} \bibnamefont{and}
  \bibinfo{author}{\bibfnamefont{M.~S.} \bibnamefont{{Turner}}},
  \bibinfo{journal}{Astrophys.J.} \textbf{\bibinfo{volume}{649}},
  \bibinfo{pages}{563} (\bibinfo{year}{2006}), \eprint{arXiv:astro-ph/0512586}.

\bibitem[{\citenamefont{{Simpson} and {Bridle}}(2006)}]{SimpBrid06}
\bibinfo{author}{\bibfnamefont{F.}~\bibnamefont{{Simpson}}} \bibnamefont{and}
  \bibinfo{author}{\bibfnamefont{S.}~\bibnamefont{{Bridle}}},
  \bibinfo{journal}{Phys.Rev.D} \textbf{\bibinfo{volume}{73}},
  \bibinfo{pages}{083001} (\bibinfo{year}{2006}),
  \eprint{arXiv:astro-ph/0602213}.

\bibitem[{\citenamefont{{Dick} et~al.}(2006)\citenamefont{{Dick}, {Knox}, and
  {Chu}}}]{DickKnoxChu06}
\bibinfo{author}{\bibfnamefont{J.}~\bibnamefont{{Dick}}},
  \bibinfo{author}{\bibfnamefont{L.}~\bibnamefont{{Knox}}}, \bibnamefont{and}
  \bibinfo{author}{\bibfnamefont{M.}~\bibnamefont{{Chu}}},
  \bibinfo{journal}{JCAP} \textbf{\bibinfo{volume}{7}}, \bibinfo{pages}{1}
  (\bibinfo{year}{2006}), \eprint{arXiv:astro-ph/0603247}.

\bibitem[{\citenamefont{{Stephan-Otto}}(2006)}]{Ste06}
\bibinfo{author}{\bibfnamefont{C.}~\bibnamefont{{Stephan-Otto}}},
  \bibinfo{journal}{Phys.Rev.D} \textbf{\bibinfo{volume}{74}},
  \bibinfo{pages}{023507} (\bibinfo{year}{2006}),
  \eprint{arXiv:astro-ph/0605403}.

\bibitem[{\citenamefont{{Huterer} and {Peiris}}(2007)}]{HutPeir07}
\bibinfo{author}{\bibfnamefont{D.}~\bibnamefont{{Huterer}}} \bibnamefont{and}
  \bibinfo{author}{\bibfnamefont{H.~V.} \bibnamefont{{Peiris}}},
  \bibinfo{journal}{Phys.Rev.D} \textbf{\bibinfo{volume}{75}},
  \bibinfo{pages}{083503} (\bibinfo{year}{2007}),
  \eprint{arXiv:astro-ph/0610427}.

\bibitem[{\citenamefont{{Kim} et~al.}(2004)\citenamefont{{Kim}, {Linder},
  {Miquel}, and {Mostek}}}]{Kimetal04}
\bibinfo{author}{\bibfnamefont{A.~G.} \bibnamefont{{Kim}}},
  \bibinfo{author}{\bibfnamefont{E.~V.} \bibnamefont{{Linder}}},
  \bibinfo{author}{\bibfnamefont{R.}~\bibnamefont{{Miquel}}}, \bibnamefont{and}
  \bibinfo{author}{\bibfnamefont{N.}~\bibnamefont{{Mostek}}},
  \bibinfo{journal}{MNRAS} \textbf{\bibinfo{volume}{347}}, \bibinfo{pages}{909}
  (\bibinfo{year}{2004}), \eprint{arXiv:astro-ph/0304509}.

\bibitem[{\citenamefont{{Hamilton} and {Tegmark}}(2000)}]{HamTeg00}
\bibinfo{author}{\bibfnamefont{A.~J.~S.} \bibnamefont{{Hamilton}}}
  \bibnamefont{and}
  \bibinfo{author}{\bibfnamefont{M.}~\bibnamefont{{Tegmark}}},
  \bibinfo{journal}{MNRAS} \textbf{\bibinfo{volume}{312}}, \bibinfo{pages}{285}
  (\bibinfo{year}{2000}), \eprint{arXiv:astro-ph/9905192}.

\bibitem[{\citenamefont{{Hu} and {Okamoto}}(2004)}]{HuOka04}
\bibinfo{author}{\bibfnamefont{W.}~\bibnamefont{{Hu}}} \bibnamefont{and}
  \bibinfo{author}{\bibfnamefont{T.}~\bibnamefont{{Okamoto}}},
  \bibinfo{journal}{Phys.Rev.D} \textbf{\bibinfo{volume}{69}},
  \bibinfo{pages}{043004} (\bibinfo{year}{2004}),
  \eprint{arXiv:astro-ph/0308049}.

\bibitem[{\citenamefont{{Leach}}(2006)}]{Leach06}
\bibinfo{author}{\bibfnamefont{S.}~\bibnamefont{{Leach}}},
  \bibinfo{journal}{MNRAS} \textbf{\bibinfo{volume}{372}}, \bibinfo{pages}{646}
  (\bibinfo{year}{2006}), \eprint{arXiv:astro-ph/0506390}.

\bibitem[{\citenamefont{{Kadota} et~al.}(2005)\citenamefont{{Kadota},
  {Dodelson}, {Hu}, and {Stewart}}}]{Kadotaetal05}
\bibinfo{author}{\bibfnamefont{K.}~\bibnamefont{{Kadota}}},
  \bibinfo{author}{\bibfnamefont{S.}~\bibnamefont{{Dodelson}}},
  \bibinfo{author}{\bibfnamefont{W.}~\bibnamefont{{Hu}}}, \bibnamefont{and}
  \bibinfo{author}{\bibfnamefont{E.~D.} \bibnamefont{{Stewart}}},
  \bibinfo{journal}{Phys.Rev.D} \textbf{\bibinfo{volume}{72}},
  \bibinfo{pages}{023510} (\bibinfo{year}{2005}),
  \eprint{arXiv:astro-ph/0505158}.

 \bibitem[{\citenamefont{{Mortonson} and {Hu}}(2007)}]{MortHu07}
\bibinfo{author}{\bibfnamefont{M.~J.} \bibnamefont{{Mortonson}}}
  \bibnamefont{and} \bibinfo{author}{\bibfnamefont{W.}~\bibnamefont{{Hu}}},
  \eprint{arXiv:0705.1132}.

\bibitem[{\citenamefont{{Huterer} and {Cooray}}(2005)}]{HutCoor05}
\bibinfo{author}{\bibfnamefont{D.}~\bibnamefont{{Huterer}}} \bibnamefont{and}
  \bibinfo{author}{\bibfnamefont{A.}~\bibnamefont{{Cooray}}},
  \bibinfo{journal}{Phys.Rev.D} \textbf{\bibinfo{volume}{71}},
  \bibinfo{pages}{023506} (\bibinfo{year}{2005}),
  \eprint{arXiv:astro-ph/0404062}.

\bibitem[{\citenamefont{{Riess {\it et al.}}}(2007)}]{Riessetal06}
\bibinfo{author}{\bibnamefont{A.~G.} \bibnamefont{{Riess {\it et al.}}}},
  \bibinfo{journal}{Astrophys.J.} \textbf{\bibinfo{volume}{659}},
  \bibinfo{pages}{98} (\bibinfo{year}{2007}), \eprint{arXiv:astro-ph/0611572}.

\bibitem[{\citenamefont{{Sullivan} et~al.}(2007)\citenamefont{{Sullivan},
  {Cooray}, and {Holz}}}]{SulCooHolz07}
\bibinfo{author}{\bibfnamefont{S.}~\bibnamefont{{Sullivan}}},
  \bibinfo{author}{\bibfnamefont{A.}~\bibnamefont{{Cooray}}}, \bibnamefont{and}
  \bibinfo{author}{\bibfnamefont{D.~E.} \bibnamefont{{Holz}}},
  \bibinfo{journal}{JCAP} \textbf{\bibinfo{volume}{9}}, \bibinfo{pages}{4}
  (\bibinfo{year}{2007}), \eprint{arXiv:0706.3730}.

\bibitem[{\citenamefont{{Efstathiou} and {Bond}}(1999)}]{EfBond99}
\bibinfo{author}{\bibfnamefont{G.}~\bibnamefont{{Efstathiou}}}
  \bibnamefont{and} \bibinfo{author}{\bibfnamefont{J.~R.}
  \bibnamefont{{Bond}}}, \bibinfo{journal}{MNRAS}
  \textbf{\bibinfo{volume}{304}}, \bibinfo{pages}{75} (\bibinfo{year}{1999}),
  \eprint{arXiv:astro-ph/9807103}.

\bibitem[{\citenamefont{{Linder} and {Huterer}}(2005)}]{LinHut05}
\bibinfo{author}{\bibfnamefont{E.~V.} \bibnamefont{{Linder}}} \bibnamefont{and}
  \bibinfo{author}{\bibfnamefont{D.}~\bibnamefont{{Huterer}}},
  \bibinfo{journal}{Phys.Rev.D} \textbf{\bibinfo{volume}{72}},
  \bibinfo{pages}{043509} (\bibinfo{year}{2005}),
  \eprint{arXiv:astro-ph/0505330}.

\bibitem[{\citenamefont{{Suzuki}}(2006)}]{Suz06}
\bibinfo{author}{\bibfnamefont{N.}~\bibnamefont{{Suzuki}}},
  \bibinfo{journal}{Astrophys.J.Supp.} \textbf{\bibinfo{volume}{163}},
  \bibinfo{pages}{110} (\bibinfo{year}{2006}).

\bibitem[{\citenamefont{{Davis} et~al.}(2007)\citenamefont{{Davis}, {James},
  {Schmidt}, and {Kim}}}]{Davetal07}
\bibinfo{author}{\bibfnamefont{T.~M.} \bibnamefont{{Davis}}},
  \bibinfo{author}{\bibfnamefont{J.~B.} \bibnamefont{{James}}},
  \bibinfo{author}{\bibfnamefont{B.~P.} \bibnamefont{{Schmidt}}},
  \bibnamefont{and} \bibinfo{author}{\bibfnamefont{A.~G.} \bibnamefont{{Kim}}},
  \bibinfo{journal}{AIP Conf. Ser.} \textbf{\bibinfo{volume}{924}},
  \bibinfo{pages}{330} (\bibinfo{year}{2007}), \eprint{arXiv:astro-ph/0701904}.

\bibitem[{}()]{glcorr}
  \bibinfo{website}{http://rkb.home.cern.ch/rkb/AN16pp/node40.html}.

\bibitem[{\citenamefont{{Huterer} and {Turner}}(2001)}]{HutTur01}
\bibinfo{author}{\bibfnamefont{D.}~\bibnamefont{{Huterer}}} \bibnamefont{and}
  \bibinfo{author}{\bibfnamefont{M.~S.} \bibnamefont{{Turner}}},
  \bibinfo{journal}{Phys.Rev.D} \textbf{\bibinfo{volume}{64}},
  \bibinfo{pages}{123527} (\bibinfo{year}{2001}),
  \eprint{arXiv:astro-ph/0012510}.

\bibitem[{\citenamefont{{Albrecht} and {Bernstein}}(2007)}]{AlbBern07}
\bibinfo{author}{\bibfnamefont{A.}~\bibnamefont{{Albrecht}}} \bibnamefont{and}
  \bibinfo{author}{\bibfnamefont{G.}~\bibnamefont{{Bernstein}}},
  \bibinfo{journal}{Phys.Rev.D} \textbf{\bibinfo{volume}{75}},
  \bibinfo{pages}{103003} (\bibinfo{year}{2007}),
  \eprint{arXiv:astro-ph/0608269}.

\bibitem[{\citenamefont{{Linder}}(2006{\natexlab{a}})}]{Linder0604}
\bibinfo{author}{\bibfnamefont{E.~V.} \bibnamefont{{Linder}}},
  \bibinfo{journal}{Astropart.Phys.} \textbf{\bibinfo{volume}{26}},
  \bibinfo{pages}{102} (\bibinfo{year}{2006}{\natexlab{a}}),
  \eprint{arXiv:astro-ph/0604280}.

\bibitem[{\citenamefont{{Linder}}(2007)}]{Linder0708}
\bibinfo{author}{\bibfnamefont{E.~V.} \bibnamefont{{Linder}}},
\eprint{arXiv:0708.0024}.

\bibitem[{\citenamefont{{Frieman} et~al.}(1995)\citenamefont{{Frieman}, {Hill},
  {Stebbins}, and {Waga}}}]{Friemanetal95}
\bibinfo{author}{\bibfnamefont{J.~A.} \bibnamefont{{Frieman}}},
  \bibinfo{author}{\bibfnamefont{C.~T.} \bibnamefont{{Hill}}},
  \bibinfo{author}{\bibfnamefont{A.}~\bibnamefont{{Stebbins}}},
  \bibnamefont{and} \bibinfo{author}{\bibfnamefont{I.}~\bibnamefont{{Waga}}},
  \bibinfo{journal}{Phys.Rev.Lett.} \textbf{\bibinfo{volume}{75}},
  \bibinfo{pages}{2077} (\bibinfo{year}{1995}),
  \eprint{arXiv:astro-ph/9505060}.

\bibitem[{\citenamefont{{Wetterich}}(2004)}]{Wett04}
\bibinfo{author}{\bibfnamefont{C.}~\bibnamefont{{Wetterich}}},
  \bibinfo{journal}{Phys.Lett.B} \textbf{\bibinfo{volume}{594}},
  \bibinfo{pages}{17} (\bibinfo{year}{2004}), \eprint{arXiv:astro-ph/0403289}.

\bibitem[{}()]{BOSS}
  \bibinfo{website}{http://cosmology.lbl.gov/BOSS}.

 \bibitem[{\citenamefont{Caldwell}()}]{Cald00}
\bibinfo{author}{\bibfnamefont{R.~R.} \bibnamefont{{Caldwell}}},
\bibinfo{journal}{in Sources and Detection of Dark Matter in the Universe 
  (DM2000), ed.\ D.\ Cline, p.\ 74 (Springer: 2001);  
  http://www.dartmouth.edu/$\sim$caldwell/index\_files/DM2000.ps}.

\bibitem[{\citenamefont{{Linder}}(2006{\natexlab{b}})}]{Linder06}
\bibinfo{author}{\bibfnamefont{E.~V.} \bibnamefont{{Linder}}},
  \bibinfo{journal}{Phys.Rev.D} \textbf{\bibinfo{volume}{73}},
  \bibinfo{pages}{063010} (\bibinfo{year}{2006}{\natexlab{b}}),
  \eprint{arXiv:astro-ph/0601052}.

\bibitem[{\citenamefont{{Ma} et~al.}(1999)\citenamefont{{Ma}, {Caldwell},
  {Bode}, and {Wang}}}]{Maetal99}
\bibinfo{author}{\bibfnamefont{C.-P.} \bibnamefont{{Ma}}},
  \bibinfo{author}{\bibfnamefont{R.~R.} \bibnamefont{{Caldwell}}},
  \bibinfo{author}{\bibfnamefont{P.}~\bibnamefont{{Bode}}}, \bibnamefont{and}
  \bibinfo{author}{\bibfnamefont{L.}~\bibnamefont{{Wang}}},
  \bibinfo{journal}{Astrophys.J.Lett.} \textbf{\bibinfo{volume}{521}},
  \bibinfo{pages}{L1} (\bibinfo{year}{1999}), \eprint{arXiv:astro-ph/9906174}.

\bibitem[{\citenamefont{{Kallosh} et~al.}(2002)\citenamefont{{Kallosh},
  {Linde}, {Prokushkin}, and {Shmakova}}}]{Kalloshetal02}
\bibinfo{author}{\bibfnamefont{R.}~\bibnamefont{{Kallosh}}},
  \bibinfo{author}{\bibfnamefont{A.}~\bibnamefont{{Linde}}},
  \bibinfo{author}{\bibfnamefont{S.}~\bibnamefont{{Prokushkin}}},
  \bibnamefont{and}
  \bibinfo{author}{\bibfnamefont{M.}~\bibnamefont{{Shmakova}}},
  \bibinfo{journal}{Phys.Rev.D} \textbf{\bibinfo{volume}{66}},
  \bibinfo{pages}{123503} (\bibinfo{year}{2002}),
  \eprint{arXiv:hep-th/0208156}.

\bibitem[{\citenamefont{{Caldwell} and {Linder}}(2005)}]{CaldLind05}
\bibinfo{author}{\bibfnamefont{R.~R.} \bibnamefont{{Caldwell}}}
  \bibnamefont{and} \bibinfo{author}{\bibfnamefont{E.~V.}
  \bibnamefont{{Linder}}}, \bibinfo{journal}{Phys.Rev.Lett.}
  \textbf{\bibinfo{volume}{95}}, \bibinfo{pages}{141301}
  (\bibinfo{year}{2005}), \eprint{arXiv:astro-ph/0505494}.

\bibitem[{\citenamefont{Abrahamse et~al.}()\citenamefont{Abrahamse, Barnard,
  Bozek, and Yashar}}]{albrechtgroup}
\bibinfo{author}{\bibfnamefont{A.}~\bibnamefont{Abrahamse}},
  \bibinfo{author}{\bibfnamefont{M.}~\bibnamefont{Barnard}},
  \bibinfo{author}{\bibfnamefont{B.}~\bibnamefont{Bozek}}, \bibnamefont{and}
  \bibinfo{author}{\bibfnamefont{M.}~\bibnamefont{Yashar}},
  \bibinfo{journal}{in preparation}.

\bibitem[{\citenamefont{{Linder}}(2003)}]{Linder03}
\bibinfo{author}{\bibfnamefont{E.~V.} \bibnamefont{{Linder}}},
  \bibinfo{journal}{Phys.Rev.Lett.} \textbf{\bibinfo{volume}{90}},
  \bibinfo{pages}{091301} (\bibinfo{year}{2003}),
  \eprint{arXiv:astro-ph/0208512}.

\end{thebibliography}

\end{document}